\documentclass{aa} 
\usepackage{txfonts}
\usepackage{newtxtext,newtxmath}

\usepackage[T1]{fontenc}
\usepackage{ae,aecompl}
\usepackage{booktabs}
\usepackage{xcolor}

\usepackage[lofdepth,lotdepth]{subfig}

\usepackage{graphicx}	
\usepackage{amsmath}	
\usepackage{amssymb}	
\usepackage{wasysym}	

\usepackage{etoolbox}
\makeatletter
\patchcmd\@combinedblfloats{\box\@outputbox}{\unvbox\@outputbox}{}{\errmessage{\noexpand patch failed}}
\makeatother
\usepackage[colorlinks]{hyperref}
\hypersetup{
 citecolor=blue}
\begin{document} 
\title{Fossil field decay due to nonlinear tides in massive binaries}

\author{
J. Vidal\inst{1,2},
D. C\'ebron\inst{2},
A. ud-Doula\inst{3},
E. Alecian\inst{4}
}

\institute{Department of Applied Mathematics, School of Mathematics, University of Leeds, Leeds, LS2 9JT, UK \\
\email{j.n.vidal@leeds.ac.uk}
         \and
            Universit\'e Grenoble Alpes, CNRS, ISTerre, F-38000 Grenoble
         \and
            Penn State Scranton, 120 Ridge View Drive, Dunmore, PA 18512, USA
         \and
            Universit\'e Grenoble Alpes, CNRS, IPAG, 38000 Grenoble, France
             }
\date{Received 10 April 2019 / Accepted 5 August 2019}
\authorrunning{J. Vidal et al.}
\titlerunning{Fossil field decay due to tides in massive binaries}

\abstract
{Surface magnetic fields have been detected in 5 to 10\% of isolated massive stars, hosting outer radiative envelopes. They are often thought to have a fossil origin, resulting from the stellar formation phase. Yet, magnetic massive stars are scarcer in (close) short-period binaries, as reported by the BinaMIcS (Binarity and Magnetic Interaction in various classes of Stars) collaboration.}
{Different physical conditions in the molecular clouds giving birth to isolated stars and binaries are commonly invoked. In addition, we propose that the observed lower magnetic incidence in close binaries may be due to nonlinear tides. Indeed, close binaries are probably prone to tidal instability, a fluid instability growing upon the equilibrium tidal flow via nonlinear effects. Yet,  stratified effects have hitherto been largely overlooked.}
{We theoretically and numerically investigate tidal instability in rapidly rotating, stably stratified fluids permeated by magnetic fields. We use the short-wavelength stability method to propose a comprehensive (local) theory of tidal instability at the linear onset, discussing damping effects. Then, we propose a mixing-length theory for the mixing generated by tidal instability in the nonlinear regime. We successfully assess our theoretical predictions against proof-of-concept, direct numerical simulations. Finally, we compare our predictions with the observations of short-period, double-lined spectroscopic binary systems.}
{Using new analytical results, cross-validated by a direct integration of the stability equations, we show that tidal instability can be generated by nonlinear couplings of inertia-gravity waves with the equilibrium tidal flow in short-period massive binaries, even against the Joule diffusion.  
In the nonlinear regime, a fossil magnetic field can be dissipated by the turbulent magnetic diffusion induced by the saturated tidal flows.}
{We predict that the turbulent Joule diffusion of fossil fields would occur in a few million years for several short-period massive binaries. 
Therefore, turbulent tidal flows could explain the observed dearth of some short-period magnetic binaries.}
\maketitle



\section{Introduction}
The magnetism of massive stars has sparked the interest of astronomers for a long time \citep{babcock1958catalog}. 
More recently, large spectropolarimetric surveys of these stars have been undertaken \citep{hubrig2014b,wade2015mimes,grunhut2016mimes}. 
They have detected surface magnetic fields in 5 to 10\% of pre-main sequence and main-sequence massive stars \citep[e.g.][]{alecian2017fossil,mathys2017ap}. In addition, a magnetic dichotomy has been evidenced between the strong magnetism of chemically peculiar A/B stars \citep[e.g.][]{auriere2007weak,sikora2018volume} and the ultra-weak magnetism of Vega-like stars \citep{lignieres2009first,petit2010rapid,petit2011detection,blazere2016discovery}.
The origin of these fields is unclear.
According to stellar evolution theory, massive stars host thick outer radiative envelopes, which are stably stratified in density. These envelopes are often assumed to be motionless in standard stellar models \citep[e.g.][]{kippenhahn1990stellar}. This severely challenges the classical dynamo mechanism \citep{parker1979cosmical}, which requires internal turbulent motions (for instance that is convection in low-mass stars). 
Some dynamo mechanisms have been proposed, such as relying on the convection of the innermost convective core \citep{brun2005simulations,featherstone2009effects} generating magnetic flux tubes rising buoyantly towards the surface \citep{macgregor2003magnetic,macdonald2004magnetic}, on differentially rotating flows \citep{spruit1999differential,spruit2002dynamo,braithwaite2006differential,jouve2015three} or on baroclinic flows \citep{simitev2017baroclinically}. 
However, the relevance of these mechanisms remains elusive and debated. 

The most accepted assumption is that magnetic fields in massive stars have a fossil origin \citep{borra1982magnetic,moss2001magnetic}, because they appear relatively stable over the observational period. The fields would be shaped in the stellar formation phase and survive into later stages of stellar evolution. The fossil theory is now well supported by the existence of magnetic configurations stable enough to survive over a stellar lifetime \citep{braithwaite2004fossil,braithwaite2006stable,reisenegger2009stable,duez2010relaxed,duez2010stability,akgun2013stability}.
Hence, the fossil theory may provide a unifying explanation for the magnetism of intermediate-mass stars \citep{braithwaite2017magnetic}.
However, the fossil hypothesis still suffers from several weaknesses. In particular, we may naively expect all massive stars to exhibit surface magnetic fields. This is not consistent with the observations \citep[e.g.][]{alecian2017fossil,mathys2017ap}. Moreover, the theory does not convincingly explain the observed magnetic bi-modality \citep[e.g.][]{auriere2007weak}. To solve these issues, different physical conditions in the star-forming regions are usually invoked \citep[e.g.][]{commercon2010protostellar,commercon2011collapse}.

An efficient way to assess this hypothesis is to survey close binaries. Although the formation of binaries is not well understood, we can reasonably assume that the two binary components were formed together, under similar physical conditions. Then, observing magnetic fields in the two components of a (short-period) binary system would provide constraints to disentangle initial condition effects from other possible physical effects.
The BinaMIcS (Binarity and Magnetic Interaction in various classes of Stars) collaboration \citep{alecian2014binamics,alecian2017fossil} surveyed short-period massive binaries, aiming at providing new constraints on the magnetic properties of massive stars. About 170 short-period, double-lined spectroscopic binary binary systems on the main-sequence have been analysed by the BinaMIcS collaboration (Alecian et al., in prep). They have typical orbital periods of $T_\text{orb} \leq 20$ days and a separation distance between the two components of $D \leq 1$ au. 

A magnetic incidence of about 1.5 \% has been measured in the BinaMIcS sample. This is much lower than what is typically found in isolated hot stars (see above). 
Therefore, radiative stars in short-period binary systems are apparently much less frequently magnetic than in isolated systems. This confirms the general trend observed in other studies, dedicated to intermediate-mass A-type stars \citep[e.g.][]{carrier2002multiplicity,mathys2017ap}. It also extends it to hotter and more massive stars. 
Note that magnetic fields have been mostly observed only in one of the two components of the close binaries \citep{alecian2017fossil}, with a notable exception in the \object{$\epsilon$ Lupi} system \citep{shultz2015detection}. 
If initial conditions were solely responsible for the presence of a fossil field, then we would naively expect fossil fields in the two components of a magnetic binary. This is clearly not a general trend. 
Thus, these puzzling observations defy the theories that are commonly invoked. They lead to scientific questions such as the following: is it due to formation
processes \citep{commercon2011collapse,schneider2016rejuvenation}, that exclude more magnetic fields in binaries than in single stars? Or is there any other mechanism in close binaries, responsible for relatively quick dissipation of magnetic fields?

An alternative scenario is to invoke mixing in radiative envelopes, that may dynamically dissipate the pervading fossil fields. Identifying mixing sources in radiative stars is a long standing issue \citep[see the review in][]{zahn2008instabilities}, because mixing also affects the transports of chemical elements and of angular momentum. 
Shear-driven turbulence, induced by the (expected) differential rotation of radiative envelopes \citep[e.g.][]{goldreich1967differential,rieutord2006dynamics}, has been largely investigated \citep[e.g.][]{zahn1974rotational,mathis2004shear,mathis2018anisotropic}. 

A more efficient mixing in short-period stellar binaries may be provided by tides. Indeed, short-periods binaries are strongly deformed \citep[e.g.][]{chandrasekhar1969ellipsoidal,lai1993ellipsoidal}. 
Tides proceed in two steps. Firstly, they generate a quasi-hydrostatic tidal bulge, known as the equilibrium tidal velocity field \citep{zahn1966marees,remus2012equilibrium}. This leads to angular momentum exchange between the orbital and spinning motions. Secondly, they induce dynamical tides \citep[e.g.][]{zahn1975dynamical,rieutord2010viscous}, that is waves propagating here within the radiative regions.
Radiative envelopes support the propagation of many waves that are continuously emitted by various mechanisms \citep[e.g.][]{gastine2008directa,gastine2008directb,mathis2014impact,edelmann2019three}. Among them, internal gravity waves \citep{dintrans1999gravito,mirouh2016gravito} do induce mixing processes in radiative regions \citep{schatzman1993transport,rogers2017chemical}.

However, the aforementioned tidal effects are only linear processes. 
They are certainly relevant for the weak tides observed in the solar system and in extra-solar planets \citep{ogilvie2009tidal}. Yet, they may be inefficient to modify fossil fields on their own. Moreover, nonlinear effects can significantly modify the outcome of the tidal response, and thereby the influence of tides on fossil fields. Indeed, the equilibrium tidal flow can be unstable against tidal instability in stars \citep[e.g.][]{rieutord2004evolution,le2010tidal,barker2013non,barker2013nonb,clausen2014elliptical,barker2016non,barker2016nonb,vidal2017inviscid,vidal2018magnetic}.
This fluid instability is the astrophysical version of the generic elliptical instability, which affects all rotating fluids with elliptically deformed streamlines  \citep{bayly1986three,pierrehumbert1986universal,waleffe1990three,gledzer1992instability,le2000three}.
The underlying physical mechanism is nonlinear triadic resonances between two waves and the background elliptical velocity \citep{kerswell2002elliptical}. 
Hence, in stellar interiors, the origin of tidal instability is a resonance between rotational waves and the underlying strain field responsible for the elliptic deformation, that is the equilibrium tidal flow.
The nonlinear saturation of tidal instability can exhibit a wide variety of nonlinear states in homogeneous fluids, such as space-filling small-scale turbulence \citep{le2017inertial,reun2019experimental} or even global mixing \citep{grannan2016tidally,vidal2018magnetic}. Interestingly, \citet{clausen2014elliptical} investigated the influence of compressibility on the stability limits of tidal instability in stars or planets. They showed that fluid compressibility has almost no effect on the onset of tidal instability.

Yet, the fate of tidal instability in stratified fluid interiors is poorly known. On the one hand, theoretical studies have shown that an axial density stratification, aligned with the spin angular velocity, has stabilising effects \citep{miyazaki1991axisymmetric,miyazaki1992three}. Moreover, in the equatorial region, radial stratification can either increase or decrease the growth rate of the instability \citep{kerswell1993elliptical,le2006thermo,cebron2013elliptical}. 
On the other hand, three-dimensional numerical simulations suggest that tidal instability is largely unaffected in stratified interiors, for a wide range of stratification \citep{cebron2010tidal,vidal2018magnetic}.
Therefore, a consistent global picture of tidal instability in stably stratified interiors is highly desirable. Indeed, this is a prerequisite to assess the astrophysical relevance of tidal instability for the stellar mixing in close massive binaries.

The present study has a twofold purpose. 
Firstly, we aim to propose a predictive global theory of tidal instability in idealised stratified interiors. Such a theory should accurately predict the onset of instability, reconciling within a single framework previous theoretical analyses \citep{miyazaki1992three,miyazaki1993elliptical,kerswell1993elliptical,le2006thermo} and numerical studies \citep{cebron2010tidal,le2018parametric,vidal2018magnetic}. Then, asymptotic predictions for the (nonlinear) tidal mixing, as found numerically in \citet{vidal2018magnetic}, must be obtained to carry out the astrophysical extrapolation. 
Secondly, we aim to propose a new physical scenario of turbulent Joule diffusion of fossil fields, compatible with the observed lower magnetic incidence in short-period massive binaries as analysed by the BinaMIcS collaboration (Alecian et al., in prep.). 
The paper is organised as follows. In Sect. \ref{sec:model}, we present the idealised model. In Sect. \ref{sec:onset}, we investigate the linear regime of tidal instability in stratified interiors. In Sect. \ref{sec:mixing}, we develop a mixing-length theory of the (turbulent) tidal mixing, which is compared with proof-of-concept simulations. Then, we attempt to propose a novel scenario for close binaries in Sect. \ref{sec:discussion}, which is applied to short-period binary systems analysed by the BinaMIcS collaboration. Finally, we end the paper with a conclusion in Sect. \ref{sec:ccl} and outline some perspectives.

\section{Formulation of the problem}
\label{sec:model}
	\subsection{Assumptions}
The full astrophysical problem is rather complex. Hence, we consider an idealised model, for which numerical simulations can be conducted and compared with theory. We describe here the main assumptions, as they will be used throughout the paper. Our model retains the essential features to study tidal instability: rotation, stratification, magnetic fields and a tidally deformed geometry. 

We consider a primary self-gravitating body of mass $M_1$ and volume $\mathcal{V}$, filled with an electrically conducting and rotating fluid. 
Radiative fluid envelopes are expected to undergo differential rotation \citep{goldreich1967differential}, for instance provided by the contraction occurring during the pre-main-sequence phase or baroclinic torques \citep{busse1981eddington,busse1982problem,rieutord2006dynamics}. However, differential rotation tends to be smoothed out by hydromagnetic effects \citep[e.g.][]{moss1992magnetic}. In particular, differential rotation may sustain magneto-rotational instability, ultimately leading a state of solid-body rotation \citep{arlt2003differential,rudiger2013astrophysical,rudiger2015angular} on a few Alfv\'en timescales \citep{jouve2015three}. 
Consequently, we assume that the radiative envelope is uniformly rotating.

Then, the primary is orbited by a companion star of mass $M_2$. We investigate here only short-period, non-coalescing binaries. 
Due to the interplay between rotation and gravitational effects, the shape of each binary component departs from the spherical geometry. We do not seek here the mutual tidal interactions between the primary and the secondary. 
Indeed, at the leading order, the primary (or the secondary) is a triaxial ellipsoid in solid-body rotation \citep[e.g.][]{chandrasekhar1969ellipsoidal,lai1993ellipsoidal}, as obtained by modelling the other component by a point-mass companion. Therefore, for the sake of simplicity, we treat the secondary as a point mass for the orbital dynamics \citep[e.g.][]{hut1981tidal,hut1982tidal}. 

The secondary rises an equilibrium tide \citep{zahn1966marees,remus2012equilibrium} on the fluid primary, with a typical equatorial amplitude denoted $\beta_0$.
An initially eccentric binary system, with non-synchronised rotating components, evolves towards an orbital configuration characterised by a circular orbit and, ultimately, the system will be synchronised \citep{hut1981tidal,hut1982tidal}. 
For weakly elliptic orbits, \citet{nduka1971roche} showed that the ellipsoidal distortion $\beta_0$ points toward the tidal companion at the leading order. 
\citet{vidal2017inviscid} also showed that weak orbital eccentricities have little effects on the internal fluid dynamics of the primary (at the leading order in the eccentricity). 
Thus, we assume that the binary system is circularised (or weakly eccentric), with an equatorial bulge aligned with the orbital companion. 

Then, we consider only the leading-order component of the tidal potential, associated with the asynchronous tides
\citep{ogilvie2014tidal}. The fluid spin and orbital angular velocities are coplanar and aligned in the inertial frame. Note that this is the expected equilibrium state of the system \citep[e.g.][]{chandrasekhar1969ellipsoidal}. The other tidal components, for instance obliquity tides, are mainly responsible for additional fluid instabilities that are superimposed on tidal instability \citep[e.g.][]{kerswell1993instability}. They can be neglected in a first attempt.

Within the fluid primary, diffusive effects appear at the second order for tidal instability, in the absence of significant surface diffusive effects at a free boundary \citep{rieutord1992ekman,rieutord1997ekman}. 
Hence, we assume that the fluid has a uniform kinematic viscosity $\nu$, a radiative (thermal) diffusivity $\kappa_T$ \citep{kippenhahn1990stellar} and a magnetic diffusivity $\eta=1/(\mu_0 \sigma)$, where $\sigma$ is the electrical conductivity and $\mu_0$ the magnetic permeability of free space. 
Finally, \citet{clausen2014elliptical} showed that compressibility has almost no effect on tidal instability. Therefore, we model density variations departing from the isentropic profile within the Boussinesq approximation \citep{spiegel1960boussinesq}. 

\subsection{Governing equations}
\begin{figure}
	\centering
	\includegraphics[width=0.35\textwidth]{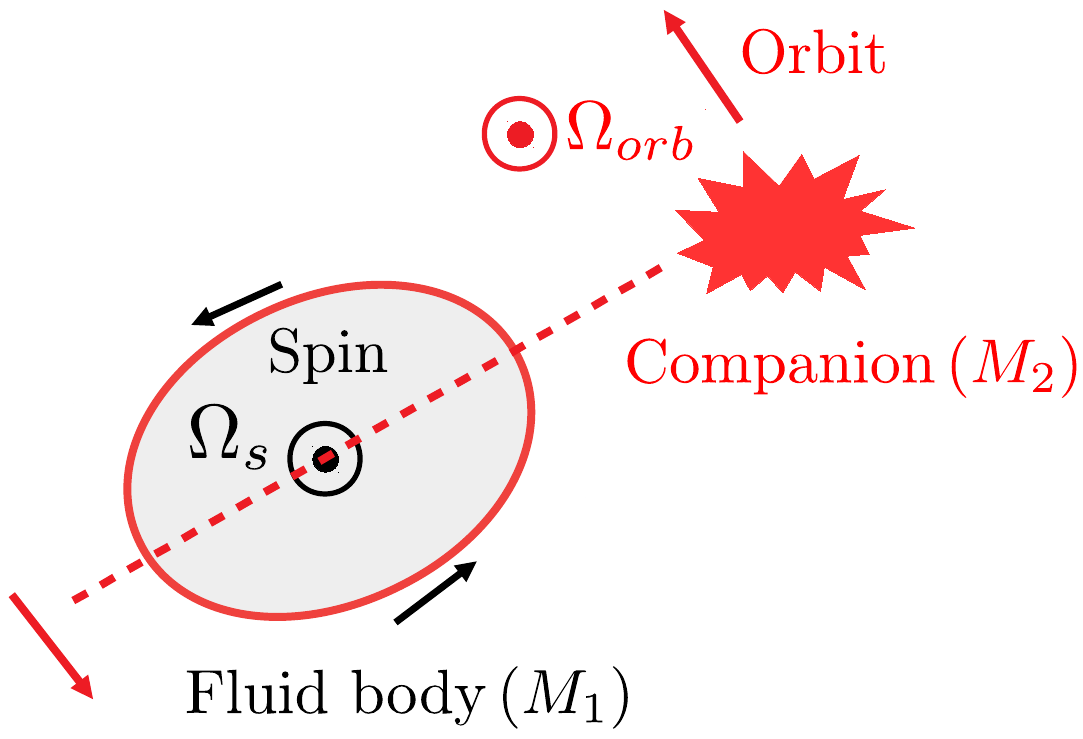}
	\caption{Sketch of idealised orbital configuration between primary body of mass $M_1$ and secondary one of mass $M_2$. View from above in the inertial frame. Coplanar and aligned spin and orbital angular velocities $[\Omega_\text{s},\Omega_\text{orb}]$.}
	\label{fig:orbit}
\end{figure}

The radiative star is modelled as a tidally deformed, uniformly rotating and stably stratified fluid domain in the Boussinesq approximation. The fluid domain, of typical density $\rho_M = M_1/\mathcal{V}$, is rotating at the angular velocity $\Omega_\text{s}$ in the inertial frame. 
The orbital configuration is illustrated in Fig. \ref{fig:orbit}.
The orbital angular velocity in the inertial frame is denoted $\Omega_\text{orb} \, \boldsymbol{1}_z$, with $\Omega_\text{orb} \neq \Omega_\text{s}$ for a non-synchronised orbit. 
In the central frame, in which the boundary shape is stationary, the outer boundary $\partial \mathcal{V}$ of the fluid domain describes an ellipsoid \citep[e.g.][]{chandrasekhar1969ellipsoidal,lai1993ellipsoidal}. Its mathematical expression in Cartesian coordinates $(x, y, z)$ is
\begin{equation}
	\left ( \frac{x}{a} \right )^2 + \left ( \frac{y}{b} \right )^2 + \left ( \frac{z}{c} \right )^2 = 1,
\end{equation}
where $(a,b,c)$ are the semi-axes. The equatorial ellipticity is defined by $\beta_0 = |{a}^2 - {b}^2|/({a}^2+{b}^2)$.

In the following, we work in dimensionless variables. To do so, we choose a typical radius $R$ of the fluid domain as unit of length, $\Omega_\text{s}^{-1}$ as a unit of time, $\Omega_\text{s}^2 R/(\alpha_T g_0)$ as unit of the temperature with $g_0$ a typical value of the gravity field at the outer boundary and $\alpha_T$ the thermal expansion coefficient (at constant pressure). For the magnetic field, we choose $\Omega_\text{s} R \sqrt{\rho_M \mu_0}$ as typical unit. 
We also introduce the dimensionless orbital frequency $\Omega_0 = \Omega_\text{orb}/\Omega_\text{s}$. 
The dimensionless variables are the velocity field $\boldsymbol{v}$, the temperature field $T$, the magnetic field $\boldsymbol{B}$ and the gravity field $\boldsymbol{g}$. They are written without $^*$, to distinguish them from their dimensional counterparts $[\boldsymbol{v}^*, T^*, \boldsymbol{B}^*, \boldsymbol{g}^*]$. 
The field variables, at the position $\boldsymbol{r}$ and time $t$, are governed in the rotating central frame by momentum, energy and induction equations. 
They read
\begin{subequations}
\label{eq:UTBtot}
\allowdisplaybreaks
\begin{align}
	\frac{\partial \boldsymbol{v}}{\partial t} &= - (\boldsymbol{v} \boldsymbol{\cdot} \nabla ) \, \boldsymbol{v} - 2 \Omega_0 \, \boldsymbol{1}_z \times \boldsymbol{v} -\nabla (P+P_m) + Ek \, \boldsymbol{\nabla}^2 \boldsymbol{v} \label{eq:Utot} \\
	&- T \boldsymbol{g} + (\boldsymbol{B} \cdot \nabla) \, \boldsymbol{B}, \nonumber\\
	\frac{\partial T}{\partial t} &= - (\boldsymbol{v} \boldsymbol{\cdot} \nabla) \, T + \frac{Ek}{Pr} \nabla^2 T + \mathcal{Q}, \label{eq:Ttot}\\
	\frac{\partial \boldsymbol{B}}{\partial t} &= \boldsymbol{\nabla} \times (\boldsymbol{v} \times \boldsymbol{B}) + Em \, \boldsymbol{\nabla}^2 \boldsymbol{B}, \label{eq:Btot} \\
	\boldsymbol{\nabla} \boldsymbol{\cdot} \boldsymbol{v} &= \boldsymbol{\nabla} \boldsymbol{\cdot} \boldsymbol{B} = 0,
\end{align}
\end{subequations}
with $P$ the hydrostatic pressure (including centrifugal effects), $P_m = |\boldsymbol{B}|^2/2$ the magnetic pressure, $\mathcal{Q}$ a heat source term and $\boldsymbol{g} = - \nabla \Phi_0$ the (imposed) gravity field in the Boussinesq approximation. 
In governing equations (\ref{eq:UTBtot}), we have introduced as dimensionless numbers the Ekman number $Ek=\nu/(\Omega_\text{s} R^2)$, the Prandtl number $Pr=\nu/\kappa_T$, the magnetic Prandtl number $Pm=\nu/\eta$ and the magnetic Ekman number $Em =Ek/Pm$. Typical values are given in Table \ref{table:DimNumbers} for stellar interiors. The latter are characterised by weakly diffusive conditions (that is $Ek, Ek/Pr, Ek/Pm \ll 1$). This regime will greatly simplify the analysis of tidal instability.

\begin{table}
	\centering
	\caption{Typical values of dimensionless numbers for stellar interiors. CZ: stellar convective zones, e.g. in the Sun \citep{charbonneau2014solar}.  RZ: (rapidly) rotating radiative zones \citep[e.g.][]{rieutord2006dynamics}.}
	{\small
		\begin{tabular}{lccc}
			\hline\hline
			Number & Symbol & \textsc{CZ} & \textsc{RZ} \\[1ex]
			\hline
			Ekman & $Ek$ & $10^{-16}$ & $10^{-18}$\\[1ex]
			Prandtl & $Pr$ & $10^{-6}$ & $10^{-6}$ \\[1ex]
			Magnetic Prandtl & $Pm$ & $10^{-6}$ & $10^{-6}$ \\[1ex]
			Magnetic Ekman & $Em$ & $10^{-10}$ & $10^{-12}$ \\[1ex]
			Brunt-V\"ais\"al\"a & $N_0/\Omega_\text{s}$ & 0 & $0-100$ \\[1ex]
			Lehnert & $Le$ & $10^{-5}$ & $\leq 10^{-4}$ \\[1ex]
			\hline
		\end{tabular}
	}
	\label{table:DimNumbers}
	\tablefoot{Note that $N_0=0$ in convective envelopes. The order of magnitude of the Lehnert number in RZ has been estimated from the typical values for the scarce short-period magnetic binaries given in Table \ref{table:databinamics2}.}
\end{table}

We do not directly solve full equations (\ref{eq:UTBtot}). Indeed, a reference ellipsoidal state is always first established, on which tidal instability grows upon and nonlinearly saturates. 
We expand the field variables as perturbations (not necessarily small) around a steady reference ellipsoidal basic state $[\boldsymbol{U}_0, T_0] (\boldsymbol{r})$ (detailed in Section \ref{subsec:refellipsoid}). Thus, the dimensionless nonlinear governing equations for the perturbations $[\boldsymbol{u}, \Theta] (\boldsymbol{r},t)$ and the magnetic field $\boldsymbol{B}(\boldsymbol{r},t)$ are
\begin{subequations}
\label{eq:UTBpert}
\allowdisplaybreaks
\begin{align}
	\frac{\mathrm{d} \boldsymbol{u}}{\mathrm{d} t} &+ (\boldsymbol{u} \boldsymbol{\cdot} \nabla) \, \boldsymbol{u} = - (\boldsymbol{u} \boldsymbol{\cdot} \nabla) \, \boldsymbol{U}_0 - 2 \Omega_0 \, \boldsymbol{1}_z \, \times \boldsymbol{u} -\nabla (p + P_m) \label{eq:Upert}  \\
	{} & + Ek \, \boldsymbol{\nabla}^2 \boldsymbol{u} - \Theta \boldsymbol{g} + (\boldsymbol{B}\boldsymbol{\cdot} \nabla) \, \boldsymbol{B} , \nonumber \\
	\frac{\mathrm{d} \Theta}{\mathrm{d} t} &+ (\boldsymbol{u} \boldsymbol{\cdot} \nabla) \, \Theta = - (\boldsymbol{u} \boldsymbol{\cdot} \nabla) \, T_0 + \frac{Ek}{Pr} \nabla^2 \Theta, \label{eq:Tpert} \\
	\frac{\partial \boldsymbol{B}}{\partial t} &+ \boldsymbol{\nabla} \times (\boldsymbol{B} \times \boldsymbol{u} ) =  \boldsymbol{\nabla} \times \left ( \boldsymbol{U}_0 \times \boldsymbol{B} \right ) + Em \, \boldsymbol{\nabla}^2 \boldsymbol{B}, \label{eq:Bpert} \\
	\boldsymbol{\nabla} \boldsymbol{\cdot} \boldsymbol{u} &= \boldsymbol{\nabla} \boldsymbol{\cdot} \boldsymbol{B} = 0, \ \, \ \boldsymbol{B}(\boldsymbol{r},t=0) = \boldsymbol{B}_0(\boldsymbol{r}),
\end{align}
\end{subequations}
with ${\mathrm{d}}/{\mathrm{d}t} = \partial/\partial t + (\boldsymbol{U}_0 \boldsymbol{\cdot} \nabla)$ the material derivative along the basic flow $\boldsymbol{U}_0$, $p$ the hydrodynamic pressure and  $\boldsymbol{B}_0(\boldsymbol{r})$ the (initial) fossil field. For the proof-of-concept simulations introduced in Sect. \ref{sec:mixing}, the equations will be supplemented by appropriate boundary conditions.

	\subsection{Reference ellipsoidal configuration}
	\label{subsec:refellipsoid}
We consider a steady reference equilibrium state, for which isopycnals coincide with isopotentials of the gravitational potential $\Phi_0$ (including centrifugal force, self-gravity and tides). This assumption is consistent with compressible models \citep{lai1993ellipsoidal}.
Hence, we assume that the background temperature profile $T_0(\boldsymbol{r})$ and the gravity field $\boldsymbol{g}$, solutions of equations (\ref{eq:Utot})-(\ref{eq:Ttot}), are in barotropic equilibrium (for a well-chosen $\mathcal{Q}$) such that $\boldsymbol{g} \times (\nabla T_0) = \boldsymbol{0}$. 
We do not consider the baroclinic part, which is known to increase the growth rate of tidal instability in the equatorial plane \citep{kerswell1993elliptical,le2006thermo}. In the nonlinear regime, a baroclinic state would certainly sustain tidal turbulence in stellar interiors. However, we focus here on the less favourable configuration for the growth of tidal instability (that is barotropic stratification). This choice is also consistent with the assumed uniform rotation of the fluid. Indeed, baroclinic torques are known to sustain differential rotation \citep[e.g.][]{busse1981eddington,busse1982problem,rieutord2006dynamics}.
Moreover, considering barotropic stratification is a relevant assumption when the isopycnals move sufficiently fast to keep track of the rotating tidal potential \citep{le2018parametric}.  This situation is expected when stratification is large enough in amplitude compared with the differential rotation $\Omega_\text{s} - \Omega_\text{orb}$ between the spin and the orbit. 

To characterise the strength of stratification, we introduce the dimensional (local) Brunt-V\"ais\"al\"a frequency $N$ in the reference state. In dimensional variables, the latter is defined by
\begin{equation}
	N^2 = - \alpha_T \, \boldsymbol{g}^* \boldsymbol{\cdot} \nabla T_0^*.
	\label{eq:Brunt}
\end{equation}
The fluid ellipsoid is assumed to be entirely stably stratified in density ($N^2 > 0$). The exact profiles in stellar interiors depend on the stellar internal processes. However, we want to compare analytical and numerical computations, which cannot be done for arbitrary profiles. 
Thus, we assume that the dimensionless total gravitational potential is quadratic, such that
\begin{equation}
	\Phi_0 = \left(\frac{x}{a}\right)^2 + \left(\frac{y}{b}\right)^2 + \left(\frac{z}{c}\right)^2.
	\label{eq:PhigravRef}
\end{equation}
Then, we consider the (dimensionless) reference temperature in barotropic equilibrium $T_0 = (N^2_0/\Omega^2_\text{s}) \, \Phi_0$, with $N_0$ a typical value of the Brunt-V\"ais\"al\"a frequency at the outer boundary. 
For intermediate-mass stars with $M_1=3 M_\odot$ (where $M_\odot$ is the solar mass), a typical value is $N_0 \sim 10^{-3} \, \text{s}^{-1}$ \citep[e.g.][]{rieutord2006dynamics}, and typical values of $\Omega_\text{s}^{-1}$ range between 1 and 100 days \citep{mathys2017ap}. This give the estimate $0 \leq N_0/\Omega_\text{s} \leq 100$ in radiative stars. Hence, a barotropic reference configuration is a reasonable starting assumption.

The ellipsoid is initially permeated by an fossil magnetic field $\boldsymbol{B}_0 (\boldsymbol{r})$ (in dimensionless form).
To measure its relative strength (with respect to rotation), we introduce the (dimensionless) Lehnert number \citep{lehnert1954magnetohydrodynamic}
\begin{equation}
	Le = \frac{B_0^*}{\Omega_\text{s} R \sqrt{\rho_M \mu_0}},
	\label{eq:Lehnert}
\end{equation}
where $B_0^*$ is the typical (dimensional) strength of the fossil field.
The Lehnert number is the ratio of the Alfv\'en and rotational velocities. 
When $Le \ll 1$, the Coriolis force dominates the Lorentz force in momentum equation (\ref{eq:Utot}). The regime $Le \ll 1$ is encountered in many magnetic stars (Table \ref{table:DimNumbers}). In the Sun, a typical value is $Le\sim 10^{-5}$ \citep{charbonneau2014solar}. For the scarce magnetic binaries which have been observed, the median field strength is $B_0^* \sim 1$ kG (see also values in Table \ref{table:databinamics2}). This gives the typical values $Le \leq 10^{-5} - 10^{-4}$. Hence, we focus on the regime $Le \ll 1$ in the following.

Finally, the orbital configuration drives the equilibrium tidal flow \citep[e.g.][]{remus2012equilibrium}. For non-synchronised orbits ($\Omega_0 \neq 1$), its leading-order flow components in the central frame are \citep[e.g.][]{cebron2012elliptical,vidal2017inviscid}
\begin{equation}
	\boldsymbol{U}_0(\boldsymbol{r}) = (1-\Omega_0) \left [-(1+\beta_0) y \, \boldsymbol{1}_x + (1-\beta_0)x \, \boldsymbol{1}_y \right ].
	\label{eq:U0TDEI_Orb}
\end{equation}
with $[\boldsymbol{1}_x,\boldsymbol{1}_y,\boldsymbol{1}_z]$ the unit Cartesian vectors. 
This is an exact incompressible solution of hydrodynamic momentum equation (\ref{eq:Utot}) without diffusion. Moreover, it satisfies the non-penetration condition $\boldsymbol{U}_0 \boldsymbol{\cdot} \boldsymbol{1}_n = 0$ at the boundary $\partial \mathcal{V}$, with $\boldsymbol{1}_n$ the unit outward normal vector. 
Note that basic flow (\ref{eq:U0TDEI_Orb}) is not rigorously a solution in the presence of an arbitrary magnetic field. Yet, the large-scale poloidal and toroidal components of $\boldsymbol{B}_0 (\boldsymbol{r})$ are unlikely to modify the equilibrium tidal flow in the weak field regime $Le \ll 1$ as often assumed \citep[e.g.][]{kerswell1993elliptical,kerswell1994tidal,mizerski2011influence}.

\section{Onset of tidal instability}
\label{sec:onset}
We present the stability analysis of tidal instability at the linear onset. Firstly, we outline the general stability method in Sect. \ref{subsec:wkb}. In Sect. \ref{subsec:theory}, we carry out an asymptotic analysis to get physical insight of the instability mechanism. The latter mechanism is compared and validated with the (numerical) solutions of the full stability equations in Sect. \ref{subsec:swan}, without making any prior assumption.  Finally, we discuss the (laminar) magnetic diffusive effects in Sect. \ref{subsec:feedbackB0}.

	\subsection{Short-wavelength perturbations}
	\label{subsec:wkb}
In the absence of any driving mechanism, a fossil field $\boldsymbol{B}_0$ slowly decays on the Ohmic diffusive timescale $(\Omega_\text{s} \, Ek/Pm)^{-1}$. This time is larger than the typical lifetime of the least massive stars on the main-sequence \citep[e.g.][]{braithwaite2017magnetic}. 
However, equations (\ref{eq:UTBpert}) support the propagation of several waves
in rotating radiative interiors, characterised by $Le \ll 1$ and $N_0/\Omega_\text{s} \gg 1$ (see Table \ref{table:DimNumbers}). 
They can strongly modify the dynamic evolution of radiative envelopes. 
These waves are continuously emitted and, in the presence of tides, they can be nonlinearly coupled with the equilibrium tidal velocity field $\boldsymbol{U}_0$ to sustain tidal instability. Tidal instability is intrinsically a local (small scale) instability \citep{kerswell2002elliptical,cebron2012elliptical,barker2013non,barker2013nonb}, but it also exists in global models \citep[e.g.][]{kerswell1993elliptical,grannan2016tidally,vidal2018magnetic}. 
The global stability analysis is beyond the scope of the present study. However, in the diffusionless regime, three-dimensional global perturbations of small enough length scales are excited \citep[e.g.][]{vidal2017inviscid}, such that they are not affected by the boundary. 
Hence, we can advantageously investigate the growth of tidal instability in stellar interiors by performing a local stability analysis. In Appendix \ref{appendix:WKB}, we have extended the general local stability theory to account for combined magnetic and buoyancy effects within the Boussinesq approximation. 

We focus on the subsonic wave spectrum (low Mach number), made of MAC (Magneto-Archimedean-Coriolis) waves. 
Indeed, high-frequency sonic waves are not involved in tidal (elliptical) instability \citep{le2001phd}, though they may be coupled with tides \citep[e.g. in coalescing binary neutron stars, see][]{weinberg2016growth}.
The properties of MAC waves have already been outlined elsewhere \citep[e.g.][]{gubbins1987magnetohydrodynamics,mathis2011low,sreenivasan2017damping}. 
Note that they have global bounded counterparts, known as Magneto-Archimedean-Coriolis (MAC) modes. The global modes are briefly discussed in Appendix \ref{appendix:MAC}. 
The wave spectrum is bounded from below by slow Magneto-Coriolis (MC) waves, sustained by the Lorentz and Coriolis forces with an angular frequency $\omega_i$ scaling as $|\omega_i|\propto Le^2$ \citep[e.g.][]{malkus1967hydromagnetic,labbe2015magnetostrophic}. The spectrum is bounded from above by internal gravity waves (modified by rotation), with an angular frequency $|\omega_i|\leq N_0/\Omega_\text{s}$ for strong stratification \citep{friedlander1982internalb}. In-between, the spectrum exhibits Coriolis waves \citep{greenspan1968theory,backus2017completeness} and inertial-gravity (or gravito-inertial) waves \citep[e.g.][]{dintrans1999gravito,mirouh2016gravito}.

In the weak field limit $Le \ll 1$, magnetic effects are negligible (at the leading order) on inertial waves \citep{schmitt2010magneto,labbe2015magnetostrophic} and gravito-inertial ones, as outlined in Appendix \ref{appendix:MAC}. 
Moreover, only nonlinear couplings of inertial and gravito-inertial waves can trigger tidal instability with significant growth rates to overcome the leading-order diffusive effects \citep{kerswell1993elliptical,kerswell1994tidal}, as we confirm in Appendix \ref{appendix:MixedMHDResonances}. This behaviour is also supported by local simulations \citep{barker2013non} and global dynamo numerical simulations in homogeneous \citep{cebron2014tidally,reddy2018turbulent} and stratified fluids \citep{vidal2018magnetic}. They showed that even a dynamo magnetic field only barely modifies the hydrodynamic tidal flows.
Therefore, we can consider only the hydrodynamic Boussinesq stability equations in relevant the weak field regime $Le \ll 1$. The leading-order magnetic effect is the Joule diffusion. From the values given in Table \ref{table:DimNumbers}, diffusive effects can be a priori neglected at the first order of the stability theory. We will confirm that this assumption is relevant by reintroducing them in Sect. \ref{subsec:feedbackB0}.

We seek three-dimensional local perturbations, solution of linearised hydrodynamic equations (\ref{eq:UTBpert}). To do so, 
we consider short-wavelength (WKB) perturbations \citep{lifschitz1991local,friedlander1991instability}. 
They are local (plane-wave) perturbations, barely sensitive to the ellipsoidal boundary $\partial \mathcal{V}$, advected along the fluid trajectories $\boldsymbol{X}(t)$ of $\boldsymbol{U}_0(\boldsymbol{r})$. Given basic tidal flow (\ref{eq:U0TDEI_Orb}), the Eulerian three-dimensional perturbations are expressed as
\begin{equation}
	[ \boldsymbol{u}, \Theta ] (\boldsymbol{r},t) = [ \widehat{\boldsymbol{u}}, \widehat{\Theta} ] (\boldsymbol{r},t) \, \exp(\mathrm{i} \boldsymbol{k} (t) \boldsymbol{\cdot} \boldsymbol{r}), \ \, \ |\boldsymbol{k}(t)| = |\boldsymbol{k}_0|,
\end{equation}
where $\boldsymbol{k}(t)$ is the local wave vector with the initial value $\boldsymbol{k}_0$. The local stability equations are solved in Lagrangian formulation, yielding the following ordinary differential equations (in dimensionless form)
\begin{subequations}
	\allowdisplaybreaks
	\label{eq:WKBstar}
	\begin{align}
		\frac{\mathrm{D} \boldsymbol{X}}{\mathrm{D} t} &= \boldsymbol{U}_0 (\boldsymbol{X}), \ \, \ \boldsymbol{X}(0) = \boldsymbol{X}_0, \label{eq:WKBstarX} \\
		\frac{\mathrm{D} \boldsymbol{k}}{\mathrm{D} t} &= - \left ( \boldsymbol{\nabla} \boldsymbol{U}_0 \right )^\top \boldsymbol{k}, \ \, \ \boldsymbol{k}(0) = \boldsymbol{k}_0, \label{eq:WKBstarK} \\
		\frac{\mathrm{D} \widehat{\boldsymbol{u}}}{\mathrm{D} t} &= \left[ \left( \frac{2\, \boldsymbol{k} \boldsymbol{k}^{T}}{|\boldsymbol{k}|^2} - \boldsymbol{I} \right) \boldsymbol{\nabla} \boldsymbol{U}_0 +2 \left( \frac{ \boldsymbol{k} \boldsymbol{k}^{T}}{|\boldsymbol{k}|^2}-\boldsymbol{I} \right) \Omega_{0} \, \boldsymbol{1}_z \times \right ]\,  \widehat{\boldsymbol{u}} \nonumber \\ 
		{} &- \, \widehat{\Theta}  \left(  \boldsymbol{I} -\frac{ \boldsymbol{k} \boldsymbol{k}^{T}}{|\boldsymbol{k}|^2} \right) \boldsymbol{g}, \label{eq:WKBstarU} \\
		\frac{\mathrm{D} \widehat{\Theta}}{\mathrm{D} t} &= - (\widehat{\boldsymbol{u}} \boldsymbol{\cdot} \nabla) \, T_0 , \label{eq:WKBstarT}
	\end{align}
\end{subequations}
with ${\mathrm{D}}/{\mathrm{D} t}$ the Lagrangian time derivative. The solenoidal condition $\widehat{\boldsymbol{u}} \cdot \boldsymbol{k} = 0$ is satisfied as long as it holds at the initial time, that is $\widehat{\boldsymbol{u}} (0) \cdot \boldsymbol{k}_0 = 0$ in the Lagrangian description. Equations (\ref{eq:WKBstar}) do depend on the fluid trajectories $\boldsymbol{X}(t)$, because the gravity field $\boldsymbol{g}$ is spatially varying. 

Equations (\ref{eq:WKBstar}) are ordinary differential equations along the Lagrangian trajectories $\boldsymbol{X}(t)$. They are also independent of the magnitude of $\boldsymbol{k}_{0}$ in the diffusionless limit. We follow \citet{le2000three}, by restricting the initial wave vector to the unit spherical surface 
\begin{equation}
	\boldsymbol{k}_0 = \sin(\theta_0) \cos(\phi_0) \, \boldsymbol{1}_x + \sin(\theta_0) \sin(\phi_0) \, \boldsymbol{1}_y + \cos(\theta_0) \, \boldsymbol{1}_z,
\end{equation}
where $\phi_0 \in [0, 2\pi]$ is the longitude and $\theta_0 \in [0, \pi]$ is the colatitude between the spin axis $\boldsymbol{1}_z$ and the wave vector $\boldsymbol{k}_{0}$. In practice, equations (\ref{eq:WKBstar}) are integrated from a range of wave vectors $\boldsymbol{k}_0$ and initial positions $\boldsymbol{X}_0$ within the reference ellipsoidal domain. 
The basic state is unstable against short-wavelength perturbations if
\begin{equation}
	\lim \limits_{t\to\infty} \left ( |\widehat{\boldsymbol{u}} (t, \boldsymbol{X}_0, \boldsymbol{k}_0)| + |\widehat{\Theta} (t, \boldsymbol{X}_0, \boldsymbol{k}_0)| \right ) = \infty.
\end{equation}
Then, we determine the maximum (diffusionless) growth rate $\sigma$ as the fastest growing solution for all initial conditions, that is the largest Lyapunov exponent. This gives a sufficient condition for instability.

\subsection{Asymptotic analysis}
\label{subsec:theory}
Equilibrium tidal flow (\ref{eq:U0TDEI_Orb}) admits analytical periodic fluid trajectories $\boldsymbol{X}(t)$ and wave vectors $\boldsymbol{k} (t)$, solution of equations (\ref{eq:WKBstarX})-(\ref{eq:WKBstarK}). 
To get physical insight of the instability mechanism, we carry out an asymptotic analysis in the limit $\beta_0 \leq 1$. We expand all quantities ($\boldsymbol{X}, \boldsymbol{k}, \widehat{\boldsymbol{u}}, \widehat{\Theta}$) in successive powers of $\beta_0$ \citep[see technical details in][]{le2000three}.

\subsubsection{Triadic (nonlinear) couplings}
It has been recognised for a long time that tidal instability is a parametric instability in homogeneous \citep[e.g.][]{bayly1986three,waleffe1990three} and stratified fluids \citep[e.g.][]{miyazaki1992three,miyazaki1993elliptical}. The instability is due to triadic interactions between pairs of waves that are coupled with the underlying tidal flow (\ref{eq:U0TDEI_Orb}). 
At the leading asymptotic order ($\beta_0 = 0$), a necessary condition for a parametric tidal instability in rotating fluids is given by the resonance condition in the central frame \citep{kerswell2002elliptical,vidal2017inviscid}
\begin{equation}
	|\omega_i - \omega_j + \delta| = 2 \, |1-\Omega_0|,
	\label{eq:resonancecond1}
\end{equation}
where $[\omega_i,\omega_j]$ are the angular frequencies of two free waves and $\delta$ a small detuning parameter, allowing for imperfect resonances \citep{kerswell1993elliptical,le2000three,lacaze2004elliptical,vidal2017inviscid}. The latter are due to either diffusive or topographic effects ($\delta \to 0$ for diffusionless fluids and weakly deformed spheres $\beta_0 \ll 1$). 
Detuning effects are negligible in the astrophysical regime (almost diffusionless and with $\beta_0 \ll 1$). 
Note that the case of synchronised orbits, characterised by $\Omega_0 = 1$ (in average), is forbidden by condition (\ref{eq:resonancecond1}). Synchronised orbits must be treated separately (see Appendix \ref{appendix:ldei}).

Among the aforementioned resonances, sub-harmonic resonances are characterised by $\omega_i=-\omega_j$. 
Then, resonance condition (\ref{eq:resonancecond1}) reduces (in the diffusionless regime) to \begin{equation}
	|\omega_i| = |1-\Omega_0|,
	\label{eq:resonancecond2}
\end{equation}
which is a necessary condition for sub-harmonic tidal instability. 
Sub-harmonic resonances have been found to be the most unstable in homogeneous fluids \citep{kerswell1993elliptical,kerswell1994tidal,le2000three,vidal2017inviscid}, that is generating the largest growth rates. 

We are now in a position to survey the possible nonlinear couplings of the different types of waves that can trigger tidal instability. The waves can be combined in several ways to satisfy the resonance condition in non-synchronised systems. For instance, from condition (\ref{eq:resonancecond2}), tidal instability traditionally exists in the orbital range $-1 \leq \Omega_0 \leq 3$ when it involves Coriolis waves \citep[e.g.][]{craik1989stability,le2000three,vidal2017inviscid}. 
We investigate in depth the coupling of hydrodynamic waves, postponing the discussion of hydromagnetic waves (unimportant for the present problem) to Appendix \ref{appendix:MixedMHDResonances}. 

\subsubsection{Hydrodynamic waves at the parametric resonance}
\begin{figure}
	\centering
	\includegraphics[width=0.4\textwidth]{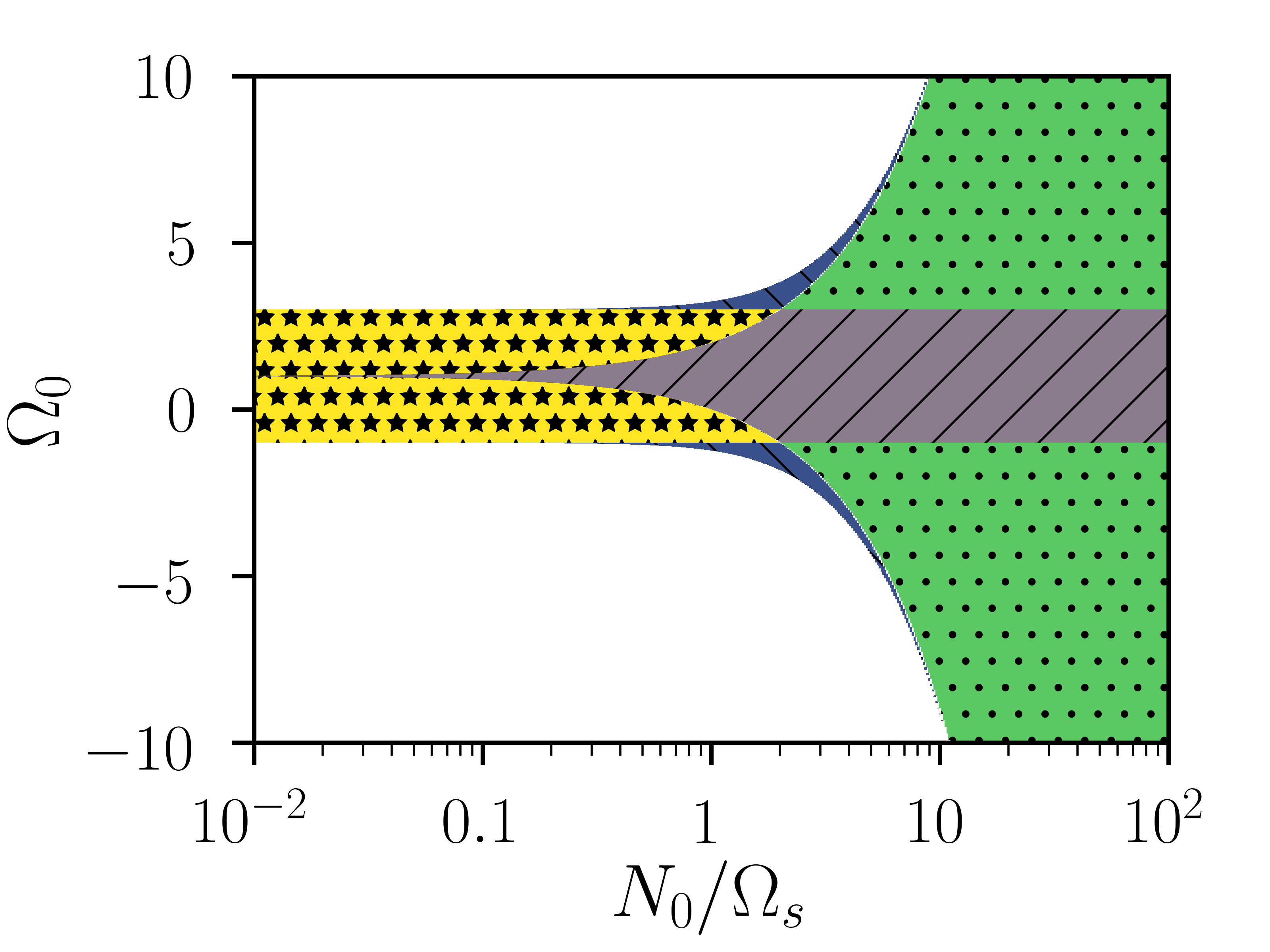}
	\caption{Domains of existence of sub-harmonic resonances (\ref{eq:resonancecond2}), as a function of $\Omega_0 = \Omega_\text{orb}/\Omega_\text{s}$ and $N_0/\Omega_\text{s}$. In white regions, no waves can satisfy sub-harmonic resonance condition (\ref{eq:resonancecond2}). Stars (yellow area): hyperbolic waves $\mathcal{H}_1$. Right slash (purple area): hyperbolic waves $\mathcal{H}_2$. Dots (green area): elliptic waves $\mathcal{E}_1$. Back slash (blue area): elliptic waves $\mathcal{E}_2$. The classical allowable region of tidal instability (for neutral fluids) is $-1 \leq \Omega_0 < 3$. wave-like  domains $[\mathcal{H}_1, \mathcal{H}_2]$ are illustrated in Fig. \ref{fig:Characteristic_wavesH}. Similarly, wave-like  domains $[\mathcal{E}_1, \mathcal{E}_2]$ are illustrated in Fig. \ref{fig:Characteristic_wavesE}.}
	\label{fig:fig2}
\end{figure}

The behaviour of tidal instability is intrinsically associated with the properties of the waves involved in the triadic resonances. 
The wave-like equation, introduced in Appendix \ref{appendix:MAC}, is a mixed hyperbolic-elliptic partial differential equation. In the general case, a wave-like hyperbolic domain coexists with an elliptic domain, in which the waves are evanescent.
At the leading asymptotic order $\beta_0=0$, the characteristic curve delimiting the two domains is \citep{friedlander1982internala} 
\begin{equation}
	z^2 + s^2 \omega_i^2/ (\omega_i^2 - 4) = \omega_i^2/(N_0/\Omega_\text{s})^2,
	\label{eqDynamo:characteristic_curveFried}
\end{equation}
with $s$ the cylindrical radius. 
The hydrodynamic wave spectrum is divided in two main regimes.
On the one hand, we have inertial waves modified by gravity, called inertial-gravity waves and denoted $\mathcal{H}$. They have hyperbolic turning surfaces given by equation (\ref{eqDynamo:characteristic_curveFried}). They are sub-divided in two families given by
\begin{subequations}
	\label{eqDynamo:H1H2_waves}
	\allowdisplaybreaks
	\begin{align}
	\mathcal{H}_1:& \ (N_0/\Omega_\text{s})^2 < \omega_i^2 < 4, \\
	\mathcal{H}_2:& \ 0 < \omega_i^2 < \min \, [4, (N_0/\Omega_\text{s})^2].
	\end{align}
\end{subequations}
On the other hand, we have gravity waves modified by rotation, called gravito-inertial waves and denoted $\mathcal{E}$. They have ellipsoidal turning surfaces given by equation (\ref{eqDynamo:characteristic_curveFried}). They are also divided in two families, characterised by
\begin{subequations}
	\label{eqDynamo:E1E2_waves}
	\allowdisplaybreaks
	\begin{align}
	\mathcal{E}_1:& \ 4 < \omega_i^2 < (N_0/\Omega_\text{s})^2, \\
	\mathcal{E}_2:& \ \max \, [4, (N_0/\Omega_\text{s})^2] < \omega_i^2 < 4 + (N_0/\Omega_\text{s})^2.
	\end{align}
\end{subequations}
These properties are quite general, because equation (\ref{eqDynamo:characteristic_curveFried}) depends solely on the reference state. Therefore, both global modes \citep[e.g.][]{dintrans1999gravito} and local waves propagating upon this reference configuration exhibit this distinction.

\begin{figure*}
	\centering
	\begin{tabular}{cc}
		\includegraphics[width=0.3\textwidth]{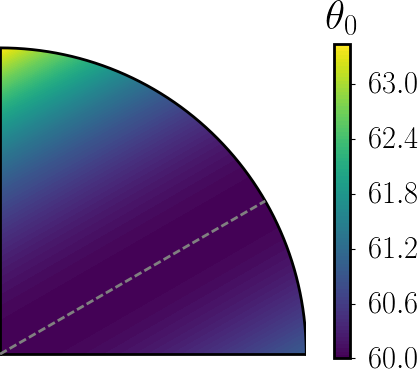} &
		\includegraphics[width=0.3\textwidth]{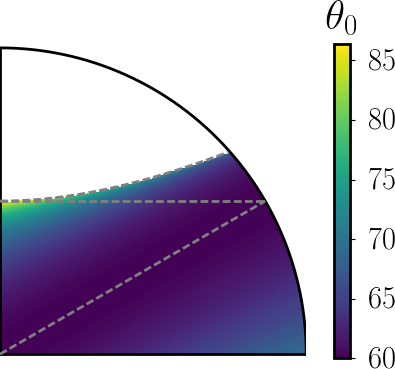} \\
		$\mathcal{H}_1$ & $\mathcal{H}_2$ \\
	\end{tabular}
	\caption{Wave-like domains and colatitude $\theta_0$ (degrees) for waves with hyperbolic turning surfaces $\mathcal{H}$ satisfying sub-harmonic resonance condition (\ref{eq:resonancecond2}). \emph{Left panel}: $\mathcal{H}_1$ wave: $\Omega_0 = 0, N_0/\Omega_\text{s} = 0.5$. \emph{Right panel}: $\mathcal{H}_2$ wave: $\Omega_0 = 0, N_0/\Omega_\text{s} = 2$. Dashed grey hyperbolic curve is given by equation (\ref{eqDynamo:characteristic_curveFried}). Tilted dashed grey line is the asymptotic curve given by $\cos \theta_0 = |1-\Omega_0|/2$. Waves at the sub-harmonic resonance disappear along the polar axis when $z \leq |1-\Omega_0|/(N_0/\Omega_\text{s})$.}
	\label{fig:Characteristic_wavesH}
\end{figure*}

\begin{figure*}
	\centering
	\begin{tabular}{cc}
		\includegraphics[width=0.3\textwidth]{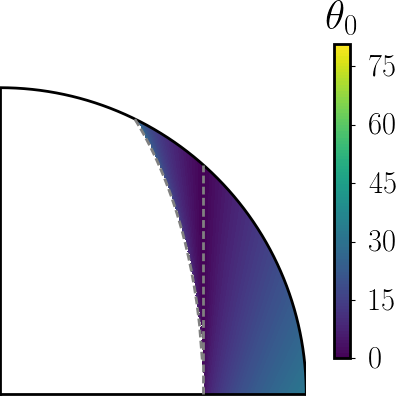} &
		\includegraphics[width=0.3\textwidth]{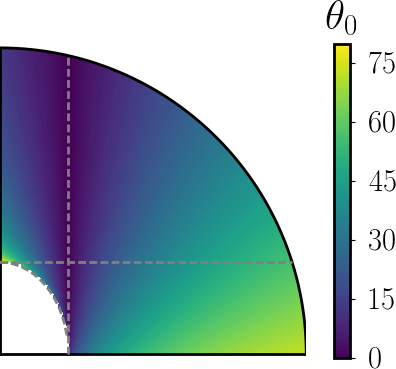} \\
		 $\mathcal{E}_1$ & $\mathcal{E}_2$ \\
	\end{tabular}
	\caption{Wave-like domains and colatitude $\theta_0$ (degrees) for waves with ellipsoidal turning surface $\mathcal{E}$ satisfying sub-harmonic resonance condition (\ref{eq:resonancecond2}). \emph{Left panel}: $\mathcal{E}_1$ wave: $\Omega_0 = 3.4, N_0/\Omega_\text{s} = 2$. \emph{Right panel}: $\mathcal{E}_2$ wave: $\Omega_0 = 4, N_0/\Omega_\text{s} = 10$. Dashed grey ellipsoidal curve is given by equation (\ref{eqDynamo:characteristic_curveFried}). Vertical dashed grey line is the asymptotic curve given by $s = (\sqrt{|1-\Omega_0|^2 - 4})/(N_0/\Omega_\text{s})$, where $s$ is the cylindrical radius from the spin axis. Waves satisfying the sub-harmonic resonance condition disappear along the polar axis when $z \leq |1-\Omega_0|/(N_0/\Omega_\text{s})$.}
	\label{fig:Characteristic_wavesE}
\end{figure*}

The different families of waves satisfying sub-harmonic resonance condition (\ref{eq:resonancecond2}) are illustrated in Fig. \ref{fig:fig2}. 
This is the main result of the linear theory, as this provides a necessary (and sufficient, see below) condition for the existence of tidal instability (in both global and local models). Two kinds of tidal instability can be obtained, depending on the value of key parameter $\Omega_0$.
At the leading asymptotic order, we have obtained a general expression for sub-harmonic resonance condition (\ref{eq:resonancecond2}) in the local theory, which can be written as
\begin{multline}
	\cos^2 (\theta_0) = \frac{\widetilde{\omega}+\widetilde{N}_0^2 r_0^2 \, [(\widetilde{N}_0^2 r_0^2 - \widetilde{\omega}) \cos^2 \alpha_0 -\cos(2 \alpha_0)]}{\widetilde{\omega}^2+\widetilde{N}_0^2 r_0^2 \, [\widetilde{N}_0^2 r_0^2-2 \widetilde{\omega} \cos (2 \alpha_0)]} \\
	+ \frac{2 \sqrt{\omega_1 \, [\widetilde{\omega} \, (1-\widetilde{N}_0^2 z_0^2)+\widetilde{N}_0^2 r_0^2 - 1]}}{\widetilde{\omega}^2+\widetilde{N}_0^2 r_0^2 \, [\widetilde{N}_0^2 r_0^2 - 2 \widetilde{\omega} \cos (2 \alpha_0)]},
\label{eqDynamo:Resonance_TDEI_zplane}
\end{multline}
with the background rotation $\widetilde{\Omega}_0=\Omega_0/(1-\Omega_0)$, $\widetilde{N}_0 = (N_0/\Omega_\text{s})/|1-\Omega_0|$, $\widetilde{\omega}=4(1+\widetilde{\Omega}_0)^2$, the initial position $\boldsymbol{X}_0 = (x_0, z_0)^\top = r_0 \, (\sin \alpha_0, \cos \alpha_0)^\top$ where $r_0$ is the initial radius and $\omega_1=\widetilde{N}_0^4 r_0^4 \cos^2 \alpha_0 \sin^2 \alpha_0$. 
The associated wave-like domains and colatitude angles $\theta_0$ are shown in Figures \ref{fig:Characteristic_wavesH} and \ref{fig:Characteristic_wavesE}. 

The classical allowable range of the instability in homogeneous fluids is $-1 \leq \Omega_0 \leq 3$ \citep{craik1989stability,le2000three}. 
Within this range, the sub-harmonic condition involves only $\mathcal{H}$ waves, as shown in Fig. \ref{fig:fig2}. For neutral stratification ($N_0 = 0$), they are inertial waves $\mathcal{H}_1$, propagating in the whole fluid cavity \citep{friedlander1982internala}. They have the colatitude angle at the sub-harmonic resonance \citep{le2000three}
\begin{equation}
	2\, \cos (\theta_0) = \frac{1}{1+\widetilde{\Omega}_0} =  1-\Omega_0.
	\label{eq:theta0_H12_N0}
\end{equation}
This remains valid in weakly stratified fluids (that is $N_0/\Omega_s \ll 1$). Indeed, $\mathcal{H}_1$ waves are only slightly modified by buoyancy. They still propagate in the whole fluid domain, as shown in Fig. \ref{fig:Characteristic_wavesH} (left panel). In addition, their colatitude angle $\theta_0$ is slightly larger than the value predicted by formula (\ref{eq:theta0_H12_N0}) on the polar axis.

When $N_0/\Omega_\text{s} \geq 1$, $\mathcal{H}_1$ waves morph into $\mathcal{H}_2$ waves made of inertia-gravity waves. These waves are strongly modified by buoyancy. Their wave-like domain is confined between hyperboloids, as shown in Fig. \ref{fig:Characteristic_wavesH} (right panel). Outside the hyperboloid volume, these waves at the sub-harmonic resonance are evanescent (in global models). The characteristic curve delimiting the wave-like and evanescent domains, given by equation (\ref{eqDynamo:characteristic_curveFried}), is hyperbolic. Along the rotation axis, local waves at the sub-harmonic resonance do not propagate in the evanescent regions for vertical positions $z_c$ satisfying
\begin{equation}
	|z_c| \geq \frac{|1-\Omega_0|}{N_0/\Omega_\text{s}}.
	\label{eqDynamo:z_H2E2waves}
\end{equation}
This shows that axial stratification has a stabilising effect. 

This behaviour is responsible for an equatorial trapping of the waves in the other directions at the sub-harmonic resonance. 
Indeed, the hyperbolic wave-like domain, bounded by  (\ref{eqDynamo:characteristic_curveFried}), converges towards the conical volume delimited by the asymptotic limit $\cos (\theta_c) = |1-\Omega_0|/2$ \citep{friedlander1982internala}, where $\theta_c$ is the critical colatitude. This is exactly formula (\ref{eq:theta0_H12_N0}). 
Therefore, expression (\ref{eq:theta0_H12_N0}) also defines the position of the critical latitudes at which the waves at the sub-harmonic resonance have a group velocity orthogonal to the gravity field (here the radial direction at the leading order in $\beta_0$), that is a wave vector $\boldsymbol{k} \propto \boldsymbol{g}$. Hence, these specific waves are insensitive to stratification. We emphasise that the presence of stratification does not alter the position of the critical latitudes \citep{friedlander1982internala,friedlander1982internalb}. When $|1-\Omega_0| \to 0$, the waves at the sub-harmonic resonance are equatorially trapped according to formula (\ref{eq:theta0_H12_N0}). 

The orbital range $\Omega_0 \leq -1$ and $\Omega_0 \geq 3$ is known as the forbidden zone. In this range, tidal instability must involve gravito-inertial waves $\mathcal{E}$ for the sub-harmonic mechanism, whatever the strength of stratification. Indeed, Fig. \ref{fig:fig2} clearly shows that the waves at the sub-harmonic resonance depend only on the value of the orbital frequency $\Omega_0$. 
When $N_0/\Omega_\text{s} \leq 1$, the sub-harmonic condition is never satisfied within this orbital range. Hence, no tidal instability is triggered. 

However, gravito-inertial waves $\mathcal{E}$  can be excited at the sub-harmonic resonance for strong stratification, typically $N_0/\Omega_\text{s} \gg 1$ when $|\Omega_0|$ increases.
Their critical characteristic surfaces, given by equation (\ref{eqDynamo:characteristic_curveFried}), are ellipsoidal. On the one hand, $\mathcal{E}_1$ gravito-inertial waves are trapped in a region that does not encompass the polar axis, as shown in Fig. \ref{fig:Characteristic_wavesE} (left panel). The minimum distance between the spin axis and the wave-like domain in the equatorial plane is given by \citep{friedlander1982internala}
\begin{equation}
x_c = \frac{\sqrt{|1-\Omega_0|^2 - 4}}{N_0/\Omega_\text{s}}.
\label{eqDynamo:s_E12waves}
\end{equation}
Therefore, the thickness of the wave-like domain increases when the ratio $N_0/\Omega_\text{s}$ increases. 
On the other hand, $\mathcal{E}_2$ waves at the sub-harmonic resonance are gravito-inertial waves, trapped in a region that excludes the central part of the fluid (right panel of Fig. \ref{fig:Characteristic_wavesE}). Along the polar axis, these waves do not propagate when $z$ is smaller than critical value (\ref{eqDynamo:z_H2E2waves}). The size of wave-like domain increases when the ratio $N_0/\Omega_\text{s}$ increases. In the limit $N_0/\Omega_\text{s} \to \infty$, these waves become almost pure internal gravity waves, propagating in the whole fluid domain at the sub-harmonic resonance. This situation has been investigated numerically in local models \citep{le2018parametric}, by assuming $\Omega_\text{s}=0$.

	\subsubsection{Asymptotic growth rate in the equatorial plane}
	\label{subsec:analyticalsigma}
At the next asymptotic order in $\beta_0$, we can obtain a concise explicit formula for the growth rate $\sigma$ of tidal instability, valid in the equatorial plane $z_0=0$. Dispersion relation (\ref{eqDynamo:Resonance_TDEI_zplane}) gives, for $\alpha_0 = \pi/2$ (after simplification),
\begin{equation}
	\sqrt{\widetilde{\omega} + \widetilde{N}_0^2 x_0^2} \, \cos (\theta_0) = \pm 1
	\label{eq:FZequat1}
\end{equation}
with $x_0 \leq 1$ the position of the initial trajectory $\boldsymbol{X}_0$ in the equatorial plane. In the particular case $\Omega_0=0$, equation (\ref{eq:FZequat1}) recovers equation (4.6) of \citet{le2006thermo}. 

Several configurations are possible, depending on the parameters. 
On the one hand, the LHS of equation (\ref{eq:FZequat1}) is purely imaginary when $-\widetilde{N}_0^2 x_0^2 > \widetilde{\omega}$, when stratification is unstably stratified (with $N_0^2/\Omega_\text{s}^2 \leq 0$). Then, a centrifugal instability grows upon the reference configuration, with a maximum (dimensionless) growth rate \citep[e.g.][]{le2006thermo}
\begin{equation}
	\frac{\sigma}{|1-\Omega_0|} = \sqrt{-\widetilde{N}_0^2 x_0^2 - \widetilde{\omega}}.
	\label{eq:sigma_centrifuge}
\end{equation} 
On the other hand, tidal instability is triggered when all terms in equation (\ref{eq:FZequat1}) are real.
Hence, no sub-harmonic instability is possible when $\widetilde{N}_0^2 x_0^2 < -3 - 4\widetilde{\Omega}_0 \, (2+\widetilde{\Omega}_0)$. This defines the forbidden zone of tidal instability in stably stratified fluids, at a given position $x_0$. For neutral fluids ($N_0 = 0$), we recover the classical allowable orbital range of tidal instability $-1 \leq\Omega_0 \leq 3$. Outside this range, we find that waves can be involved in triadic resonances in stratified fluids. Thus, (sub-harmonic) tidal instability could be triggered in stratified fluids when $\Omega_0 \leq -1$ and $\Omega_0 \geq 3$ (range known as the forbidden zone in neutral fluids). 
Then, the dimensionless growth rate in the equatorial plane is 
\begin{equation}
	\frac{\sigma}{|1-\Omega_0|} = \frac{(2 \widetilde{\Omega}_0+3)^2}{16 \, (1+\widetilde{\Omega}_0)^2+4\widetilde{N}_0^2 x_0^2} \beta_0.
	\label{eq:sigma_equat}
\end{equation}
Hence, the growth rate $\sigma$ is weakened by stratification when $\widetilde{N}_0^2 x_0^2$ increases. This effect was already discussed in the conclusion of \citet{le2006thermo}. They found that elliptical equipotentials are stabilising contrary to circular equipotentials. However, in this former case, their equation slightly differs from equation (\ref{eq:sigma_equat}). Actually, their formula is erroneous because we will confirm the validity of expression (\ref{eq:sigma_equat}) by direct numerical integration of the local stability equations (see below). Note also that equation (\ref{eq:sigma_equat}) does not recover equation (24) of \citet{cebron2013elliptical}, obtained in the limit of a buoyancy force of order $\beta_0$. In this limit, we recover their approximate formula (24) if we use their value for $\theta_0$, artificially set to its hydrodynamic value $\widetilde{\omega} \cos^2 \theta_0=1$ instead of its exact value given by expression (\ref{eq:FZequat1}).

\begin{figure}
	\centering
	\includegraphics[width=0.49\textwidth]{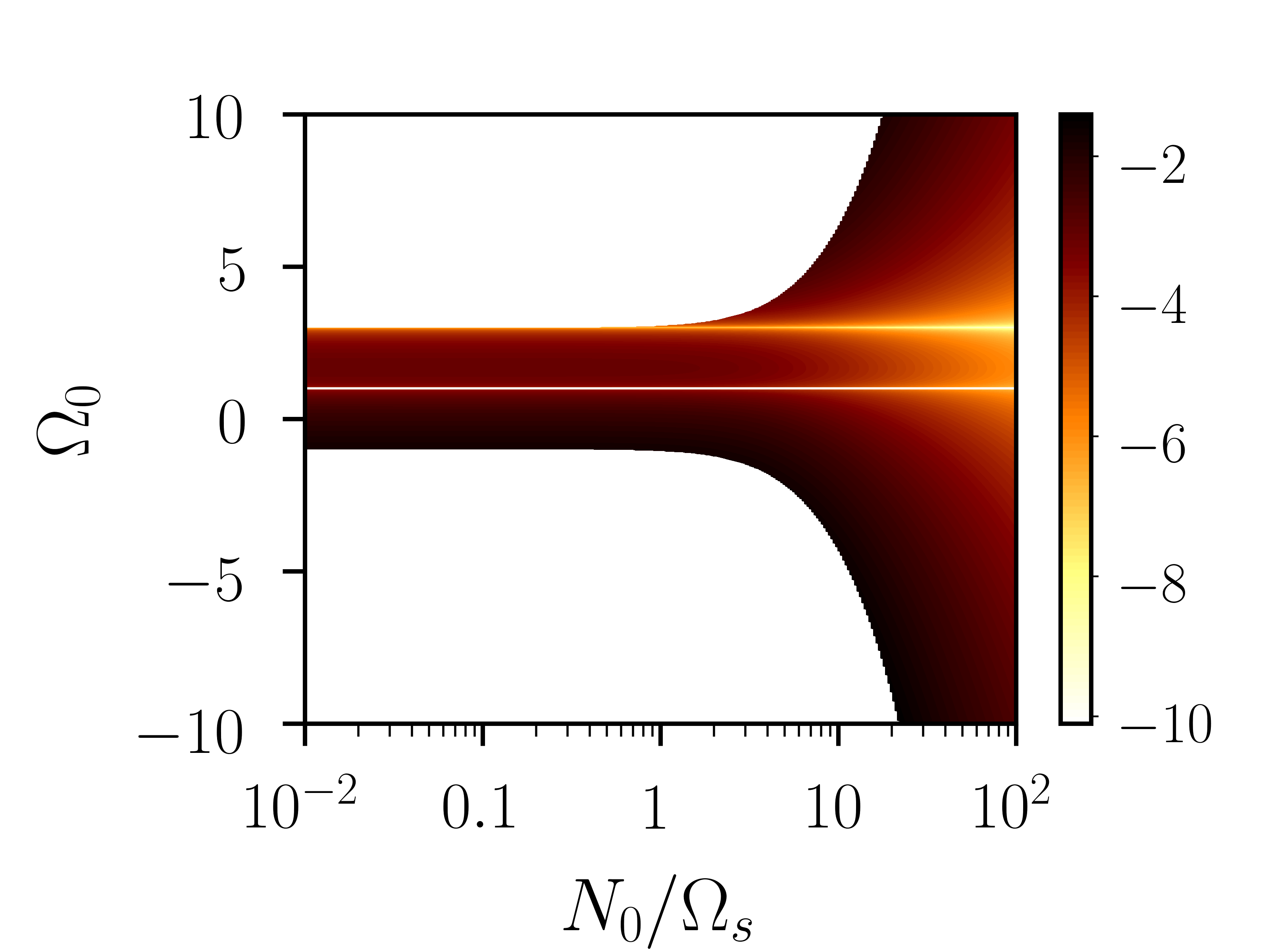}
	\caption{Growth rate $\sigma$ of tidal instability, predicted by formula (\ref{eq:sigma_equat}) in equatorial plane ($x_0=0.5,z_0=0$), as a function of $N_0/\Omega_\text{s}$ and $\Omega_0$. Colour bar shows the normalised ratio $\log_{10}(\sigma/\beta_0)$. White areas correspond to marginally stable areas. For neutral fluids, tidal instability is restricted to the allowable range $-1 \leq \Omega_0 \leq 3$ when $\beta_0 \ll 1$. When $\Omega_0=1$ (horizontal white line), the basic state is synchronised (see Appendix \ref{appendix:ldei}).}
	\label{fig:wkbequat}
\end{figure}

We show in Fig. \ref{fig:wkbequat} the maximum growth rate, computed from formula (\ref{eq:sigma_equat}), for different orbital configurations $\Omega_0$. Several points are worthy of comment. 
Firstly, tidal instability is excited in the equatorial region when $-1 \leq \Omega_0 \leq 3$ (in the diffusionless limit), that is in the classical orbital range of tidal instability \citep{le2000three}. This mechanism occurs for any realistic value of $N_0/\Omega_\text{s} \leq 100$ (see Table \ref{table:DimNumbers}). 
In this orbital range, the maximum growth rate is always obtained for neutral fluids ($N_0=0$), yielding the usual (dimensionless) growth rate \citep{le2000three}
\begin{equation}
	\frac{\sigma}{|1-\Omega_0|} = \frac{(2 \widetilde{\Omega}_0+3)^2}{16 \, (1+\widetilde{\Omega}_0)^2} \beta_0.
	\label{eq:sigma_equatN0}
\end{equation}
Secondly, outside the classical orbital range (in the forbidden zone), we unravel new tidal instabilities, triggered for large enough values of the Brunt-V\"ais\"al\"a frequency ($N_0/\Omega_\text{s} \gg 1$). Their growth rate can be larger than one in our dimensionless units (not shown), because their typical timescale is $N_0^{-1}$ (rather than $\Omega_\text{s}^{-1}$). Note that these sub-harmonic instabilities have been reported in local stratified simulations \citep{le2018parametric}.

Therefore, in the equatorial region, we have shown that barotropic stratification has (i) a destabilising effect within the usual forbidden zone ($\Omega_0 \leq -1$ and $\Omega_0 \geq 3$), and (ii) a stabilising effect when $-1 \leq \Omega_0 \leq 3$. However, we emphasise that a baroclinic state (that is $\boldsymbol{g} \times \nabla T_0 \neq \boldsymbol{0}$) has the opposite effect when $-1 \leq \Omega_0 \leq 3$ \citep{kerswell1993elliptical,le2006thermo}. 
This behaviour can be recovered by our asymptotic analysis, by assuming an imposed gravity field with a different equatorial ellipticity $\beta_1 \neq \beta_0$. For such a reference ellipsoidal configuration, formula (\ref{eq:sigma_equat}) becomes
\begin{equation}
	\frac{\sigma}{|1-\Omega_0|} = \frac{ ( 2 \widetilde{\Omega}_0+3 )^2}{16 \, ( 1+\widetilde{\Omega}_0 )^2 + 4 \, \widetilde{N}_0^2 x_0^2} \left |  \beta_0 + \widetilde{N}_0^2 x_0^2 \, \frac{\beta_0-\beta_1}{2\widetilde{\Omega}_0+3} \right |.
	\label{eq:sigma_equatbarocline}
\end{equation}
This corrects misprints in equation (D.1) of \citet{cebron2012elliptical}, obtained with a different unit of time. For circular isopotentials ($\beta_1=0$), formula (\ref{eq:sigma_equatbarocline}) clearly shows that the growth rate of tidal instability is enhanced in the equatorial plane. This is the configuration considered by \citet{kerswell1993elliptical} and \citet{le2006thermo}. Besides, equation (\ref{eq:sigma_equatbarocline}) recovers formula (4.7) of \citet{le2006thermo} in their particular case $\Omega_0=0$.

	\subsubsection{Along rotation axis}
	\label{subsec:analyticalsigma_polar}
Similarly, we can obtain an analytical formula along the axis of rotation. To do so, we consider initial fluid trajectories close to the spin axis (that is $s_0 = \beta_0 \ll 1$). Dispersion relation (\ref{eqDynamo:Resonance_TDEI_zplane}) simplifies along the polar axis into (with $\alpha_0 = 0$)
\begin{equation}
	\cos^2 (\theta_0) = \frac{1 - \widetilde{N}_0^2 z_0^2}{\widetilde{\omega} - \widetilde{N}_0^2 z_0^2}.
	\label{eq:FZzplane2}
\end{equation}
Condition (\ref{eq:FZzplane2}) shows that the forbidden zone of tidal instability coincides with the one for neutral fluid, that is $\Omega_0 \leq -1$ and $\Omega_0 \geq 3$. Outside this range, the asymptotic (dimensionless) growth rate is
\begin{equation}
	\frac{\sigma}{|1-\Omega_0|} = \frac{ ( 2 \widetilde{\Omega}_0 + 3 )^2 \left ( 1 - \widetilde{N}_0^2 z_0^2 \right )}{16 \, ( 1+\widetilde{\Omega}_0 )^2 - 4 \widetilde{N}_0^2 z_0^2} \beta_0.
	\label{eq:sigma_zplane}
\end{equation}
Formula (\ref{eq:sigma_zplane}) is identical to the diffusionless growth rate devised by \citet{miyazaki1993elliptical}, denoting $\widetilde{N}_0 z_0$ their local value of stratification. Hence, an axial stratification is uniformly stabilising along the polar axis.

\subsection{Numerical solutions in the orbital range $-1 \leq \Omega_0 \leq 3$}
\label{subsec:swan}

\begin{figure*}
	\centering
	\captionsetup[subfigure]{labelformat=empty}
	\subfloat[$\Omega_0 = -0.5$]{
		\begin{tabular}{ccc}
			\includegraphics[width=0.23\textwidth]{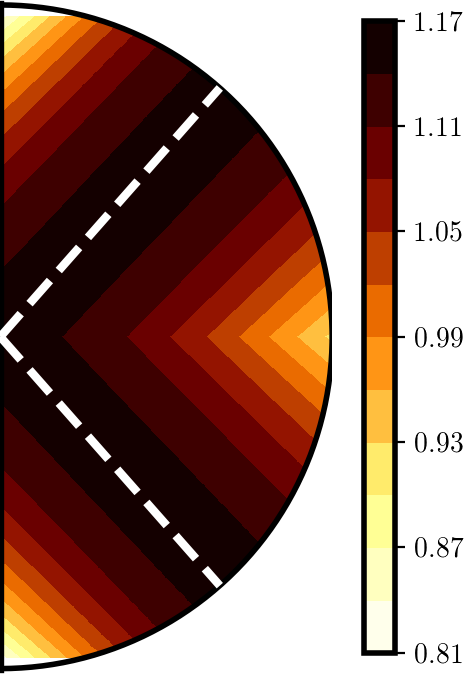} & 
            \includegraphics[width=0.23\textwidth]{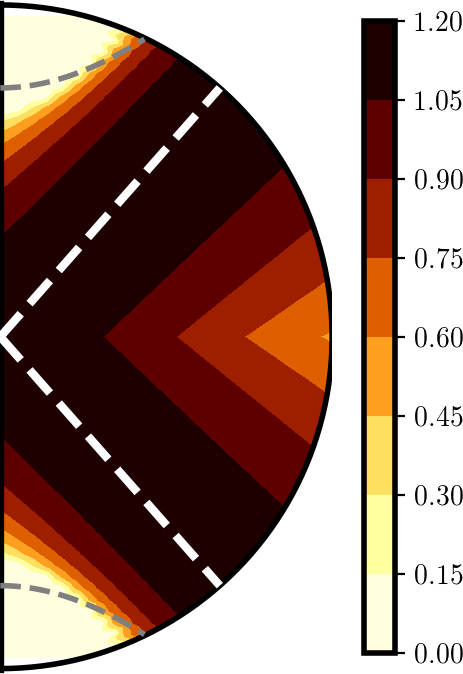} & 
            \includegraphics[width=0.23\textwidth]{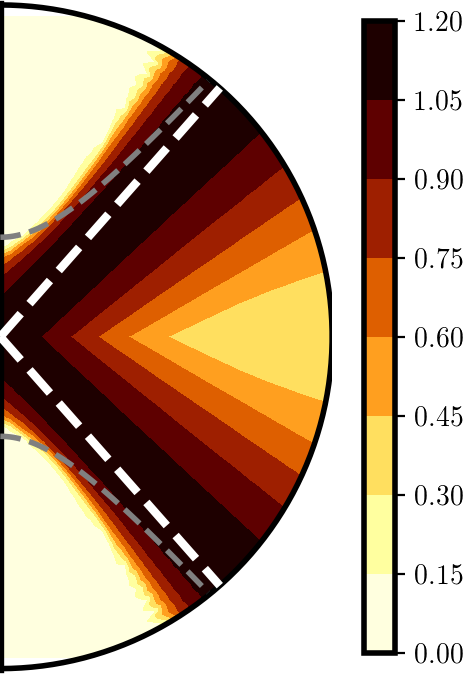} \\ 
			$N_0/\Omega_\text{s}=1$ ($\mathcal{H}_1$) & $N_0/\Omega_\text{s}=2$ ($\mathcal{H}_2$) & $N_0/\Omega_\text{s}=5$ ($\mathcal{H}_2$) \\
		\end{tabular}
	}
	\\
	\subfloat[$\Omega_0=0$]{
		\begin{tabular}{ccc}
            \includegraphics[width=0.23\textwidth]{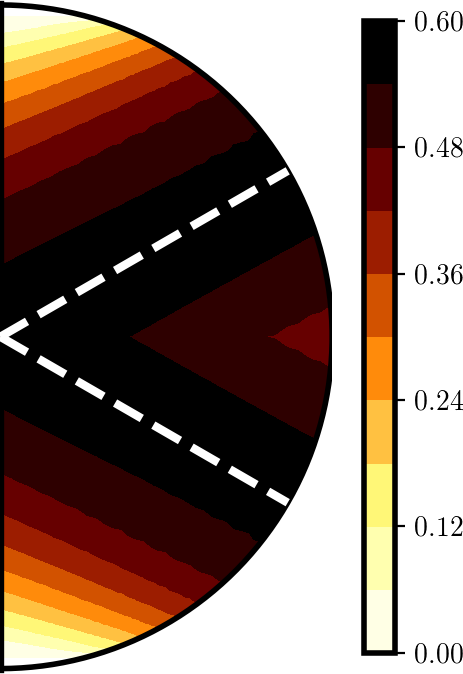} &
			\includegraphics[width=0.23\textwidth]{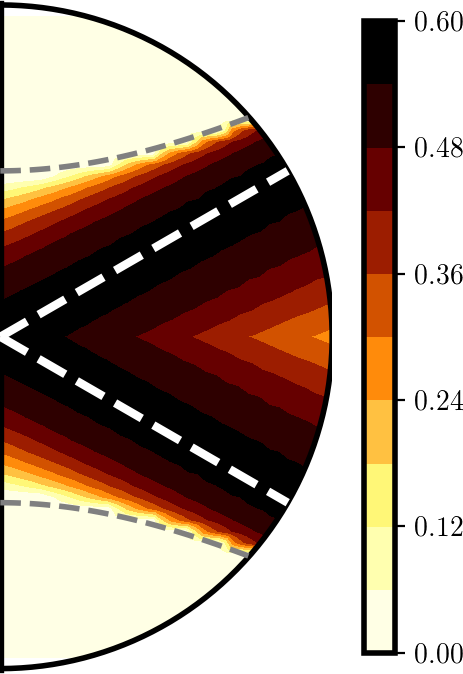} &
			\includegraphics[width=0.23\textwidth]{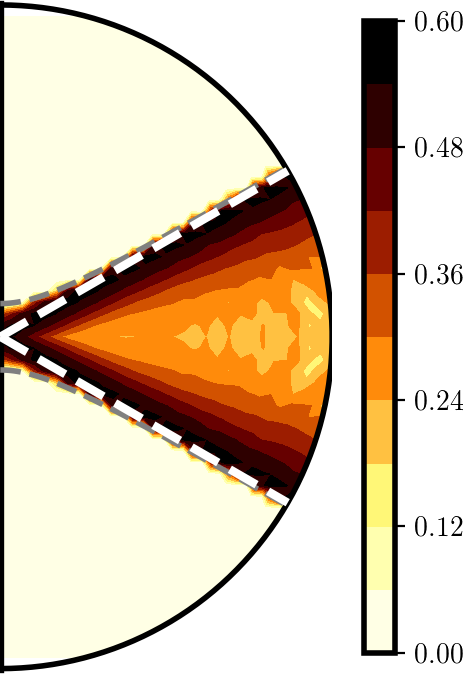} \\
			$N_0/\Omega_\text{s}=1$ ($\mathcal{H}_1$) & $N_0/\Omega_\text{s}=2$ ($\mathcal{H}_2$) & $N_0/\Omega_\text{s}=10$ ($\mathcal{H}_2$)\\
		\end{tabular}
	}
	\\
	\subfloat[$\Omega_0=0.5$]{
		\begin{tabular}{ccc}
            \includegraphics[width=0.23\textwidth]{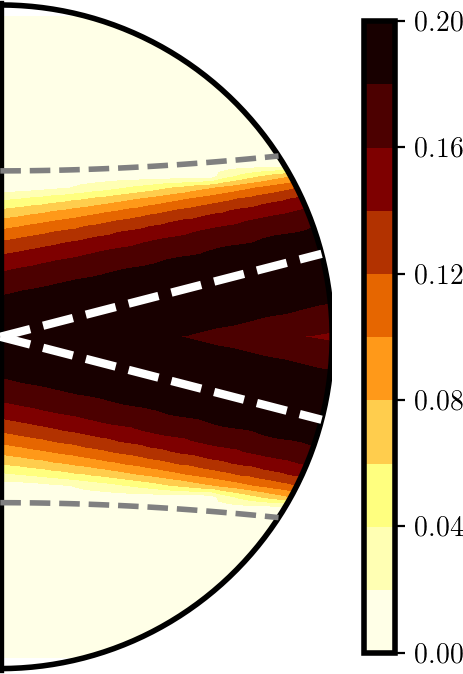} &
			\includegraphics[width=0.23\textwidth]{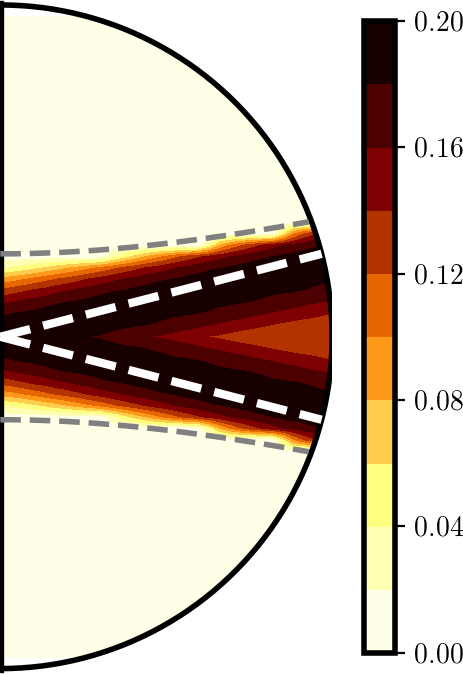} & 
			\includegraphics[width=0.23\textwidth]{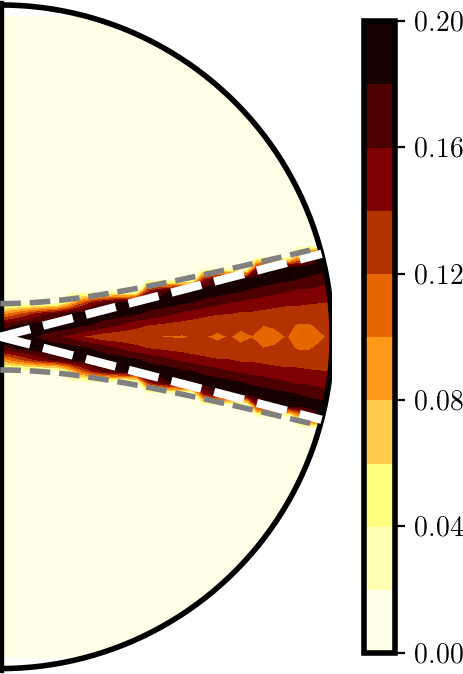} \\
			$N_0/\Omega_\text{s}=1$ ($\mathcal{H}_2$) & $N_0/\Omega_\text{s}=2$ ($\mathcal{H}_2$) & $N_0/\Omega_\text{s}=5$ ($\mathcal{H}_2$) \\
		\end{tabular}
	}
	\caption{Largest normalised growth rate $\sigma/\beta_0$ for several configurations, computed with SWAN for equatorial ellipticity $\beta_0=0.2$. Visualisations in a meridional section using the normalised axes $x/a$ and $z/c$, with $a=\sqrt{1+\beta_0}$, $b=\sqrt{1-\beta_0}$ and $c=1/(ab)$. White dashed lines, given by formula (\ref{eq:theta0_H12_N0}), show the critical latitudes on which the growth rate is maximum as predicted by (\ref{eq:sigma_equatN0}). For each case, the type of waves involved in parametric mechanism is specified between brackets. Dashed (grey) curves illustrate the domain of existence of $\mathcal{H}_2$ waves at the resonance (in the regime $\beta_0 \ll 1$).}
	\label{fig:SWAN_Wm05_0_05_2}
\end{figure*}

The previous asymptotic analysis shows that stable stratification ($N_0/\Omega_\text{s} \geq 0$) has indubitably a stabilising behaviour. 
In particular, axial stratification is responsible for a trapping of the instability in the equatorial region.
These observations agree with existing local analyses \citep{miyazaki1992three,miyazaki1993elliptical,kerswell1993elliptical,le2006thermo,cebron2012elliptical}. 
However, this is barely consistent with three-dimensional numerical simulations \citep{vidal2018magnetic},  showing that the growth rate at the onset is largely unaffected by stratification. To reconcile these approaches, we investigate the onset of tidal instability in the whole reference fluid domain.

To go beyond the analytical formulas in the equatorial plane and on the polar axis, we solve numerically local stability equations (\ref{eq:WKBstar}). To do so, we have used the local stability code SWAN \citep{vidal2017inviscid}. We have updated it to handle the general local stability equations, which are described in Appendix \ref{appendix:WKB}. Moreover, by solving numerically the full local equations, we do not assume a priori sub-harmonic condition (\ref{eq:resonancecond2}). Hence, we emphasise that the numerical solutions will assess the general validity of sub-harmonic condition (\ref{eq:resonancecond2}) in stratified fluids, which has already been confirmed in homogeneous fluids \citep{kerswell1993elliptical,kerswell1994tidal,le2000three,vidal2017inviscid}.  

In the astrophysical regime $\beta_0 \ll 1$, the resonance condition (\ref{eq:resonancecond1}) or (\ref{eq:resonancecond2}) (if valid), are satisfied numerically for only a few initial wave vectors $\boldsymbol{k}_0$. Numerically, this is too expansive to survey all the possible configurations for $\boldsymbol{k}_0$. Thus, we set the equatorial ellipticity to the value $\beta_0=0.2$. 
This does not change in any way the relevance of the following numerical results, because $\sigma$ is proportional to $\beta_0$ (when $\beta_0\ll 1$).
However, for large values of $\beta_0$, the general resonance condition (\ref{eq:resonancecond1}) can be satisfied for a wider range of initial wave vectors $\boldsymbol{k}_0$, due to geometrical detuning effects \citep{le2000three,vidal2017inviscid}. Hence, the computations are more tractable numerically. In practice, we have considered a large enough number of fluid trajectories $\boldsymbol{X}(t)$ and $\boldsymbol{k}_0$, sampling the whole ellipsoidal domain to get representative results. 

We have validated the code against analytical formulas (\ref{eq:sigma_equat}) and (\ref{eq:sigma_zplane}), obtaining a perfect agreement and cross-validating the asymptotic analysis (not shown). Then, we only investigate the stability of equilibrium tidal flow (\ref{eq:U0TDEI_Orb}) within the orbital range $-1 \leq \Omega_0 \leq 3$, representative of the binary systems considered in Sect. \ref{sec:discussion}. 
When stratification is neutral ($N_0=0$), the whole domain is unstable as expected (not shown), with a homogeneous growth rate predicted by formula (\ref{eq:sigma_equatN0}). 
We survey illustrative stably stratified configurations $N_0/\Omega_0 \geq 0$ in Fig. \ref{fig:SWAN_Wm05_0_05_2}. 
Several aspects are worthy of comment. 
We clearly recover the trapping of the instability due to axial stratification, outlined by the weakening of the growth rate in formula (\ref{eq:sigma_zplane}). In the bulk, the weakening first occurs near the polar regions, and then spreads out towards lower latitudes when $N_0/\Omega_s$ increases (from top to bottom panels in Fig. \ref{fig:SWAN_Wm05_0_05_2}). Along the polar axis, it turns out that the transition between unstable and stable areas occurs at position (\ref{eqDynamo:z_H2E2waves}). In addition, the equatorial region is still unstable for the range of $N_0/\Omega_\text{s}$ considered, as observed in Fig. \ref{fig:wkbequat}. 
Then, the numerical analysis unravels an unexpected feature compared to the asymptotic analysis. When $N_0/\Omega_\text{s}$ increases, tidal instability is always triggered in the bulk. Non-vanishing growth rates exist as long as waves can be nonlinearly coupled, according to the resonance condition that is valid when $\beta_0 \ll 1$ (bounded from below and above by the grey dashed curves). An exception appears here for $\Omega_0 = -0.5$ and $N_0/\Omega_\text{s} = 5$ (top panel of Fig. \ref{fig:SWAN_Wm05_0_05_2}). This is due the finite value $\beta_0 = 0.2$ used in the numerics, which is responsible for imperfect resonances in condition (\ref{eq:resonancecond1}) due to geometric detuning effects \citep[e.g.][]{le2000three,lacaze2004elliptical,vidal2017inviscid}. 
Moreover, the striking feature is that stratification tends to confine tidal instability along critical (conical) latitudes (white dashed lines), tilted from the spin (polar) axis. The tilt angle in the numerics is exactly the colatitude angle $\theta_0$ (given our numerical resolution, not shown), predicted by formula (\ref{eq:theta0_H12_N0}) and which maximises the classical tidal instability for neutral fluids ($N_0=0$). 
This shows that the equatorial trapping does not affect similarly all the orbits. When $-1\leq \Omega_0 \leq 1$, the tilt angle $\theta_0$ given by formula (\ref{eq:theta0_H12_N0}) goes from $\theta_0=0$ to $\theta_0=\pi/2$. Hence, the instability on retrograde orbits (with small values of $\theta_0$) is less weakened than on prograde orbits. When $N_0/\Omega_\text{s} \gg 1$, tidal instability is equatorially trapped between the conical layers, with growth rates in the equatorial plane predicted by formula (\ref{eq:sigma_equat}). However, on these conical layers, it turns out that the largest growth rate $\sigma$ is unaffected by stratification, for any value of $N_0/\Omega_s$. Hence, the maximum growth rate of tidal instability in stratified fluids is always given by formula (\ref{eq:sigma_equatN0}), for any orbit in the orbital range $-1 \leq \Omega_0 \leq 3$.

Therefore, the numerical analysis has confirmed and extended the asymptotic analysis. 
In stably stratified interiors, resonance condition (\ref{eq:resonancecond2}) illustrated in Fig. \ref{fig:fig2} is a necessary and sufficient condition for tidal instability (when $\beta_0 \ll 1$). Indeed, we have not found any other resonance yielding larger growth rates than the ones at the sub-harmonic resonance. In the orbital range $-1 \leq \Omega_0 \leq 3$, tidal instability is triggered by sub-harmonic resonances of inertia-gravity waves. Moreover, there is an equatorial trapping of tidal instability between conical latitudes, depending on the orbital configuration according to formula (\ref{eq:theta0_H12_N0}). At these latitudes, the wave vector is parallel to the gravity field, such that the maximum growth rate is unaffected by the stable stratification. 

\subsection{Leading-order (laminar) diffusive effects}
\label{subsec:feedbackB0}
We reintroduce now the leading-order (laminar) diffusive effects at the onset of tidal instability. In the diffusive regime, tidal instability is triggered if the largest diffusionless growth rate $\sigma$ overcomes the (negative) laminar damping rates due to viscosity $\tau_\nu$, radiative diffusivity $\tau_\kappa$ and Joule diffusion $\tau_\Omega$. Hence, the diffusionless growth rate $\sigma$ ought to be reduced by the laminar damping rates, yielding the diffusive growth rate 
\begin{equation}
    \sigma_\mathcal{D} = \sigma + \left ( \tau_\nu + \tau_\kappa + \tau_\Omega \right ).
    \label{eq:sigmadiff}
\end{equation}
We have confirmed in Sect. \ref{subsec:swan} that tidal instability is a parametric instability, involving only inertial and/or gravito-inertial waves in radiative interiors. Consequently, we can simply estimate the laminar damping rates by computing the damping rates of the inertial and gravito-inertial waves involved in the triadic couplings. Indeed, triadic couplings can only give non-vanishing growth rates (\ref{eq:sigmadiff}) if the waves individually exist, that is if they are not damped by any diffusive effect before being efficiently nonlinearly coupled. 
We have shown in Sect. \ref{subsec:swan} that the diffusionless growth rate $\sigma$ is maximum on critical latitudes, where the wave vector satisfies $\boldsymbol{k}_0 \times \boldsymbol{g} = \boldsymbol{0}$ (when $\beta_0 \ll 1$). Then, in the local plane-wave model, the buoyancy term in the local vorticity equation (which is proportional to $\boldsymbol{k}_0 \times \boldsymbol{g}$) vanishes such that vorticity and energy equations are uncoupled (in the local formalism). 
This means that these waves are locally insensitive to stratification on the critical latitudes, yielding $\tau_\kappa=0$. Thus, in the absence of background turbulent motions (see the discussion in Sect. \ref{subsec:dissip2}), the waves are individually damped by viscosity and Joule diffusion (in the weak field regime $Le \ll 1$). 

For the stability computations, we rewrite here the magnetic field as
\begin{equation}
    \boldsymbol{B} = \boldsymbol{B}_0 + \boldsymbol{b},
\end{equation}
where the fossil field $\boldsymbol{B}_0$ is assumed to be steady here. The pervading fossil magnetic fields are nearly axisymmetric and dipole-dominated at the leading order, as observed in magnetic binaries \citep[e.g.][]{alecian2016magnetic,landstreet2017bd,kochukhov2018hd,shultz2017hd,shultz2018magnetic}. For the stability computations, we assume a fossil field of the form $\boldsymbol{B}_0 \propto \boldsymbol{1}_z$, with a dimensionless strength measured by the Lehnert number $Le$. 
The presence of other field components only slightly modifies the frequencies of inertial and inertial-gravity waves at the onset. We also expect the damping rates to have a similar behaviour in the laminar regime. 
In the weak field regime $Le \ll 1$, the damping rates have been devised by \citet{sreenivasan2017damping} in the local theory and by \citet{kerswell1994tidal} in the global one. They depend on the wave properties, that is here the wave vector. Notably, we explain in Appendix \ref{appendix:MixedMHDResonances} why the mixed couplings between inertial waves and slow MC waves cannot lead to tidal instability in short-period binaries (in the presence of Joule diffusion). Hence, we remind the reader that only parametric resonances of inertial and gravito-inertial waves can generate tidal instability in the presence of magnetic fields. 

Then, the viscous and the Joule damping rates in the weak field regime ($Le \ll 1$) in any $z$-plane read
\begin{subequations}
    \begin{align}
    \tau_\nu & = - |\boldsymbol{k}_0|^2 \, Ek, \label{eq:DecayViscous_TDEI} \\
    \tau_\Omega &= - \frac{\cos^2 (\theta_0) \, |\boldsymbol{k}_0|^4 Em \, Le^2}{4 \cos^2 (\theta_0) + |\boldsymbol{k}_0|^4 \, Em^2} \, |1-\Omega_0|,
	\label{eq:DecayFMC_TDEI}
    \end{align}
\end{subequations}
with $|\boldsymbol{k}_0|$ the norm of the wave vector at the resonance (and at the initial time) and $\cos (\theta_0)$ given by condition (\ref{eq:theta0_H12_N0}). 
Expression (\ref{eq:DecayFMC_TDEI}) is quantitatively valid when $\boldsymbol{B}_0 \propto \boldsymbol{1}_z$ \citep{sreenivasan2017damping}. 
In the regime $Pm \ll 1$, laminar Joule diffusion is the leading-order dissipative effect ($|\tau_\Omega| \gg |\tau_\nu|)$. The Joule damping has already been considered for homogeneous fluids  \citep{kerswell1994tidal,herreman2009effects,herreman2010elliptical,cebron2012elliptical}. 
Note that formula (\ref{eq:DecayFMC_TDEI}) is exactly the Joule damping rate of tidal instability in neutral fluids ($\widetilde{N}_0=0$). Besides, formulas of \citet{herreman2009effects} and \citet{cebron2012elliptical} are recovered in the limit $|\boldsymbol{k}_0| \gg 1 $, by using the resonance condition $2 \cos \theta_0=\pm 1$ to set $\theta_0$ for $\widetilde{N}_0=0$. Formula (\ref{eq:DecayFMC_TDEI}) has two asymptotic behaviours, depending on the value of $\boldsymbol{k}_0$. They are separated by the condition
\begin{equation}
	|\boldsymbol{k}_0| = \sqrt{2 \cos (\theta_0) /Em} \sim Em^{-1/2}.
	\label{eq:ktworeginesMHD}
\end{equation}
On the one hand, we obtain a wave-dominated regime when $|\boldsymbol{k}_0| \leq Em^{-1/2}$, in which the Joule damping rate scales as $\tau_\Omega \propto - Em \, Le^2 |\boldsymbol{k}_0|^4/4$. On the other hand, we get a diffusion-dominated regime when $|\boldsymbol{k}_0| \geq Em^{-1/2}$. In the latter regime, the damping rate is independent of the wave vector and scales as $\tau_\Omega \propto - Le^2/Em$. 

\begin{figure}
	\centering
	\includegraphics[width=0.49\textwidth]{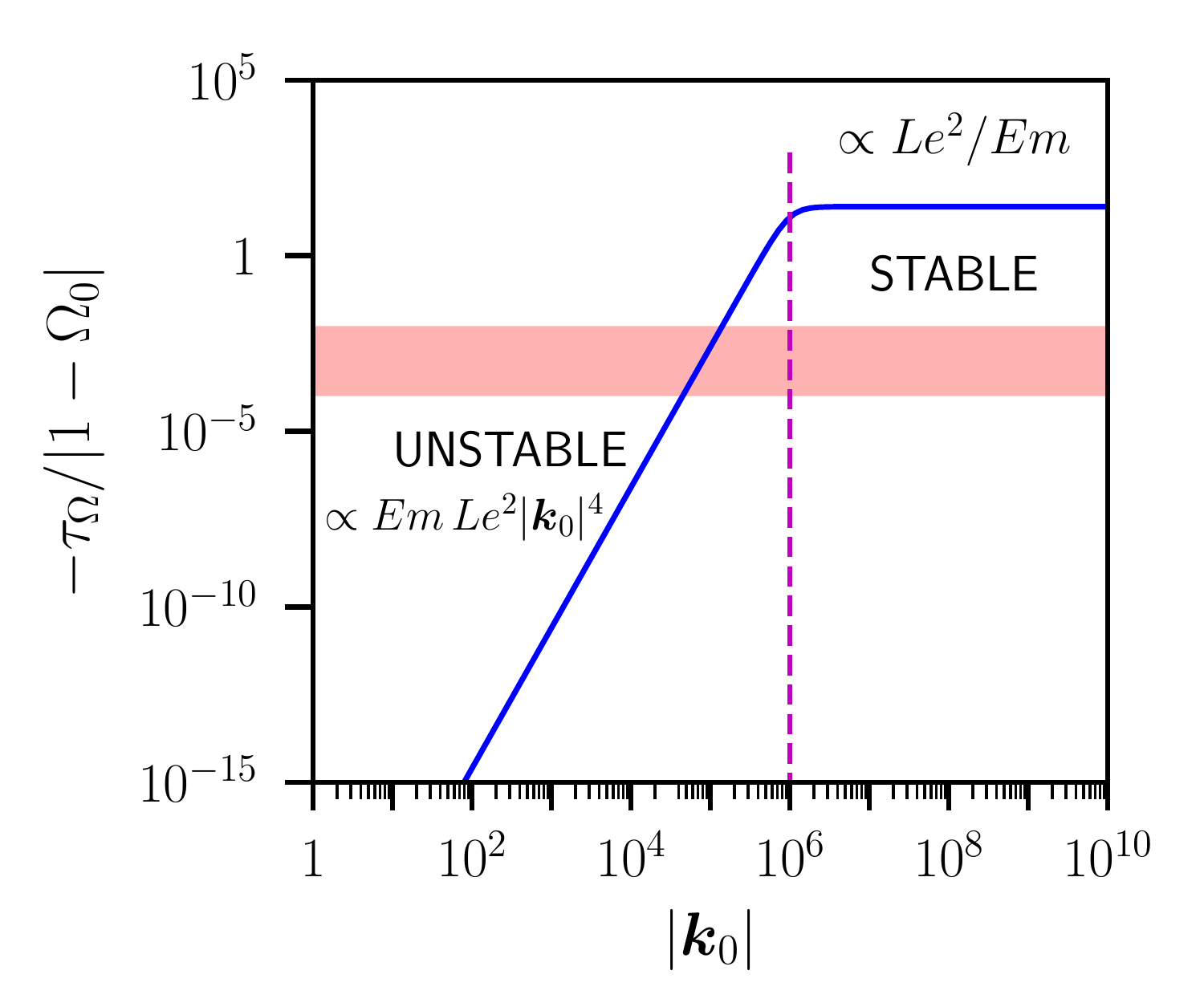}
	\caption{Dimensionless Joule damping $-\tau_\Omega/|1-\Omega_0|$ of tidal instability (solid blue line), as a function of magnitude $|\boldsymbol{k}_0|$. Dashed magenta line is given by formula (\ref{eq:ktworeginesMHD}), delimiting the two hydromagnetic regimes. Red shaded areas show the typical strength of the diffusionless growth rate of tidal instability $\sigma \sim \mathcal{O}(\beta_0)$, with $\beta_0 \in [10^{-4}, 10^{-2}]$ for close binaries. Computations at $Le=10^{-5}$ and $Ek/Pm=10^{-12}$ for the dimensionless fossil field $\boldsymbol{B}_0 = \boldsymbol{1}_z$ aligned with the spin axis.}
	\label{fig:DampingMHDWaves}
\end{figure}

We illustrate in Fig. \ref{fig:DampingMHDWaves} the evolution of Joule damping rate (\ref{eq:DecayFMC_TDEI}) in the different regimes. 
Tidal instability will survive in the presence of magnetic fields if $\sigma \gg |\tau_\Omega|$. 
Typical values of the diffusionless growth rate, given by formula (\ref{eq:sigma_equatN0}), are $\sigma \sim \mathcal{O}(\beta_0)$ with $\beta_0 \in [10^{-4}, 10^{-2}]$ in close binaries.
We clearly observe that tidal instability does survive against Joule diffusion, for short-wavelength perturbations with $|\boldsymbol{k}_0| \leq 10^{4} - 10^5$. For larger values of the wave number, the Joule damping rate always overcomes the diffusionless growth rate, such that no instability is triggered. 

\subsection{Other dissipative mechanisms}
\label{subsec:dissip2}
At the linear onset, the laminar diffusive effects discussed in Sect. \ref{subsec:feedbackB0} are always present, but we have shown that they are smaller than the largest diffusionless growth rate $\sigma$. Hence, these effects can be reasonably neglected at the onset, yielding $\sigma_\mathcal{D} \sim \sigma$.
However, other diffusive effects do exist in stellar interiors, which may weaken the growth of tidal instability.

Phase mixing is known to provide a significant source of Joule heating, by dissipating Alfv\'en (and magneto-sonic) waves in stellar atmospheres \citep[e.g.][]{heyvaerts1983coronal} or stellar interiors \citep{spruit1999differential}. Yet, phase-mixing is probably irrelevant for tidal instability in the weak field regime ($Le \ll 1$), notably because Aflv\'en waves are not involved in tidal instability (see Appendix \ref{appendix:MixedMHDResonances}). Whether phase-mixing could increase the dissipation of inertial and gravito-inertial waves in stellar interiors remains unknown and is largely beyond the scope of the present study.

In the presence of an innermost convective envelope, inertial and gravito-inertial waves can exhibit singular shear layers, reminiscent of wave attractors \citep[e.g.][]{dintrans1999gravito,rieutord2010viscous,mirouh2016gravito,lin2017tidal,rieutord2018axisymmetric}. These global wave patterns are not directly involved in the parametric mechanism of tidal instability, but they fill the whole fluid domain and may provide an additional bulk damping rate for tidal instability. 
Indeed, these structures can be destabilised in the nonlinear regime  \citep{jouve2014direct}, possibly yielding small-scale instabilities. \citet{brunet2019attractor} showed that the resulting small-scale turbulence in the bulk could be well modelled by a turbulent eddy diffusion. In particular, anisotropic shear-driven turbulence may be generated \citep[e.g.][]{zahn1992circulation}. 
In such a case, \citet{garaud2017turbulent} and \citet{gagnier2018turbulent} proposed to model the local shear-driven turbulence by introducing the turbulent viscosity
\begin{equation}
    \nu_\text{t} \propto 0.08 \, \kappa_T/J, \ \, \ J = N_0^2/S^2,
    \label{eq:mixingShear}
\end{equation}
with $\kappa_T$ the radiative diffusivity, $J$ the local gradient Richardson number and $S$ the local shearing rate (responsible for the shear instabilities). The stability criterion for shear instabilities is apparently $J Pr \simeq 0.007$ \citep{garaud2017turbulent}. Then, prediction (\ref{eq:mixingShear}) would yield an upper-bound effective turbulent Ekman number $Ek_\text{t} \leq 10^{-10}$ for speculative stellar values, to use in expression (\ref{eq:DecayViscous_TDEI}) for the viscous damping rate. For the range of wave numbers $|\boldsymbol{k}_0|$ given in Fig. \ref{fig:DampingMHDWaves}, we find that the associated turbulent damping rate is smaller than the diffusionless growth rate $\sigma$ (not shown). Therefore, even in the presence of shear-driven instabilities, the associated turbulent damping can be ignored at the onset of tidal instability for the (strong enough) tidal deformations considered in this work ($\beta_0 \sim 10^{-3} - 10^{-2}$, see Table \ref{table:databinamics}).

\section{Turbulent mixing due to nonlinear tidal flows}
\label{sec:mixing}
At this stage, we have shown that tidal instability can be triggered within stably stratified interiors, even against the stabilising effect of a background (fossil) magnetic field in the weak field regime ($Le \ll 1$). 
The next step is to characterise the saturated regime of tidal flows. 
Modelling turbulent mixing in radiative interiors is one of the enduring problems in stellar dynamics \citep[e.g.][]{zahn1974rotational}.
Several studies have examined the turbulence in radiative zones \citep[e.g.][]{zahn1992circulation,mathis2004shear,garaud2017turbulent,gagnier2018turbulent,mathis2018anisotropic}. Yet, these models focus on shear-driven turbulence. Hence, tidally driven turbulence in binaries remains to be described. 
Numerical simulations have shown that small-scale turbulence can be excited by tidal instability \citep{barker2013non,barker2013nonb,le2017inertial}, possibly leading to global tidal mixing \citep{vidal2018magnetic}. Thus, tidal mixing is expected in radiative interiors. 
We motivate our assumptions in Sect. \ref{subsec:dissipB}. Then, we use dimensional-type arguments in Sect. \ref{subsec:mlt} to develop a phenomenological description of the nonlinear tidal mixing in radiative interiors in Sect. \ref{subsec:eddy}, valid in the orbital range $-1 \leq \Omega_0 \leq 3$. Finally, we assess its validity by using proof-of-concept simulations in Sect. \ref{subsec:simuB}.

\subsection{Assumptions}
\label{subsec:dissipB}
As shown in Sect. \ref{sec:onset}, magnetic effects play a minor role at the onset of instability in the orbital range $-1 \leq \Omega_0 \leq 3$. They essentially weaken the growth rate of tidal instability, due to the laminar Joule damping.
In the (transient) linear growth, the fossil field $\boldsymbol{B}_0$ is not much affected by tidal flows, which are not expected to generate significant mixing. It only decays on the slow (laminar) Joule diffusion time, which is much larger than the timescale for the onset of tidal instability for stellar parameters. This phenomenon is well-known in global models of resistive magnetohydrodynamics, also known as free-decay of magnetic fields \citep[e.g.][]{moffatt1978field}. However, in the saturated regime, the fossil field would interact nonlinearly with the nonlinear tidal flows, as governed by induction equation
\begin{subequations}
\allowdisplaybreaks
\label{eq:Bfossilmixing}
\begin{align}
	\frac{\partial \boldsymbol{B}}{\partial t} &=  \boldsymbol{\nabla} \times \left [ (\boldsymbol{U}_0 + \boldsymbol{u}) \times \boldsymbol{B} \right ] + \frac{Ek}{Pm} \, \boldsymbol{\nabla}^2 \boldsymbol{B}, 
	\label{eq:Bpertmixing} \\
	\boldsymbol{\nabla} \boldsymbol{\cdot} \boldsymbol{B} &= 0, \ \, \ \boldsymbol{B} (\boldsymbol{r},t=0) = \boldsymbol{B}_0 (\boldsymbol{r}), 	\label{eq:Bpertmixing2}
\end{align}
\end{subequations}
in which the initial time $t=0$ refers now to an initial time just after the growth of the instability. In equation (\ref{eq:Bpertmixing}), the nonlinear velocity field $\boldsymbol{u}$ is governed by momentum equation (\ref{eq:Upert}). In the relevant weak field regime $Le \ll 1$, nonlinear numerical simulations of the coupled problems showed that magnetic effects do not weaken the turbulent tidal flows \citep{barker2013nonb,cebron2014tidally,vidal2018magnetic}. These turbulent flows generate mixing, that would ultimately increase the Ohmic diffusion of the fossil field $\boldsymbol{B}_0$. Therefore, Ohmic diffusion ought to be increased (a priori). This is often modelled by introducing a turbulent magnetic diffusivity \citep[e.g.][]{kitchatinov1994turbulent,yousef2003turbulent,kapyla2019turbulent}. In this configuration, the initial fossil field is expected to decay on somehow faster timescales, due to the presence of mixing generated by tidal instability. This situation strongly differs from the picture of ideal magnetohydrodynamics, in which the laminar decay of the fossil field is small (and so can be sometimes neglected). 
Note that an initial fossil field may still be in quasi-equilibrium with tidal flows, if the dissipated field is continuously regenerated by some kind of dynamo action. However, dynamo action of tidal flows in strongly stratified interiors remains elusive  \citep{vidal2018magnetic} and will not be investigated here. Consequently, to estimate the fossil field decay due to tidal instability, we must estimate the turbulent magnetic diffusivity generated by the saturation of tidal instability.

	\subsection{Mixing-length theory}
	\label{subsec:mlt}
Estimating a realistic turbulent magnetic diffusivity is challenging, because no numerical model cannot probe accurately the stellar conditions. This makes the relevance of numerical results sometimes elusive. Therefore, we aim to build asymptotic scaling laws for the tidal mixing, based on dimensional-type arguments that embrace both numerical and stellar conditions. 
To estimate the local tidal mixing in stratified interiors, we develop a mixing-length theory, by analogy with mixing-length arguments commonly used for shear-driven turbulence in radiative interiors of stars \citep[e.g.][]{zahn1992circulation,mathis2004shear,mathis2018anisotropic}. 

In turbulent flows, the laminar viscosity is often replaced by an effective eddy (turbulent) viscosity, usually modelled by using mixing-length theory in stellar contexts. In hydromagnetic turbulence, \citet{yousef2003turbulent} and \citet{kapyla2019turbulent} argued that in the weak field regime ($Le \ll 1$) the turbulent magnetic Prandtl number is not far from unity. Hence, the turbulent magnetic diffusivity can be a priori modelled by mixing-length type predictions. This is supported by local hydromagnetic simulations of the three-dimensional turbulence generated by tidal instability \citep{barker2013nonb}. They showed that weak magnetic fields can even favour the small-scale tidal turbulence. Global tidal mixing has also been found in global stratified models \citep{vidal2018magnetic}. 
Thus, we may replace any laminar diffusivity (denoted $\mathcal{D}$) by an effective eddy diffusivity (denoted $\mathcal{D}_\text{t}$), induced by the nonlinear tidal flows. 
Then, mixing-length theory  \citep[e.g.][]{tennekes1972first} predicts in dimensional form (up to a unknown proportional constant)
\begin{equation}
	\mathcal{D}_\text{t} \propto \frac{1}{3} u_\text{t} \, l_\text{t},
	\label{eq:etaturmlt}
\end{equation}
where $u_\text{t}$ and $l_\text{t}$ are respectively the typical (dimensional) local velocity and length scale of the turbulent motions. Note that $u_\text{t}$ is the typical amplitude of the nonlinear tidal flows. This must not be confused with the amplitude $u_w$ of the waves that are excited by the forcing mechanism \citep[see the case of internal gravity waves in][]{rogers2017chemical}. 
Here, $u_w$ is much smaller than $u_\text{t}$ in amplitude. Hence, the eddy diffusivity $\mathcal{D}_\text{t}$ is a local property of the nonlinear flows, rather than a property of the fluid (or of the wave amplitude). The key point to apply formula (\ref{eq:etaturmlt}) is to find accurate predictions for $u_\text{t}$ and $l_\text{t}$ in the nonlinear regime of tidal instability. 

On the one hand, we have shown in Sect. \ref{sec:onset} that tidal instability is generated by sub-harmonic resonances of inertial waves, more or less modified by the gravity field in the orbital range $-1 \leq \Omega_0 \leq 3$. 
This mechanism holds whatever the strength of stratification, measured by the ratio $N_0/\Omega_\text{s}$. Therefore, the turbulent velocity scale $u_\text{t}$ should not depend (strongly) on the local strength of stratification $N_0/\Omega_\text{s}$. This is supported by proof-of-concept simulations \citep[see Fig. 2b in][]{vidal2018magnetic}, showing that nonlinear tidal flows exhibit the scaling devised in homogeneous fluids \citep{barker2013non,grannan2016tidally}. This reads
\begin{equation}
	u_\text{t} \sim \alpha_1 \beta_0 r_l \, \Omega_\text{s} (1 - \Omega_0)
	\label{eq:uturbmlt}
\end{equation}
with $r_\text{l} \leq R$ the local position and $\alpha_1 \sim 0.3-0.5$ a dimensionless pre-factor obtained numerically both in homogeneous \citep[][estimated from Fig. 4d]{grannan2016tidally} and strongly stratified tidal flows \citep[][estimated from Fig. 2b]{vidal2018magnetic}. Hence, we reasonably estimate the turbulent velocity $u_\text{t}$ by using prescription (\ref{eq:uturbmlt}). 
On the other hand, $l_\text{t}$ should depend on the local ratio $N_0/\Omega_\text{s}$. Several regimes have been found in forced stratified turbulence \citep[e.g.][]{brethouwer2007scaling}. 

	\subsection{Phenomenological prescriptions}
	\label{subsec:eddy}
	\subsubsection{Weakly stratified regime ($N_0 /\Omega_\text{s} \leq 1$)}
In the weakly stratified regime, characterised by $N_0/\Omega_\text{s} \leq 1$, $\mathcal{H}_1$ waves satisfying the sub-harmonic resonance condition are barely affected by stratification. 
We estimate $l_\text{t}$ by balancing the nonlinear term $(\boldsymbol{u} \boldsymbol{\cdot} \nabla) \, \boldsymbol{u}$ with the injection term $(\boldsymbol{u} \boldsymbol{\cdot} \nabla) \, \boldsymbol{U}_0$ in momentum equation (\ref{eq:Upert}). This yields the typical turbulent length scale in dimensional form $l_\text{t} \propto \alpha_1 r_\text{l}$. 
Then, the weakly stratified regime is characterised by the eddy diffusivity  (in dimensional form)
\begin{equation}
	\mathcal{D}_\text{t} \propto \frac{1}{3} \alpha_1^2 \beta_0 r_\text{l}^2 \, \Omega_\text{s} (1 - \Omega_0).
	\label{eq:etaturmltlowN}
\end{equation}
Formula (\ref{eq:etaturmltlowN}) predicts a roughly homogeneous mixing in the weakly stratified regime, as found in global models \citep{grannan2016tidally,vidal2018magnetic} in which $r_\text{l} \simeq R$. This explains why the tidal mixing computed in \citet{vidal2018magnetic} is roughly constant as a function of stratification, when $N_0/\Omega_\text{s} \leq 1$ (see their Fig. 9). However, estimate (\ref{eq:etaturmltlowN}) may be reduced in this regime due to (compressible) density variations (close to the isentropic profile when $N_0/\Omega_\text{s} \ll 1$).

Finally, formula (\ref{eq:etaturmltlowN}) provides a good estimate of the leading-order term in the eddy diffusivity tensor \citep[e.g.][]{dubrulle1991eddy,wirth1995eddy}. In addition, note that rotation would also support small anisotropic diffusion in the axial direction \citep{tilgner2004small,elstner2007can}.

	\subsubsection{Stratified regimes ($N_0 /\Omega_\text{s} \geq 1$)}
We now investigate the stratified regimes $N_0/\Omega_\text{s} \geq 1$.
Stratified turbulence is highly anisotropic. Indeed, a commonly observed feature of strongly stratified flows is the formation of quasi-horizontal layers, often described as pancake structures \citep[e.g.][]{billant2001self}. Such layers are conspicuous in simulations of tidal flows in strongly stratified fluids, both in non-rotating \citep{le2018parametric} and rotating fluids \citep{vidal2018magnetic}. 
Hence, $l_\text{t}$ depends on both the direction and the strength of stratification. We introduce two turbulent length scales, respectively $l_\text{t}^\parallel$ in the normal direction (that is along the gravity field) and $l_\text{t}^\perp$ in the other horizontal directions. 

Several regimes of stratified turbulence have been devised in fundamental fluid mechanics \citep{billant2001self,brethouwer2007scaling}. They are characterised by the buoyancy Reynolds number
\begin{equation}
	\mathcal{R} \sim \frac{u_\text{t}^3}{l_\text{t}^\perp N_0^2 \nu}.
\end{equation}
\citet{le2018parametric} investigated the small-scale turbulence sustained by tides in the regime $\mathcal{R} \leq 1$, in which vertical viscous shearing is significant.  
However, radiative interiors are in the opposite regime $\mathcal{R} \gg 1$ \citep{mathis2018anisotropic}.
Moreover, they neglected rotation, by setting $\Omega_\text{s}=0$. 
In such a configuration, the subspaces of waves $[\mathcal{H}_1,\mathcal{H}_2]$ at the sub-harmonic resonance are empty, according to dispersion relations (\ref{eqDynamo:H1H2_waves}). Hence, tidal instability can only involve sub-harmonic resonances of internal waves $\mathcal{H}_2$ in the limit $N_0 /\Omega_\text{s} \to \infty$ and $|\Omega_0| \to \infty$. 
Therefore, their results do not apply for our astrophysical problem, for any orbit in the range $-1 \leq \Omega_0 \leq 3$.
In the relevant strongly stratified regime ($\mathcal{R} \gg 1$), diffusion is unimportant and the turbulence is three-dimensional \citep{brethouwer2007scaling}. The general scalings of this regime have been confirmed by turbulence simulations \citep[e.g.][]{godeferd2003statistical,maffioli2016dynamics}. Thus, they can be applied to the tidal problem.
In addition, rotational effects are also significant within the orbital range $-1 \leq \Omega_0 \leq 3$, even for large values of $N_0/\Omega_\text{s} \geq 10$. 
Hence, the resulting turbulence undergoes the combined action of stratification and rotation. 

In rotating stratified turbulence, the two turbulent length scales are related by \citep{billant2001self}
\begin{equation}
	l_\text{t}^\perp \sim \alpha_2 \frac{N_0}{\Omega_\text{s}} l_\text{t}^\parallel.
	\label{eq:lturb2mlthighN}
\end{equation}
with $\alpha_2 \sim 0.6$ a (numerical) pre-factor constrainted from local turbulent simulations in rapidly rotating and strongly stratified turbulent regime \citep{reinaud2003shape,waite2006transition}. This regime is expected to be valid for radiative interiors, notably to describe shear-driven turbulence \citep{mathis2018anisotropic}. 
For strong stratification ($N_0/\Omega_\text{s} \geq 10$), we combine the two balances obtained by equating (i) the nonlinear term with the buoyancy force in momentum equation (\ref{eq:Upert}) and (ii) the injection term $(\boldsymbol{u} \boldsymbol{\cdot} \nabla) \, T_0$ and the nonlinear term $(\boldsymbol{u} \boldsymbol{\cdot} \nabla) \, \Theta$ in energy equation (\ref{eq:Tpert}). These balances yield respectively
\begin{equation}
	\frac{u_\text{t}^2}{l_\text{t}^\parallel} \sim \alpha_T g_0 \, \Theta_\text{t} \ \, \ \text{and} \ \, \ \alpha_T g_0 \, \Theta_\text{t} \sim N_0^2 \, l_\text{t}^\parallel,
	\label{eq:highNtemp}
\end{equation}
where $\Theta_\text{t}$ is the typical dimensional turbulent buoyancy perturbation. We recover from balances (\ref{eq:highNtemp}) the classical scaling for the turbulent length scale in the normal direction, that is $u_\text{t} \sim l_\text{t}^\parallel N_0$ \citep[e.g.][]{billant2001self,brethouwer2007scaling}. Hence, the turbulent length scale along the gravity direction is
\begin{equation}
	l_\text{t}^\parallel \sim \alpha_1 \beta_0 \, r_\text{l} \, (1-\Omega_0) \frac{\Omega_\text{s}}{N_0} \ \, \ (\text{with} \ \,\ \alpha_1 \sim 0.3 - 0.5).
	\label{eq:lturbmlthighN}
\end{equation}
Scaling (\ref{eq:lturbmlthighN}) shows that tidal mixing falls in the asymptotic regime of strongly stratified turbulence \citep{brethouwer2007scaling}. 
Then, we obtain two prescriptions for the eddy diffusivity, the first one $\mathcal{D}_\text{t}^\parallel$ valid in the gravity direction and the second one $\mathcal{D}_\text{t}$ in the perpendicular (horizontal) directions. They yield
\begin{subequations}
\allowdisplaybreaks
\label{eq:etaturmlthighN}
\begin{align}
	\mathcal{D}_\text{t}^\parallel &\propto \frac{1}{3} \alpha_1^2 \, \beta_0^2 \, r_\text{l}^2 \, \Omega_\text{s} (1 - \Omega_0)^2 \frac{\Omega_\text{s}}{N_0}, \label{eq:etaturmlthighN_parallel} \\
	\mathcal{D}_\text{t}^\perp &\propto \frac{1}{3} \alpha_1^2 \alpha_2 \, \beta_0^2 \, r_\text{l}^2 \, \Omega_\text{s} (1 - \Omega_0)^2, \label{eq:etaturmlthighN_perp}
\end{align}
\end{subequations}
with $\alpha_1 \sim 0.3 - 0.5$ and $\alpha_2 \sim 0.6$ (see above). Prescriptions (\ref{eq:etaturmlthighN}) show that the eddy diffusivity should have a quadratic dependence with the equatorial ellipticity, in any spatial direction. Another interesting prediction in this regime is that the turbulent potential and kinetic energies, defined by (in dimensional variables) 
\begin{equation}
	E_\text{t}(\Theta^*) \sim \frac{1}{2} \frac{\alpha^2_T g_0^2}{N_0^2} \Theta_\text{t}, \ \, \ E_\text{t}(\boldsymbol{u}^*) \sim \frac{1}{2} u_\text{t}^2,
	\label{eq:ratioPEKEbillant}
\end{equation}
are comparable in magnitude \citep{billant2001self}. This can be checked in the numerical simulations (see below).

In-between the two aforementioned stratified regimes, when $1 \leq N_0/\Omega_\text{s} \leq 10$, the situation is unclear. Indeed, \citet{vidal2018magnetic} found that $\boldsymbol{u} \boldsymbol{\cdot} \boldsymbol{g}$, which is responsible for tidal mixing in the normal direction, is largely unaffected by stratification when $N_0/\Omega_\text{s} \leq 10$ (see their Fig. 4). Hence, we may extend prescription (\ref{eq:etaturmltlowN}) for the turbulent mixing up to $N_0/\Omega_\text{s} \leq 10$. Yet, this behaviour is not conspicuous in the numerics \citep[see Fig. 9b in][]{vidal2018magnetic}. This may be due to the rather specific numerical method, which inaccurately probed the intermediate regime $1 \leq N_0/\Omega_\text{s} \ll 10$. Thus, a transition may be also expected between the two regimes (\ref{eq:etaturmltlowN}) and (\ref{eq:etaturmlthighN}) when $1 \leq N_0/\Omega_\text{s} \leq 10$.

	\subsection{Validation against numerical simulations}
	\label{subsec:simuB}
We assess the relevance of predictions (\ref{eq:etaturmltlowN}) and (\ref{eq:etaturmlthighN}) by using direct numerical simulations. To do so, we solve nonlinear and diffusive equations (\ref{eq:UTBpert}) in a global model. We supplement the governing equations by considering the stress-free conditions
\begin{equation}
	\boldsymbol{u} \boldsymbol{\cdot} \boldsymbol{1}_n = 0, \ \, \ \boldsymbol{1}_n \times \left [ (\boldsymbol{\nabla} \boldsymbol{u} + (\boldsymbol{\nabla} \boldsymbol{u})^\top) \, \boldsymbol{1}_n \right ] = \boldsymbol{0},
	\label{eq:stressfree}
\end{equation}
and assuming a fixed temperature $\Theta=0$ at the boundary. Stress-free conditions (\ref{eq:stressfree}) are known to lead to spurious numerical behaviours, associated with the evolution of angular momentum in weakly deformed spheres \citep{guermond2013remarks}. To circumvent this numerical problem, we follow \citet{cebron2014tidally} and \citet{vidal2018magnetic} by imposing a zero-angular momentum for the velocity perturbation. Moreover, the external region is assumed to be electrically insulating, such that the magnetic field $\boldsymbol{b}$ matches a potential field at the boundary. 

For the computations, we use the proof-of-concept global numerical model introduced in \citet{vidal2018magnetic}.
Briefly, the reference ellipsoidal configuration (described in Sect. \ref{subsec:refellipsoid}) is approximated in spherical geometry by an spatially varying equatorial ellipticity profile $\epsilon(\boldsymbol{r},\beta_0)$, depending of the ellipticity $\beta_0$ of the ellipsoidal configuration. This profile is chosen such that the reference configuration satisfies all the aforementioned boundary conditions in the spherical geometry. The simulations have been performed with the open-source nonlinear code XSHELLS (\url{https://nschaeff.bitbucket.io/xshells/}), described in \citet{schaeffer2017turbulent} and validated against standard spherical benchmarks \citep{marti2014full,matsui2016performance}. A second-order finite difference scheme is used in the radial direction. The angular directions are discretised using a pseudo-spectral spherical harmonic expansion, provided by the SHTns library \citep{schaeffer2013efficient}. The time-stepping scheme is of second order in time and treats the diffusive terms implicitly, while the nonlinear and Coriolis terms are handled explicitly. We refer the reader to \citet{vidal2018magnetic} for additional methodological details of the tidal problem. 

To estimate the turbulent magnetic diffusivity in a global model, we measure the decay of an initial large-scale magnetic field \citep{yousef2003turbulent, kapyla2019turbulent} in the presence of nonlinear tides, to  compare it with the free decay rate of the same magnetic configuration in laminar diffusive models \citep[e.g.][]{moffatt1978field}. 
We compute the (dimensionless) decay rate $\sigma_\eta \leq 0$ of the volume average of the magnetic energy over the computational integration time $T$ as
\begin{equation}
	\sigma_\eta = \lim \limits_{T\to\infty} \frac{1}{T} \log \left ( \int_{\mathcal{V}} \frac{1}{2} |\boldsymbol{B}|^2 \, \mathrm{d} \mathcal{V} \right ).
	\label{eq:sigmaturb1}
\end{equation}
Decay rate (\ref{eq:sigmaturb1}) is a global estimate in the simulations of the effective diffusivity $\mathcal{D}_\text{t}$. \citet{kapyla2019turbulent} measured in a similar way the turbulent diffusivity, obtaining a good quantitative agreement with mean-field analyses. Then, global decay rate (\ref{eq:sigmaturb1}) should have the same scaling law in $\beta_0$ for all the initial magnetic fields $\boldsymbol{B}_0$, even if the (numerical) pre-factors will be different. 
Indeed, all the magnetic components will not obey the same scaling law in the strongly stratified regime (due to the anisotropic mixing). Notably, we expect toroidal magnetic fields, satisfying $\boldsymbol{B}_0 \boldsymbol{\cdot} \boldsymbol{1}_n = 0$ (at any position), to be preferentially dissipated in the normal direction. Thus, scaling (\ref{eq:etaturmlthighN_parallel}) should apply predominantly for toroidal fields. On the contrary, we expect the dissipation of poloidal magnetic fields (with predominant components in the normal direction) to obey scaling (\ref{eq:etaturmlthighN_perp}) in the horizontal directions. 
However, we emphasise that the pre-factors obtained from numerical simulations, performed for conditions far-removed from the astrophysical regimes, are often irrelevant for astrophysical problems (compared to mixing-length predictions). We only focus on the dependence in $\beta_0$, which should be generic whatever the topology of the initial magnetic field in the numerics. 
Thus, we aim at recovering (i) $\sigma_\eta \propto \beta_0$ for weakly stratified regime (\ref{eq:etaturmltlowN}) and (ii) $\sigma_\eta \propto \beta_0^2$ for strongly stratified regime (\ref{eq:etaturmlthighN}). 

In magnetic radiative stars, the initial fossil field is unlikely force-free \citep[e.g.][]{duez2010relaxed,duez2010stability}, except possibly close to the stellar surface. The exact topology of the field does depend on the Lorentz force, and only magnetic equilibria involving poloidal and toroidal components have been found \citep[e.g.][]{braithwaite2017magnetic}.
Then, in addition to the slow laminar Joule diffusion, \citet{braithwaite2012weak} showed that an initial fossil field can decay due to the propagation of (slow) Magneto-Coriolis waves (see Appendix \ref{appendix:MAC}) in the presence of rotation. Such a magnetic decay occurs
on the (rather slow) dynamic timescale
\begin{equation}
    \tau_\text{MC} \sim (\Omega_s Le^2)^{-1}.
    \label{eq:braithwaite2012MC}
\end{equation}
Moreover, the field can be also dissipated by the turbulent mixing generated by nonlinear tidal flows. Thus, the initial field can be dissipated simultaneously by several mechanisms if we neglect in-situ dynamo mechanisms, that would regenerate the field against laminar and turbulent diffusion but are highly debated.

However, we would like the magnetic decay to be insensitive to dynamic evolution (\ref{eq:braithwaite2012MC}) in the numerics, to investigate only the turbulent effects in a well controlled set-up. Hence, we aim to find a magnetic configuration in which the initial field would decay solely by laminar Joule diffusion in the absence of tides. To do so, we can reasonably switch-off the Lorentz force in momentum equation, to estimate turbulent magnetic diffusivity (\ref{eq:sigmaturb1}) for a given initial magnetic field. Without magnetic forces, MC waves are no longer sustained in the system. Moreover, as explained above, the Lorentz force surprisingly plays a negligible role\footnote{Even though it is essential for the self-sustained generation of dynamo magnetic fields.} on the turbulent mixing generated by nonlinear tidal flows in the (relevant) weak field regime $Le \ll 1$ \citep{barker2013nonb,cebron2014tidally,vidal2018magnetic}. Consequently, for this particular problem of tidal instability, neglecting the Lorentz force is advisable in the numerics. 

As a reference configuration, we have assumed $\Omega_0 = 0$. Indeed, we have shown theoretically in Sect. \ref{sec:onset} that the underlying mechanism of tidal instability does not change in the range $-1 \leq \Omega_0 \leq 3$, and similarly the turbulent scalings \citep[e.g.][]{grannan2016tidally,vidal2018magnetic}. Hence, investigating only one orbital configuration is necessary. 
Then, problem (\ref{eq:Bfossilmixing}) reduces here to a kinematic (linear) initial value problem for the initial field. We emphasise that the exact topology of the initial field will not be essential here for the numerical model. Indeed, without the Lorentz force, induction equation (\ref{eq:Bpertmixing}) is uncoupled to the momentum equation. To mimic the slow magnetic decay on the laminar Joule diffusion (in the absence of tides), we have chosen for the initial fossil field the least-damped, poloidal free decay magnetic mode of the sphere \citep[see][p. 36-40]{moffatt1978field}. This particular magnetic field is an exact solution of the purely diffusive induction equation. It has the smallest laminar Ohmic free decay rate $\sigma_\Omega$ (in dimensionless form), given by 
\begin{equation}
	\sigma_\Omega = \pi^2 {Ek}/{Pm}.
	\label{eq:sigmaFreeDecayB}
\end{equation}
Thus, this is the most suited initial magnetic field to assess the validity of the turbulent scaling laws.
Indeed, slow laminar Joule diffusion (\ref{eq:sigmaFreeDecayB}) should not be coupled with the expected faster turbulent diffusion in the numerics to get robust results. 
In practice, we conducted the simulations at the fixed dimensionless numbers $Ek=10^{-4}, Pr=1$ and $Pm=0.1$. The latter value ensures that no dynamo magnetic field can grow exponentially. Our spatial discretisation is $N_r = 224$ radial points, $l_{\max} = 128$ spherical harmonic degrees and $m_{\max} = 100$ azimuthal wave numbers. We have integrated the equations on one (dimensionless) Ohmic diffusive time $(Ek/Pm)^{-1}$, to determine accurately the turbulent decay rate $\sigma_\eta$.

\begin{figure}
	\centering
	\begin{tabular}{c}
		\includegraphics[width=0.46\textwidth]{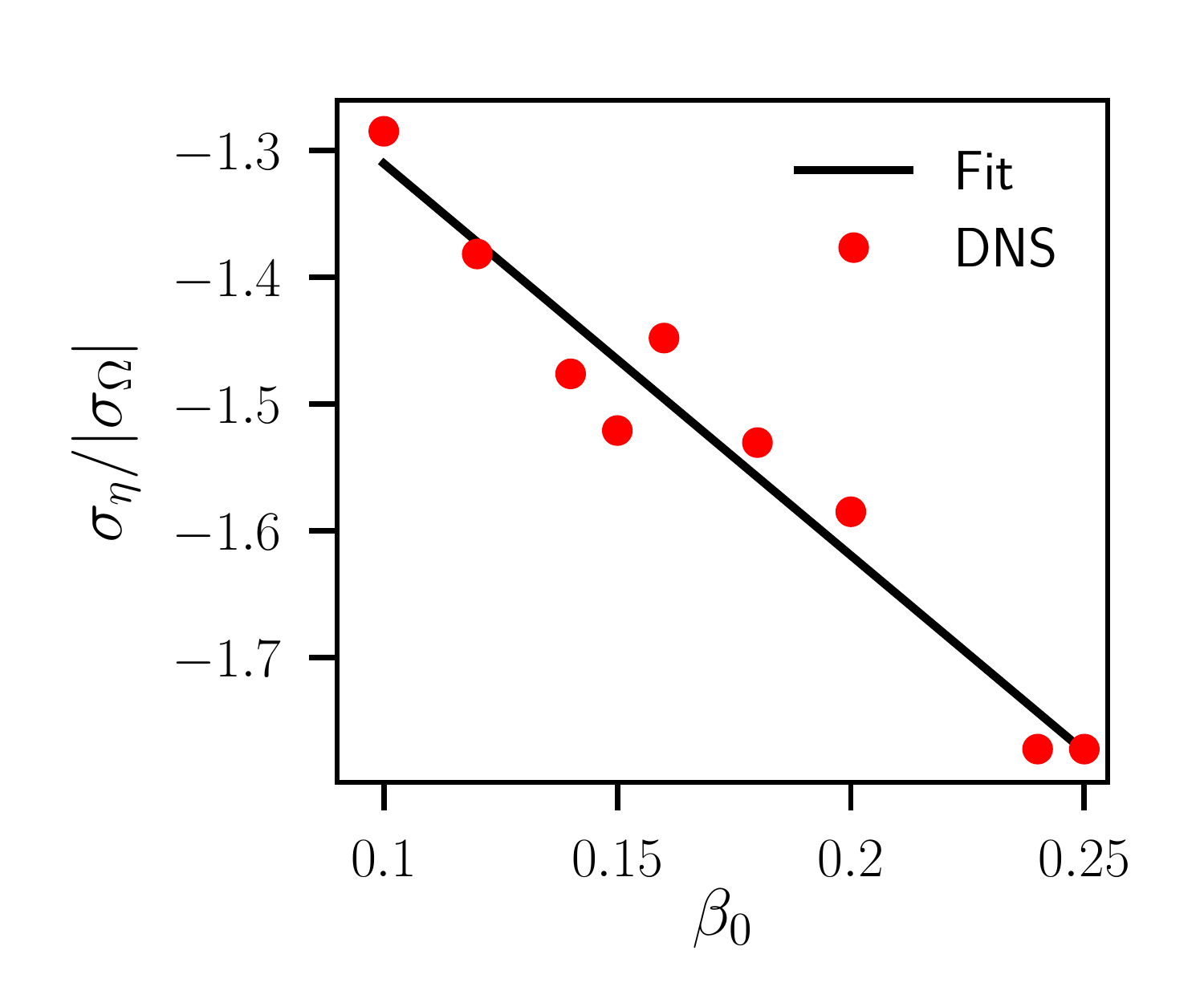} \\
		\includegraphics[width=0.46\textwidth]{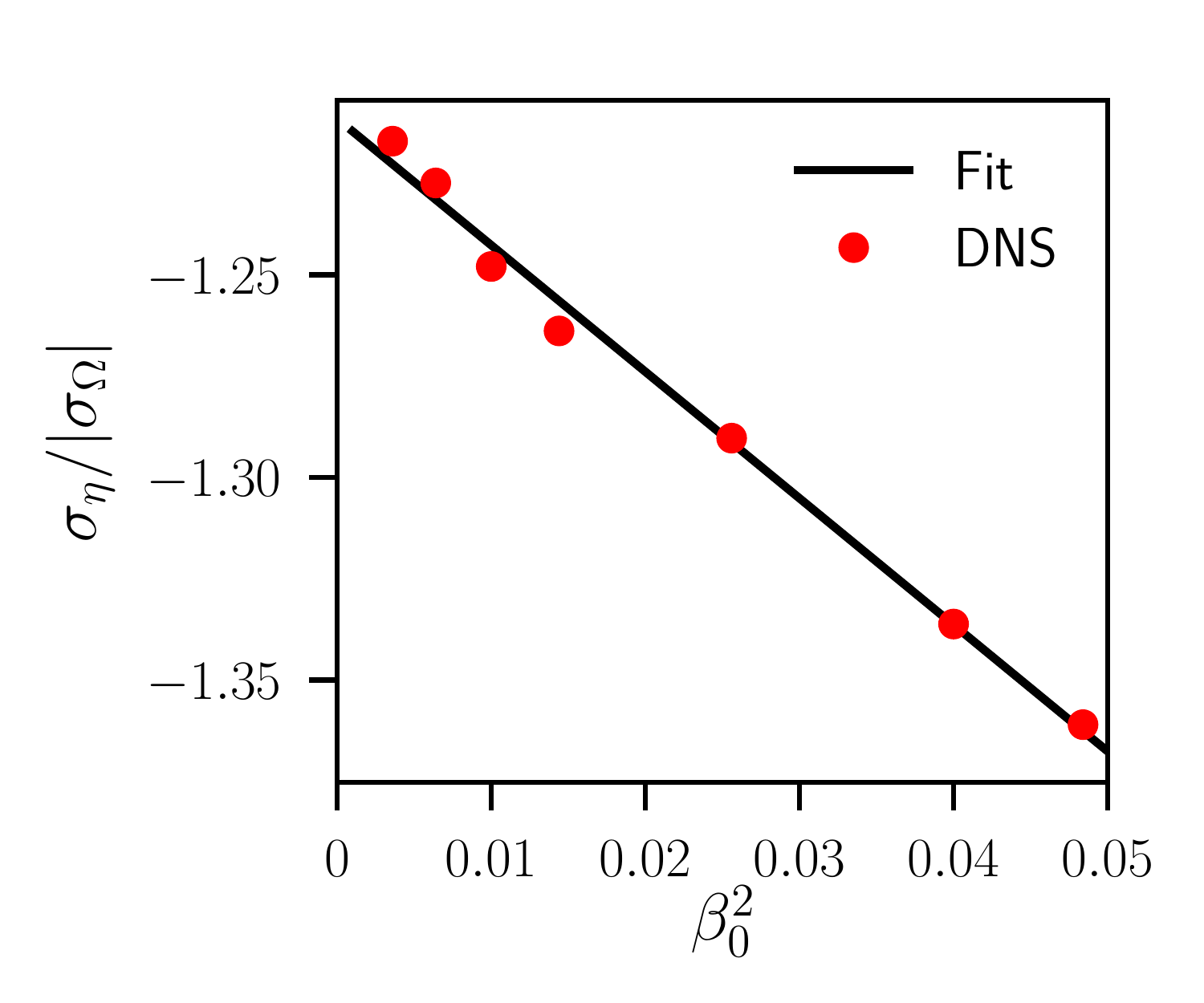} \\
	\end{tabular}
	\caption{Turbulent diffusion of magnetic field by tidal instability, as a function of equatorial ellipticity $\beta_0$. Ratio $\sigma_\eta/|\sigma_\Omega|$, with $\sigma_\eta$ the global decay rate (\ref{eq:sigmaturb1}) and $\sigma_\Omega$ the free decay rate (\ref{eq:sigmaFreeDecayB}) without tides. Simulations at $\Omega_0 = 0$, $Ek=10^{-4}$, $Pr=1$ and $Pm=0.1$. Solid lines are the least-squares fits.
	\emph{Top panel}: Weakly stratified regime ($N_0/\Omega_\text{s} = 0$), with $\sigma_\eta/|\sigma_\Omega| = -3.09 \, \beta_0 -1.00$. \emph{Bottom panel}: Strongly stratified regime ($N_0/\Omega_\text{s} = 10$) with $\sigma_\eta/|\sigma_\Omega| = -3.13 \, \beta_0^2 -1.21$.}
	\label{fig:mixingB}
\end{figure}

Figure \ref{fig:mixingB} shows the representative results for the two stratified regimes. We observe that the decay rate $\sigma_\eta$ is always larger than the free decay rate $\sigma_\Omega$ of the initial fossil field. Then, the striking feature is that we recover the two scalings as a function of the ellipticity, as predicted by our mixing-length theory. In the weakly stratified regime (top panel), numerical decay (\ref{eq:sigmaturb1}) agrees well with the linear scaling $\sigma_\eta \propto \beta_0$, consistent with mixing-length formula (\ref{eq:etaturmltlowN}). The agreement is even much better in the strongly stratified regime (bottom panel), obtaining the quadratic scaling $\sigma_\eta \propto \beta_0^2$ expected from (\ref{eq:etaturmlthighN}). 

We note that the observed enhancement generated by tidal instability is rather weak in the simulations. This is not due to the tidal amplitude, which is already two orders of magnitude larger than the typical values for binaries ($\beta_0 \simeq 10^{-1}$ in the numerics and $\beta_0 \simeq 10^{-3} - 10^{-2}$, see Table \ref{table:databinamics} below). This simply comes from the over-estimated value of the laminar Joule diffusion in the simulations (that is $Ek/Pm = 10^{-3}$). This makes the laminar and turbulent decay rates roughly comparable in amplitude. Simulations in the astrophysical regime (that is $Ek/Pm \leq 10^{-10}$) would show a stronger tidal effect. Yet, our simulations already support the trend predicted by mixing-length theory (\ref{eq:etaturmlthighN}). For stellar conditions, the latter predicts that the tidal decay rate would be much stronger than the laminar Joule decay rate (see the discussion in Sect. \ref{sec:discussion}).

Finally, the typical ratio of the volume averaged thermal and kinetic (dimensionless) energies, for the simulations in the strongly stratified regime (bottom panel of Fig. \ref{fig:mixingB}), is $E(\Theta)/E(\boldsymbol{u}) = 8.1 \pm 3.5$. 
This numerical value agrees very well with the theoretical scaling (\ref{eq:ratioPEKEbillant}) in the strongly stratified regime \citep{billant2001self}, yielding $E(\Theta)/E(\boldsymbol{u}) \sim N_0/\Omega_\text{s} = 10$ in dimensionless variables. This is another evidence of the validity of the mixing-length theory.

\section{Astrophysical discussion}
\label{sec:discussion}
We have obtained a consistent picture of tidal instability in an idealised set-up of radiative interiors. This predicts the linear onset (Sect. \ref{sec:onset}) and the nonlinear mixing induced by the saturated flows (Sect. \ref{sec:mixing}). For the sake of theoretical and numerical validations, we have only considered rather idealised stellar models, described in Sect. \ref{sec:model}. 
Then, the predictions have been successfully compared with proof-of-concept numerical simulations, paving the way for astrophysical applications. 

Indeed, we emphasise that the theory can a priori embrace more realistic stellar conditions. 
In particular, the mixing-length theory is only based on local dimensional arguments, that should remain valid for more realistic conditions. 
Therefore, we discuss now our findings in the context of tidally deformed and stably stratified (radiative) interiors. Notably, we are in the position to build a new physical scenario, that may explain the lower incidence of fossil fields in some short-period and non-synchronised binaries (Alecian et al., in prep.).

\subsection{A new scenario?}
We consider a close binary system with a radiative primary of mass $M_1$ and a secondary of mass $M_2$. 
The primary is pervaded by an initial fossil field $\boldsymbol{B}_0$.
Note that distinction between the primary and secondary is only made for convenience, such that the situation can be reversed in the scenario (if we are interested in the secondary). 
The orbital and spin angular velocities are respectively $\Omega_\text{orb}$ and $\Omega_\text{s}$. 
We focus on non-synchronised binaries in the orbital range $-1 \leq \Omega_0 \leq 3$, where $\Omega_0 = \Omega_\text{orb}/\Omega_\text{s}$ is the dimensionless orbital frequency. The orbits are almost circularised, but small orbital eccentricities $e \ll 1$ do not strongly modify the fate of tidal flows \citep{vidal2017inviscid}. We also focus on binaries with short-period systems, with typical periods of $T_\text{s} = 2 \pi/\Omega_\text{s} \leq 10$ days. 
Due to the combined action of the tides and the spin, the star is deformed into an triaxial ellipsoid \citep{chandrasekhar1969ellipsoidal,lai1993ellipsoidal,barker2016non}. 
The latter is characterised by a typical equatorial ellipticity $\beta_0$, estimated from the static bulge theory \citep{cebron2012elliptical,vidal2018magnetic}. For the bulge generated onto the primary, this reads 
\begin{equation}
	\beta_0 \sim \frac{3}{2} \frac{M_2}{M_1} \left ( \frac{R}{D} \right )^3,
	\label{eq:beta0bulge}
\end{equation}
where $R$ is the typical radius of the primary and $D$ is the typical distance separating the two bodies. 
The density stratification of the radiative envelope is measured by the typical dimensionless ratio $N_0/\Omega_\text{s}$, where $N_0$ is the typical Brunt-V\"ais\"al\"a frequency. A representative value for intermediate-mass stars is $N_0 \sim 10^{-3} \, \text{s}^{-1}$ \citep[e.g.][]{rieutord2006dynamics}, yielding a typical ratio $N_0/\Omega_\text{s} \gg 10$.

The tidal forcing sustains an equilibrium tidal velocity field \citep{remus2012equilibrium,vidal2017inviscid} in the primary fluid body. This equilibrium tidal flow can be nonlinearly coupled with inertial-gravity waves, triggering tidal instability. The dimensional growth rate $\sigma^*$ of tidal instability, which does not depend on stratification, is given by
\begin{equation}
	\sigma^* = \frac{(2 \widetilde{\Omega}_0+3)^2}{16(1+\widetilde{\Omega}_0)^2} |\Omega_\text{s}-\Omega_\text{orb}| \, \beta_0,
	\label{eq:sigmadimensional}
\end{equation}
with $\widetilde{\Omega}_0 = \Omega_0/(1-\Omega_0)$. 
In the saturated regime, tidal instability increases the internal mixing (due to turbulence).
In strongly stratified radiative interiors ($N_0/\Omega_\text{s} \gg 10$), the turbulent mixing generated by tidal instability is anisotropic, characterised by an eddy turbulent diffusivity $\mathcal{D}_\text{t}^\parallel$ in the direction of the self-gravity and by $\mathcal{D}_\text{t}^\perp \, (\gg \mathcal{D}_\text{t}^\parallel)$ in the other (horizontal) directions.

\begin{figure}
	\centering
	\includegraphics[width=0.4\textwidth]{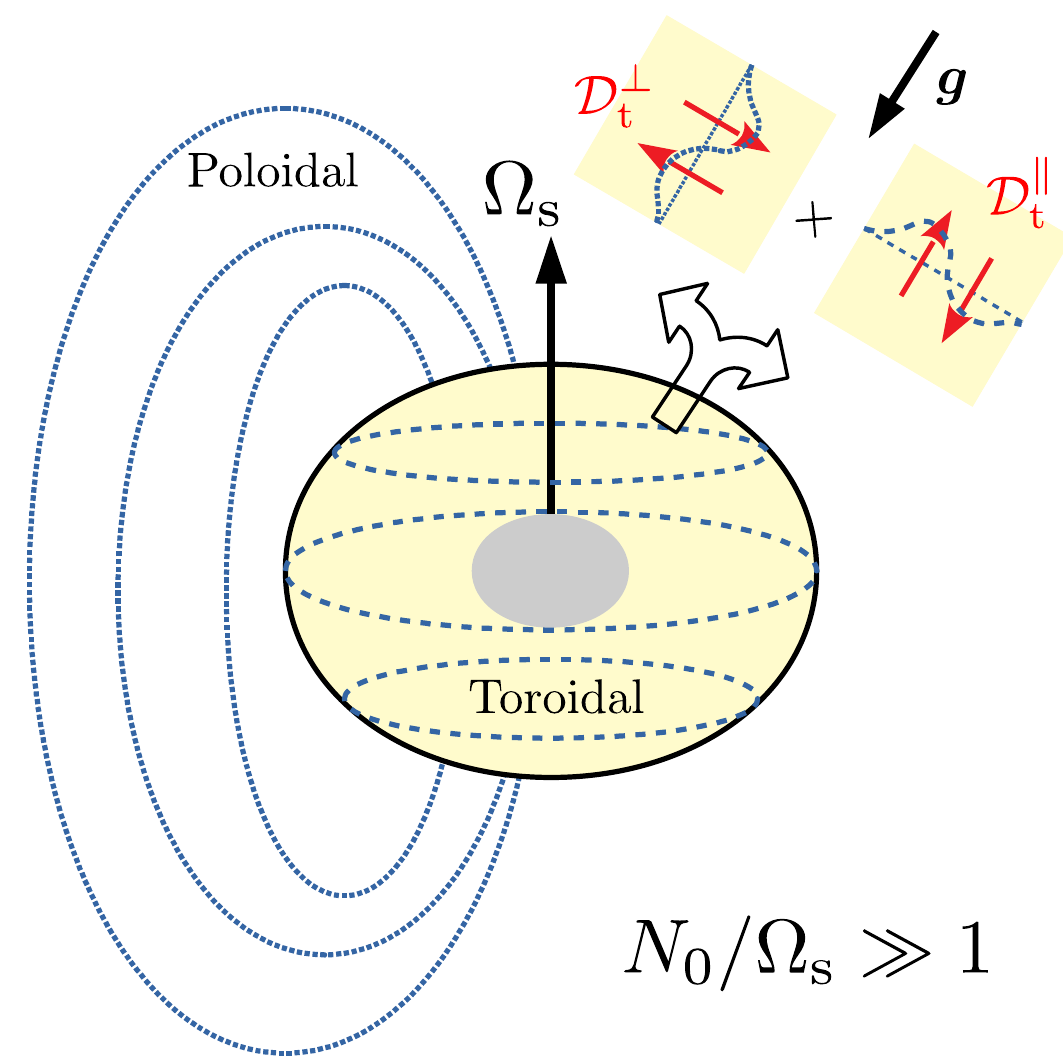}
	\caption{Anisotropic turbulent diffusion, generated by tidal instability, of poloidal (dotted) and toroidal (dashed) field lines of fossil field $\boldsymbol{B}_0$. A possible innermost convective core is represented.}
	\label{fig:scenarioB}
\end{figure}

Then, the turbulent mixing will dynamically increase the Joule decay of the fossil field $\boldsymbol{B}_0$. 
However, the latter field, containing both poloidal and toroidal components (to be in quasi-static magnetic equilibrium in the initial stage), will undergo an enhanced anisotropic turbulent Joule diffusion. The mechanism is illustrated in Fig. \ref{fig:scenarioB}. 
On the one hand, the poloidal components, which are mainly along the normal direction, would be preferentially dissipated by the (large) eddy diffusivity $\mathcal{D}_\text{t}^\perp$ in the horizontal directions. On the other hand, the toroidal components, trapped in the stellar interior because they have only horizontal components, are preferentially mixed by the (small) eddy diffusivity $\mathcal{D}_\text{t}^\parallel$ in the normal direction. 
Thus, poloidal and toroidal field lines are dissipated on different turbulent timescales. For the poloidal components which can be observed at the stellar surface, tidal instability would yield a global magnetic dissipation within the stellar interior on a few turbulent timescales $\tau_\text{t}$ (at the position $r_\text{l} \leq R$), given by
\begin{equation}
	\tau_\text{t} \propto \frac{r_\text{l}^2}{\mathcal{D}_\text{t}^\perp} \sim \frac{K_\alpha}{\beta_0^2 \, \Omega_\text{s} (1 - \Omega_0)^2}
	\label{eq:timeturmlt}
\end{equation}
with the pre-factor $K_\alpha \sim 30 - 50$ estimated from the numerical pre-factors in formulas (\ref{eq:etaturmlthighN}). 
Timescale (\ref{eq:timeturmlt}) is the (fast) turbulent timescale in the perpendicular (horizontal) directions. 
In addition, the magnetic field would also die out in the presence of rotation on dynamic timescale (\ref{eq:braithwaite2012MC}) of the (slow) Magneto-Coriolis waves, as shown by \citet{braithwaite2012weak}. 

\begin{table*}
    \centering
	\caption{Physical and orbital characteristics of non-synchronised and non-magnetic binary systems, surveyed by the BinaMIcS collaboration (Alecian et al., in prep.).}
    {\small
	\begin{tabular}{lccccccccccccc}
	\hline\hline
	\multicolumn{2}{c}{System} & $M_1$ & $M_2$ & $R$ ($M_1$) & $R$ ($M_2$) & $D$ & $T_\text{s}$ ($M_1$) & $T_\text{s}$ ($M_2$) & $T_\text{orb}$ & $e$ & \multicolumn{2}{c}{$\beta_0$}\\
	{} & {} & ($M_\odot$) & ($M_\odot$) & ($R_\odot$) & ($R_\odot$) & ($R_\odot$) & (days) & (days) & (days) & &  Body 1 &  Body 2 \\
	\hline 
	$\circ$ & \object{HD 23642} & $2.22$ & $1.57$ & $1.84$ & $1.57$ & $11.96$ & $2.49$ & $2.45$ & $2.46$ & $0.00$ & $3.9 \times 10^{-3}$ & $4.8 \times 10^{-3}$ & \\
	$\bigtriangledown$ & \object{HD 24133} & $1.39$ & $1.31$ & $1.78$ & $1.49$ & $5.042$ & $0.827$ & $0.783$ & $0.80$ & $0.00$ & $6.2 \times 10^{-2}$ & $4.1 \times 10^{-2}$ \\
	$\bigtriangleup$ & \object{HD 24909} & $3.53$ & $1.72$ & $2.47$ & $1.53$ & $10.59$ & $1.8$ & $1.8$ & $1.74$ & $0.07$ & $9.3 \times 10^{-3}$ & $9.3 \times 10^{-3}$ \\
	$\triangleleft$ & \object{HD 25638} & $14.3$ & $10.7$ & $8.91$ & $6.70$ & $23.97$ & $3.01$ & $2.76$ & $2.70$ & $0.00$ & $5.8 \times 10^{-2}$ & $4.4 \times 10^{-2}$\\
	$\triangleright$ & \object{HD 25833} & $5.36$ & $4.90$ & $2.99$ & $2.60$ & $14.67$ & $2.0$ & $1.7$ & $2.03$ & $0.07$ & $1.2 \times 10^{-2}$ & $9.1 \times 10^{-3}$ \\
	$\octagon$ & \object{HD 32964} & $2.63$ & $2.57$ & $1.95$ & $1.92$ & $22.90$ & $5.57$ & $5.55$ & $5.52$ & $0.08$ & $9.0 \times 10^{-4}$ & $9.0 \times 10^{-4}$ \\
	$\square$ & \object{HD 34364} & $2.48$ & $2.29$ & $1.78$ & $1.82$ & $18.24$ & $3.90$ & $4.01$ & $4.13$ & $0.00$ & $1.3 \times 10^{-3}$ & $1.6 \times 10^{-3}$ \\
	$\pentagon$ & \object{HD 36486} & $24.0$ & $8.40$ & $16.5$ & $6.50$ & $43.00$ & $6.24$ & $2.13$ & $5.73$ & $0.11$ & $3.0 \times 10^{-2}$ & $1.5 \times 10^{-3}$ \\
	$\hexagon$ & \object{HD 150136} & $62.6$ & $39.5$ & $13.1$ & $9.54$ & $38.00$ & $2.9$ & $2.7$ & $2.67$ & $0.00$ & $3.9 \times 10^{-2}$ & $3.8 \times 10^{-2}$ \\
	\hline
	\end{tabular}
	}
\tablefoot{The masses $[M_1, M_2]$ of the primary and the secondary bodies are given in Sun mass unit $M_\odot$. The typical stellar radius $R$ and the typical distance $D$ between the two bodies is given in Sun radius unit $R_\odot$. Spin and orbital periods $[T_\text{s}, T_\text{orb}]$ are expressed in days. Spin and angular velocities are defined as $\Omega_\text{s} = 2\pi/T_\text{s}$ and $ \Omega_\text{orb} = 2\pi/T_\text{orb}$. Note that $T_\text{s}$ has been estimated by assuming aligned spin-orbit systems. Symbols refer to Fig. \ref{fig:DisruptionB}. \object{HD 23642}: \citet{groenewegen2007pleiades}; \object{HD 24133}:  \citet{clausen2010absolute}; \object{HD 24909}: \citet{deugirmenci1997photometry}; \object{HD 25638}: \citet{tamajo2012asiago}; \object{HD 25833}: \citet{gimenez1994ag}; \object{HD 32964}: \citet{makaganiuk2011chemical}; \object{HD 34364}: \citet{nordstrom1994radii}; \object{HD 36486}: \citet{shenar2015coordinated}; \object{HD 150136}: \citet{mahy2012evidence}.}
	\label{table:databinamics}
\end{table*}

\subsection{Non-magnetic binaries}
We assess here the relevance of the tidal scenario for short-period massive binary systems. Non-magnetic and non-synchronised ($\Omega_0 \neq 1$) binaries are given in Table \ref{table:databinamics}. They have been surveyed by the BinaMIcS collaboration (Alecian et al., in prep.). 
The predictions of the tidal scenario for these binary systems are given in Table \ref{table:dataextrapolation}. 
All these close-binaries are rapidly rotating and undergo strong tidal effects (in the two bodies), as measured by the large values of the ellipticity $\beta_0 \sim 10^{-3} - 10^{-2}$. The strong tides should trigger quickly tidal instability, growing on the typical timescale $(\sigma^*)^{-1} \simeq \mathcal{O}(10^3)$ years. This is much shorter than the lifetime of these stars, about $\tau_\text{MS} \sim 10^9$ years for a star of mass $M_1 = 2 M_\odot$ on the main sequence. Hence, tidal instability is likely to be present in these non-synchronised binaries. 

\begin{figure}
    \centering
    \includegraphics[width=0.48\textwidth]{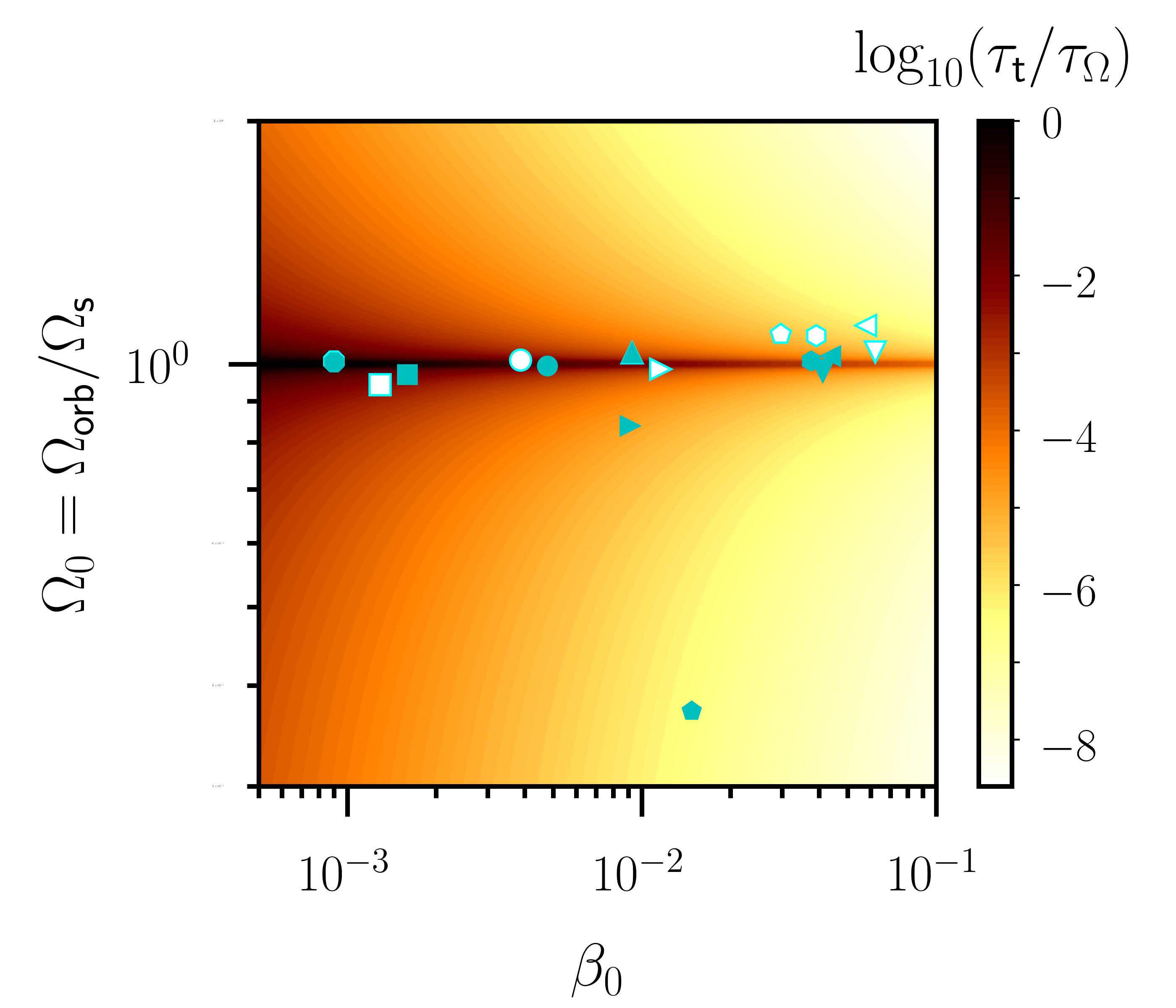}
    \caption{Turbulent magnetic decay $\tau_\text{t}$ (\ref{eq:timeturmlt}) of fossil fields , normalised by laminar Ohmic timescale $\tau_\Omega \sim (\Omega_\text{s} \, Ek/Pm)^{-1}$, as a function of equatorial ellipticity $\beta_0$ and dimensionless orbital angular frequency $\Omega_0 = \Omega_\text{orb}/\Omega_s$. Non-magnetic close binaries are illustrated by the symbols given in Table \ref{table:databinamics}. Large (white) symbols refer to body 1 of the considered binary, whereas small (cyan) symbols refer to body 2. Computations at $Ek/Pm = 10^{-12}$ and $K_\alpha = 30$.}
    \label{fig:DisruptionB}
\end{figure}

Then, typical values for turbulent timescale (\ref{eq:timeturmlt}) are $\tau_\text{t} \in [10^3, 10^7]$ years, except for \object{HD 23642} and \object{HD 32964} which are less affected by tidal instability (smaller $\beta_0$). Thus, the turbulent Joule diffusion of the initial fossil fields may occur on timescales much shorter than the stellar lifetime, typically $\tau_\text{t}/\tau_\text{MS} \ll 10^{-3}$ for the most favourable systems. 
Turbulent timescale (\ref{eq:timeturmlt}) is also often smaller that the timescale for the laminar Ohmic diffusion of the magnetic field in the absence of turbulence $\tau_\Omega \propto (\Omega_\text{s} \, Ek/Pm)^{-1}$. As illustrated in Fig. \ref{fig:DisruptionB}, we get $\tau_\text{t}/\tau_\Omega \leq 10^{-2}$ (except for \object{HD 23642} and \object{HD 32964}). Similarly, for several systems, $\tau_\text{t}$ is smaller than the dynamic timescale $\tau_\text{MC}$ proposed by \citet{braithwaite2012weak}, given by expression (\ref{eq:braithwaite2012MC}). 

Therefore, nonlinear tidal flows generated by tidal instability in non-synchronised close binaries may sustain an enhanced turbulent Joule diffusion of the fossil fields, occurring on timescales that are often shorter than the stellar lifetime. This may explain the scarcity of significant magnetic fields at the surface of some massive stars in short-period binaries.

\begin{table*}
	\caption{Predictions of tidal scenario for (non-magnetic) close binaries described in Table \ref{table:databinamics}.}
    \centering
    {\small
	\begin{tabular}{lcccccccc}
	\hline\hline
	System & \multicolumn{2}{c}{$\sigma^*$ (1/year)} & \multicolumn{2}{c}{$\tau_\text{t}$ (years)} & \multicolumn{2}{c}{$\tau_\text{t}/\tau_\Omega$} & \multicolumn{2}{c}{$\tau_\text{t}/\tau_\text{MC}$} \\
	\cmidrule{2-9}
	{} & Body 1 & Body 2 & Body 1 & Body 2 & Body 1 & Body 2 & Body 1 & Body 2 \\
	\hline 
	\object{HD 23642} & $1.03 \times 10^{-2}$ & $5.07 \times 10^{-3}$ & $1.58 \times 10^{7}$ & $6.88 \times 10^{7}$ & $1.46 \times 10^{-2}$ & $6.44 \times 10^{-2}$ & $1.5 \times 10^{+0}$ & $6.4 \times 10^{+0}$ \\
	\object{HD 24133} & $1.46 \times 10^{+0}$ & $6.11 \times 10^{-1}$ & $2.26 \times 10^{3}$ & $1.53 \times 10^{4}$ & $6.26 \times 10^{-6}$ & $4.48 \times 10^{-5}$ & $6.3 \times 10^{-4}$ & $4.5 \times 10^{-3}$ \\
	\object{HD 24909} & $9.25 \times 10^{-2}$ & $9.26 \times 10^{-2}$ & $2.61 \times 10^{5}$ & $2.61 \times 10^{5}$ & $3.33 \times 10^{-4}$ & $3.32 \times 10^{-4}$ & $3.3 \times 10^{-2}$ & $3.3 \times 10^{-2}$ \\
	\object{HD 25638} & $1.12 \times 10^{+0}$ & $2.03 \times 10^{-1}$ & $8.91 \times 10^{2}$ & $3.60 \times 10^{4}$ & $6.79 \times 10^{-7}$ & $2.99 \times 10^{-5}$ & $6.8 \times 10^{-5}$ & $3.0 \times 10^{-3}$ \\
	\object{HD 25833} & $4.78 \times 10^{-2}$ & $5.83 \times 10^{-1}$ & $9.68 \times 10^{5}$ & $1.02 \times 10^{4}$ & $1.11 \times 10^{-3}$ & $1.37 \times 10^{-5}$ & $1.1 \times 10^{-1}$ & $1.4 \times 10^{-3}$ \\
	\object{HD 32964} & $7.89 \times 10^{-4}$ & $4.61 \times 10^{-4}$ & $1.22 \times 10^{9}$ & $3.61 \times 10^{9}$ & $5.01 \times 10^{-1}$ & $1.49 \times 10^{+0}$ & $5.0 \times 10^{+1}$ & $1.5 \times 10^{+2}$ \\
	\object{HD 34364} & $1.14 \times 10^{-2}$ & $7.12 \times 10^{-3}$ & $9.53 \times 10^{6}$ & $2.25 \times 10^{7}$ & $5.60 \times 10^{-3}$ & $1.28 \times 10^{-2}$ & $5.6 \times 10^{-1}$ & $1.3 \times 10^{+0}$ \\
	\object{HD 36486} & $2.20 \times 10^{-1}$ & $4.32 \times 10^{+0}$ & $1.18 \times 10^{4}$ & $3.22 \times 10^{2}$ & $4.35 \times 10^{-6}$ & $3.47 \times 10^{-7}$ & $4.3 \times 10^{-4}$ & $3.5 \times 10^{-5}$ \\
	\object{HD 150136} & $5.98 \times 10^{-1}$ & $7.53 \times 10^{-2}$ & $3.49 \times 10^{3}$ & $7.37 \times 10^{2}$ & $2.76 \times 10^{-6}$ & $6.26 \times 10^{-7}$ & $2.8 \times 10^{-4}$ & $6.3 \times 10^{-5}$ \\
	\hline
	\end{tabular}
	}
	\label{table:dataextrapolation}
	\tablefoot{We have taken as representative value for the dimensional Brunt-V\"ais\"al\"a frequency  $N_0 = 10^{-3} \, \text{s}^{-1}$ \citep[e.g.][]{rieutord2006dynamics}. The equatorial ellipticity $\beta_0$ is given by expression (\ref{eq:beta0bulge}). The dimensional growth rate $\sigma^*$ is given by formula (\ref{eq:sigmadimensional}). The timescale of turbulent Joule diffusion $\tau_\text{t}$ is given by formula (\ref{eq:timeturmlt}) with $K_\alpha = 30$. The laminar Ohmic diffusive timescale is $\tau_\Omega \sim (\Omega_\text{s} \, Ek / Pm)^{-1}$ (in dimensional units of $\Omega_\text{s}$) with $Ek/Pm \sim 10^{-12}$. The dynamic timescale associated with the propagation of (slow) Magneto-Coriolis waves is $\tau_\text{MC} \sim (\Omega_\text{s} \,Le^2)^{-1}$ \citep{braithwaite2012weak}, with $Le \sim 10^{-5}$.}
\end{table*}

\subsection{Magnetic binaries}
\begin{table*}
    \centering
	\caption{Physical and orbital characteristics of magnetic binary systems surveyed by the BinaMIcS collaboration \citep{folsom2013orbital,shultz2015detection,shultz2017hd,shultz2018magnetic}.}
{\small
	\begin{tabular}{lcccccccccccc}
	\hline\hline
	System & $M_1$ & $M_2$ & $R$ ($M_1$) & $R$ ($M_2$) & $D$ & $T_\text{s}$ ($M_1$) & $T_\text{s}$ ($M_2$) & $T_\text{orb}$ & Eccentricity & $B_0^*$ ($M_1$) & $B_0^*$ ($M_2$) \\
	{} & ($M_\odot$) & ($M_\odot$) & ($R_\odot$) & ($R_\odot$) & ($R_\odot$) & (days) & (days) & (days) & $e$ & (kG) & (kG) \\
	\hline 
	\object{HD 156324} & $8.5$ & $4.1$ & $3.8$ & $2.3$ & $13.2$ & $1.58$ & $1.58$ & $1.58$ & $0.0$ & $14$ & $<2.6$ \\
	\object{HD 98088} & $2.19$ & $1.67$ & $2.76$ & $1.77$ & $21.7$ & $5.905$ & $5.905$ & $5.905$ & $0.18$ & $3.9$ & $<1.6$ \\ 
	\\
	\object{$\epsilon$ Lupi} (corot) & $8.7$ & $7.3$ & $4.7$ & $3.8$ & $29.2$ & $2.30$ & $2.5$ & $4.56$ & $0.277$ & $0.9$ & $0.6$\\
	\object{$\epsilon$ Lupi} (slow) & $8.7$ & $7.3$ & $4.7$ & $3.8$ & $29.2$ & $6.4$ & $7.1$ & $4.56$ & $0.277$ & $0.9$ & $0.6$\\
	\object{$\epsilon$ Lupi} (fast) & $8.7$ & $7.3$ & $4.7$ & $3.8$ & $29.2$ & $0.40$ & $0.32$ & $4.56$ & $0.277$ & $0.9$ & $0.6$\\
	\hline
	\end{tabular}
	}
	\label{table:databinamics2}
	\tablefoot{Masses $[M_1, M_2]$ of primary and secondary bodies are given in Sun mass unit $M_\odot$. The typical stellar radius $R$ and the typical distance $D$ between the two bodies is given in Sun radius unit $R_\odot$. The spin and orbital periods $[T_\text{s}, T_\text{orb}]$ are expressed in days. They yield the spin and angular velocities $[\Omega_\text{s} = 2\pi/T_\text{s}, \Omega_\text{orb} = 2\pi/T_\text{orb}]$. The typical surface magnetic field $B_0^*$, believed to be of fossil origin, is given in kiloGauss (kG) for the two components. \object{HD 156324} and \object{HD 98088} are synchronised systems (see Appendix \ref{appendix:ldei}), whereas {$\epsilon$ Lupi} system is not synchronised.}
\end{table*}

We give in Table \ref{table:databinamics2} the orbital properties of some scarce magnetic binaries, analysed by the BinaMIcS collaboration. 
They were already known to be magnetic, such as \object{HD 98088} \citep{babcock1958catalog,abt1968mass,carrier2002multiplicity}, 
\object{$\epsilon$ Lupi} \citep{shultz2015detection} and \object{HD 156324} \citep{alecian2014discovery}.
The aforementioned tidal scenario would suggest that (strong) magnetic fields may be anomalies in short-period massive binaries. However, their existence does not necessarily challenge the tidal scenario. 

We note that \object{HD 156324} and \object{HD 98088} are synchronised. The fate of tidal instability in synchronised orbits ($\Omega_0 = 1$) is discussed in Appendix \ref{appendix:ldei}. 
On the one hand, system \object{HD 156324} is nearly circularised \citep{shultz2017hd}, whereas non-circular orbits are required for the tidal mechanism to operate in synchronised systems \citep[e.g.][]{vidal2017inviscid}. Hence, the tidal mechanism is not currently relevant for \object{HD 156324}. This may explain why the fossil field is still observed.  
On the other hand, \object{HD 98088} is not circularised such that nonlinear tidal mixing would be expected. However, as shown in Appendix \ref{appendix:ldei}, formula (\ref{eq:timeturmlt}) for the typical turbulent timescale ought to be reduced in synchronised systems, such that $(1-\Omega_0)^2 \sim \epsilon_l^2$ where $\epsilon_l \ll 2e$ is the dimensionless amplitude of differential rotation due to the elliptical orbit \citep{cebron2012elliptical,vidal2017inviscid}. 
Based on the accuracy of the measured periods in Table \ref{table:databinamics2}, we may assume $\epsilon_l \leq 10^{-3}$, such that the turbulent timescale $\tau_\text{t}$, given by formula (\ref{eq:timeturmlt_LDEI}), is expected to be much larger in \object{HD 98088} than for the systems of Table \ref{table:dataextrapolation} (for similar values of the equatorial ellipticity $\beta_0 \sim 10^{-3}$). Therefore, the existence of the (synchronised) magnetic binaries \object{HD 156324} and \object{HD 98088} appears to be consistent with the tidal scenario.
However, the tidal mechanism may have occurred before the synchronisation and/or the circularisation of the systems. Indeed, observations show that circularisation and synchronisation processes are effective for radiative stars \citep[e.g.][]{giuricin1984synchronizationa,giuricin1984synchronizationb,zimmerman2017pseudosynchronization}.
On the one hand, the radiative damping of the dynamical tide has received  attention in radiative stars \citep[e.g.][]{zahn1975dynamical,zahn1977tidal}. 
On the other hand, synchronisation mechanisms have been much less studied in radiative interiors \citep[e.g.][]{rocca1989tidal,rocca1987forced,witte1999tidal,witte2001tidal}, and the comparison with the observations is less satisfactory \citep[e.g.][]{goupil2008observational,zimmerman2017pseudosynchronization}.
Understanding these two processes in radiative stars still deserves further work, notably to consider the overlooked effects of tidal instability in short-period binaries.

Finally, the case of \object{$\epsilon$ Lupi} system \citep[e.g.][]{uytterhoeven2005orbit,shultz2015detection} is more intricate. Nonlinear tidal mixing should occur within these stars, with a typical turbulent timescale $\tau_\text{t} \sim 10^3$ years. 
The fossil field may be currently dissipated by the tidal turbulence, but the process may have not last long enough to yield vanishing observable fields. Another possibility is that these magnetic fields are internally regenerated by dynamo action, to balance the decay due to the nonlinear tidal flows. Such a (currently speculative) mechanism may be particularly relevant for the rapidly rotating component of \object{$\epsilon$ Lupi} in Table \ref{table:databinamics2}. Several dynamo mechanisms may be advocated, for instance driven by differentially rotating flows \citep{braithwaite2006differential}, baroclinic flows \citep{simitev2017baroclinically} or even tidal instability \citep{vidal2018magnetic}. Though the dynamo action of tides in strongly stratified interiors remains elusive, the scaling law for the magnetic field strength at the stellar surface, proposed by \citet{vidal2018magnetic}, would yield $|\boldsymbol{B}_0| \sim 0.1 - 1$ kG. This is the order of magnitude of the observed surface fields. Thus, understanding the origin of the magnetic fields in the \object{$\epsilon$ Lupi} system deserves future studies.

\section{Conclusion}
\label{sec:ccl}
\subsection{Summary}
In this work, we have investigated nonlinear tides in short-period massive binaries, motivated by the puzzling lower magnetic incidence of close binaries compared to isolated stars \citep{alecian2017fossil}. To do so, we have adopted an idealised model for rapidly rotating stratified fluids within the Boussinesq approximation. 
This model consistently takes into account the ingredients encountered in massive binaries, namely the combination of rotation and non-isentropic stratification, the tidal distortion (on coplanar and aligned orbits) and the leading-order magnetic effects.
We have revisited the fluid instabilities triggered by the nonlinear tides in the system \citep{vidal2018magnetic}, by combining analytical computations and proof-of-concept simulations. 

Firstly, we have studied the linear onset of tidal instability in non-synchronised, stratified fluid masses. Within a single framework, we have unified all the previous existing stability analyses and we have unravelled new phenomena. We have shown that tidal instability in radiative stratified interiors is due to parametric resonances between inertial-gravity waves and the underlying equilibrium tidal flow, for any orbit in the range $-1 \leq \Omega_0 \leq 3$. Within this orbital range, tidal instability is weakened by barotropic stratification on the polar axis \citep{miyazaki1991axisymmetric,miyazaki1993elliptical} and in the equatorial plane. On the contrary, baroclinic stratification does increase the growth rate of tidal instability \citep{kerswell1993elliptical,le2006thermo}. However, the striking feature is that tidal instability onsets with a maximum growth rate which is unaffected by stratification. The instability is triggered in volume along three-dimensional conical layers, whose position depends solely on the orbital parameter $\Omega_0$.
In the other orbital range $\Omega_0 \leq -1$ and $\Omega_0 \geq 3$, that is in the forbidden zone of tidal instability in homogeneous fluids \citep[e.g.][]{le2000three}, tidal instability can be generated by parametric resonances of gravito-inertial waves, provided that stratification is strong enough for the considered orbital configuration. This provides a theoretical explanation of the instability mechanism investigated numerically in \citet{le2018parametric}.

Secondly, we have developed a mixing-length theory \citep[e.g.][]{tennekes1972first} of the anisotropic turbulent mixing, sustained by tidal instability in the orbital regime $-1 \leq \Omega_0 \leq 3$. For strongly stratified interiors, we have modelled the anisotropic turbulent mixing by introducing two turbulent eddy diffusivities, one describing the mixing in the direction of the gravity field and the second in the other (horizontal) directions. We have shown that these two turbulent diffusivities should scale as $\beta_0^2$, where $\beta_0$ is the equatorial ellipticity of the equilibrium tide. We have assessed these scalings against proof-of-concept simulations, by using the numerical method introduced in \citet{vidal2018magnetic}.

Finally, we have used the mixing-length theory to extrapolate the numerical results towards more realistic stellar conditions. 
We have built a new physical scenario, predicting an enhanced Joule diffusion of the fossil fields due to the turbulent mixing induced by tidal instability in short-period (non-coalescing) massive binaries. We have applied it to a subset of short-period binaries, analysed by the BinaMIcS collaboration (Alecian et al., in prep.). This scenario may (partially) explain the lower incidence of surface magnetic fields in some short-period binaries (compared to isolated stars). Indeed, we predict a turbulent Joule diffusion of the fossil fields occurring in a few million years for the most favourable systems. This is much shorter than the (laminar) Joule diffusion timescale of the fossil fields, and similarly than the typical lifetime of these stars. 
Therefore, we cannot rule out a priori the tidal mechanism to explain the scarcity of massive magnetic stars in close binary systems.

\subsection{Perspectives}
We have shown that the tidal mechanism is plausible, because close binaries are known to be strongly deformed by tides. Then, future studies should strive to assess the likelihood of this new mechanism with more realistic physical models. 
Indeed, we have only handled the key physical ingredients. Many improvements are worth doing on the numerical and theoretical fronts.

Firstly, the validity of mixing-length predictions for the magnetic diffusivity is questionable. Though they are commonly used in hyromagnetic turbulence \citep[e.g.][]{yousef2003turbulent,kapyla2019turbulent}, \citet{vainshtein1991turbulent} proposed that even weak large-scale magnetic fields may suppress the turbulent magnetic diffusion. 
This behaviour has been obtained in simulations of non-rotating, two-dimensional turbulence \citep[e.g.][]{cattaneo1991suppression,cattaneo1994effects,kondic2016decay}. 
However, the relevance of this inhibiting mechanism for three-dimensional, rotating and tidally driven turbulence remains unclear, notably because Alfv\'en waves do not play (a priori) a significant role in the tidal turbulent mixing (contrary to inertial waves). Indeed, this seems in contradiction with the turbulent hydromagnetic simulations of \citet{barker2013nonb}, who showed that a weak magnetic field can instead sustain small-scale tidal turbulence. Thus, investigating this effect in tidally forced turbulence seems necessary, by performing demanding simulations of the consistent rotating hydromagnetic set-up.

Secondly, it would be interesting to examine if (secondary) shear instabilities are sustained by nonlinear tides in the strongly stratified regime. Shear instabilities are common in radiative interiors \citep[e.g.][]{mathis2004shear,mathis2018anisotropic}, which undergo differential rotation \citep{goldreich1967differential}. To do so, the usual diffusionless instability condition for shear instabilities ought to be modified in radiative interiors, to take the thermal diffusivity into account  \citep{townsend1958effects,zahn1974rotational}.
In the presence of turbulent tidal flows, secondary shear instabilities may exist if
\begin{equation}
	Ri_\text{t} \, Pe_\text{t} \leq 1,
	\label{eq:richardson}
\end{equation}
with $Ri_\text{t} = N_0^2/(u_\text{t}/l_\text{t}^\parallel)^2$ the turbulent Richardson number and $Pe_\text{t} = u_\text{t} l_\text{t}^\parallel / \mathcal{D}_\text{t}^\parallel$ the turbulent P\'eclet number. By using our mixing-length predictions, a typical estimate would be $Ri_\text{t} Pe_\text{t} \sim 1$ in the strongly stratified regime. Thus, such secondary shear instabilities might be triggered by the nonlinear tidal flows. This may increase the turbulent diffusion coefficients.

Then, a natural extension would be to investigate consistently the interplay between tidal instability and differential rotation, which would result from in-situ baroclinic torques \citep[e.g.][]{busse1981eddington,busse1982problem,rieutord2006dynamics}. 
Whether differential rotation is important for the tidal mixing is elusive, for instance because differential rotation is damped by several hydromagnetic effects \citep{moss1992magnetic,spruit1999differential,arlt2003differential,rudiger2013astrophysical,rudiger2015angular,jouve2015three}. 
Nonetheless, elliptical (tidal) instability does exist in differentially rotating elliptical flows, as shown in fundamental fluid mechanics \citep{eloy1999three,lacaze2007elliptic}. The properties of the waves for more astrophysically relevant profiles of differential rotation can be investigated in global models \citep{friedlander1989hydromagnetic,mirouh2016gravito}, such that extending the present theory seems achievable. 
Closely related to the study of differential rotation is the study of baroclinic flows \citep[e.g.][]{kitchatinov2014baroclinic,caleo2016radiative,simitev2017baroclinically}. We have shown that baroclinic stratification does enhance tidal instability, as first noticed by \citet{kerswell1993elliptical} and \citet{le2006thermo}. Thus, we may even expect a stronger turbulent tidal mixing in baroclinic radiative interiors. 

Radiative stars also host innermost convective cores. Thus, the outcome of tidal instability in shells should be considered. 
The tidal (elliptical) instability does exist in shells, as confirmed experimentally and numerically for homogeneous fluids \citep{aldridge1997elliptical,seyed2000numerical,lacaze2005elliptical,seyed2004elliptical,lemasquerier2017libration}. Indeed, the local stability theory we have presented remains formally valid in shells. 
Hence, we do not expect any significant difference for stratified fluids at the onset. Yet, boundary effects on the turbulent tidal mixing remain to be determined.

Another daunting perspective is to account for compressibility. Using the Boussinesq approximation seems exaggerated in global models of stellar interiors \citep{spiegel1960boussinesq}. However, the influence of compressibility is apparently negligible at the onset of tidal instability \citep{clausen2014elliptical}. This is one of the reasons why we have adopted the Boussinesq approximation. 
Moreover, our mixing-length theory only invokes local estimates. In particular, we may naively expect radial turbulent diffusion (\ref{eq:etaturmlthighN_parallel}) to be only governed by the local value of stratification (rather independently of its origin). Moreover, compressibility would barely modify the (strongest) horizontal mixing (\ref{eq:etaturmlthighN_perp}), because horizontal motions are less inhibited by compressibility. Therefore, our typical turbulent timescale (\ref{eq:timeturmlt}) may still be relevant in compressible interiors. Clarifying the effects of compressibility deserves future works, both in the linear and nonlinear regimes.

Finally, the scarce non-synchronised magnetic binaries \citep{carrier2002multiplicity,shultz2015detection,alecian2017fossil,kochukhov2018hd} seem to challenge the general trend of the tidal scenario, predicting a lack of magnetic massive stars in short-period binaries.
These fields do not appear to be strongly dissipated by the nonlinear tidal flows. If the tidal mechanism remains valid by including the aforementioned proposed improvements, they might be dynamically regenerated in situ by dynamo action. 
For instance, tides do sustain dynamo action at small-scale \citep{barker2013nonb} and large-scale \citep{cebron2014tidally,reddy2018turbulent} in homogeneous fluids, and also in weakly stratified interiors \citep{vidal2018magnetic}. Yet, the dynamo capability of tides remains elusive in strongly stratified interiors \citep{vidal2018magnetic}. Baroclinic flows are another possible candidate, because they are dynamo capable \citep{simitev2017baroclinically}. They may also favour the radial mixing  generated by tidal instability, which is a necessary ingredient for dynamo action \citep{kaiser2017robustness}.
This certainly deserves future works to investigate dynamo magnetic fields in more realistic models of radiative stars. 

\begin{acknowledgements}
JV was initially supported by a Ph.D grant from the French \emph{Minist\`ere de l'Enseignement Sup\'erieur et de la Recherche} and later partly by STFC Grant ST/R00059X/1. 
DC was funded by the French {\it Agence Nationale de la Recherche} under grant ANR-14-CE33-0012 (MagLune) and by the 2017 TelluS program from CNRS-INSU (PNP) AO2017-1040353. AuD acknowledges support from NASA through Chandra Award number TM7-18001X issued by the Chandra X-ray Observatory Center, which is operated by the Smithsonian Astrophysical Observatory for and on behalf of NASA under contract NAS8-03060. 
EA and the BinaMIcS collaboration acknowledges financial support from "Programme National de Physique Stellaire" (PNPS) of CNRS/INSU (France).
JV and DC kindly aknowledges Dr N. Schaeffer (ISTerre, UGA) for several suggestions improving the quality of the paper and for fruitful discussions on the mixing observed in the numerical simulations performed with the XSHELLS code. XSHELLS is developed and maintained by Dr N. Schaeffer at \url{https://bitbucket.org/nschaeff/xshells}. 
JV aknowledges EA for the invitation to the BinaMIcS Workshop \#5, where came the idea to explain the lack of magnetic binaries by using tidal instability. 
AuD and EA aknowledge Dr S. Mathis (CEA, Paris Saclay) and the BinaMIcS collaboration for fruitful discussions.
The authors acknowledge Dr F. Gallet (IPAG, UGA), who validated the typical estimate of the Brunt-V\"ais\"al\"a frequency in massive stars by using a stellar evolution code. 
The XSHELLS code is freely available at \url{https://bitbucket.org/nschaeff/}.
Computations were performed on the Froggy platform of CIMENT (\url{https://ciment.ujf-grenoble.fr}), supported by the Rh\^one-Alpes region (CPER07${}_{}$13 CIRA), OSUG\@2020 LabEx (ANR10 LABX56) and Equip\@Meso (ANR10 EQPX-29-01). ISTerre is also part of Labex OSUG@2020 (ANR10 LABX56). 
SWAN is described at \url{https://bitbucket.org/vidalje/}, and most figures were produced using matplotlib (\url{http://matplotlib.org/}).
\end{acknowledgements}




\bibliographystyle{aa}
\bibliography{bib_tidal} 

\begin{appendix}

\section{Local (WKB) stability equations}
\label{appendix:WKB}
We present the local Wentzel-Kramers-Brillouin (WKB) stability method. In the local analysis, the unbounded growth of the perturbations gives sufficient conditions for local instability \citep{friedlander1991instability,lifschitz1991local}. 
The original WKB hydrodynamic stability theory has been extended by several authors, for instance to take buoyancy effects into account within the Boussinesq approximation \citep{kirillov2017short}.

In the following, we derive the coupled (WKB) stability equations for arbitrary, spatially varying Boussinesq and magnetic background states. 
We emphasise that their derivation is intrinsically different from the one of Kelvin wave stability equations \citep{craik1986evolution,craik1989stability}, also accounting for magnetic fields \citep{craik1988class,fabijonas2002secondary,lebovitz2004magnetoelliptic,herreman2009effects,mizerski2011influence,cebron2012elliptical,mizerski2012mean,mizerski2012connection,bajer2013elliptical} and buoyancy effects \citep{cebron2012elliptical}. Indeed, the Kelvin wave method cannot investigate the stability of arbitrary background states, contrary to the WKB method. 

\subsection{Linearised stability equations}
We use in the following dimensional variables to devise the general stability equations in the diffusionless limit. Contrary to the main text, the dimensional variables are written here without $^*$, to keep concise mathematical expressions. 
We consider a fluid rotating at the angular velocity $\boldsymbol{\Omega}$ and stratified in density under the arbitrary gravity field $\boldsymbol{g}$. The fluid has a typical density $\rho_M$ and is pervaded by an imposed magnetic field $\boldsymbol{B}_0(\boldsymbol{r},t)$. 
We expand the velocity, the magnetic field and the temperature as small Eulerian perturbations $[\boldsymbol{u}, \boldsymbol{b}, \Theta] (\boldsymbol{r},t)$ around a spatially varying and time-dependent background state $[\boldsymbol{U}_0, \boldsymbol{B}_0, T_0] (\boldsymbol{r},t)$. In unbounded fluids, the perturbations are governed by the linearised hydromagnetic, Boussinesq equations 
\begin{subequations}
\label{appendixWKB:UBTCEQN}
\begin{align}
\frac{\mathrm{d} \boldsymbol{u}}{\mathrm{d} t} &= - (\boldsymbol{u} \boldsymbol{\cdot} \nabla) \, \boldsymbol{U}_0 -2 \, \boldsymbol{\Omega} \times \boldsymbol{u} - \nabla (p + p_b) \label{appendixWKBU}  \\
& -\alpha_T \, \Theta \, \boldsymbol{g} + \alpha_B \left [ (\boldsymbol{B}_0 \boldsymbol{\cdot} \nabla) \, \boldsymbol{b} + (\boldsymbol{b} \boldsymbol{\cdot} \nabla) \, \boldsymbol{B}_0 \right ], \nonumber  \\
\frac{\mathrm{d} \boldsymbol{b}}{\mathrm{d} t} &= (\boldsymbol{b} \boldsymbol{\cdot} \nabla) \, \boldsymbol{U}_0  - ( \boldsymbol{u} \boldsymbol{\cdot} \nabla ) \, \boldsymbol{B}_0 + ( \boldsymbol{B}_0 \boldsymbol{\cdot} \nabla ) \, \boldsymbol{u}, \label{appendixWKBB} \\
\frac{\mathrm{d} \Theta}{\mathrm{d} t} &= - (\boldsymbol{u} \boldsymbol{\cdot} \nabla) \, T_0, \label{appendixWKBT} \\
\boldsymbol{\nabla} \boldsymbol{\cdot} \boldsymbol{u} &= 0, \ \, \ \boldsymbol{\nabla} \boldsymbol{\cdot} \boldsymbol{b} = 0 \label{appendixWKBdiv},
\end{align}
\end{subequations}
where $\mathrm{d}/\mathrm{d}t = \partial/\partial t + (\boldsymbol{U}_0 \boldsymbol{\cdot} \nabla)$ is the material derivative along the basic flow, $p$ is the hydrodynamic pressure and $p_b = \alpha_B (\boldsymbol{B}_0 \boldsymbol{\cdot} \boldsymbol{b})$ the magnetic pressure. 
In equations (\ref{appendixWKB:UBTCEQN}), $\alpha_T$ is the coefficient of thermal expansion (at constant pressure) in the Boussinesq equation of state (EoS) $\delta \rho/\rho_M  = -\alpha_T \, \Theta$, with $\delta \rho$ the Eulerian perturbation in density. 

\subsection{Short-wavelength perturbations}
We seek short-wavelength perturbations in Eulerian description, with respect to the small asymptotic parameter $0 <\varepsilon \ll 1$. We introduce the formal asymptotic series 
\begin{subequations}
    \allowdisplaybreaks
	\label{appendixWKB:WKBforms}
	\begin{align}
    \boldsymbol{u} (\boldsymbol{r},t) &= \left [ \boldsymbol{u}^{(0)} + \varepsilon \boldsymbol{u}^{(1)} \right ] (\boldsymbol{r},t) \, \exp(\mathrm{i} \varPhi (\boldsymbol{r},t) / \varepsilon) + \dots, \\
	\boldsymbol{b} (\boldsymbol{r}, t) &= \left [ \boldsymbol{b}^{(0)} + \varepsilon \boldsymbol{b}^{(1)} \right ] (\boldsymbol{r}, t) \, \exp ( \mathrm{i} \varPhi (\boldsymbol{r}, t) / \varepsilon ) + \dots, \\
	\Theta (\boldsymbol{r},t) &= \left [ \Theta^{(0)} + \varepsilon \Theta^{(1)} \right ] (\boldsymbol{r},t) \, \exp(\mathrm{i} \varPhi (\boldsymbol{r},t) / \varepsilon) + \dots, \\
	p (\boldsymbol{r},t) &= \left [ p^{(0)} + \varepsilon p^{(1)} \right ] (\boldsymbol{r},t) \, \exp(\mathrm{i} \varPhi (\boldsymbol{r},t) / \varepsilon) + \dots,
	\end{align}
\end{subequations}
where $\varPhi$ is a real-valued scalar function that represents the rapidly varying phase of oscillations and $[\boldsymbol{u}^{(i)}, \Theta^{(i)}, p^{(i)}]$ are slowly varying complex-valued amplitudes. Note that we have omitted in expansions (\ref{appendixWKB:WKBforms}) the reminder terms, assumed to be uniformly bounded in $\varepsilon$ on any fixed time interval \citep{lifschitz1991local,lebovitz1992short,lifschitz1993local}. 
We further introduce the local wave vector, defined by $\boldsymbol{k} = \nabla \varPhi$. 
The small asymptotic parameter $\varepsilon \ll 1$ is actually related to the typical scale of the instability $l$, which must be much smaller to the typical length scale of the large-scale background flow $L_0$. This requires $\varepsilon = l/L_0 \ll 1$ \citep{nazarenko1999wkb}. In the hydrodynamic and diffusionless case, its value is arbitrary small. 

However, in hydromagnetics, $\varepsilon$ does affect the magnetic field because the Lorentz force depends on the length scale. The general magnetic configuration leads to a set of partial differential equations  \citep{friedlander1995stability,kirillov2014local}, which must be solved locally in Eulerian description. However,
by assuming \citep[see also for uniform fields][]{mizerski2011influence}
\begin{equation}
	\boldsymbol{B}_0 (\boldsymbol{r}) = \varepsilon \, \widetilde{\boldsymbol{B}}_0 (\boldsymbol{r}),
\end{equation} 
the partial differential equations simplify into ordinary differential equations (even for spatially varying magnetic fields). 
This is the central approximation of the hydromagnetic stability theory, which is not required in the non-magnetic case. For tidal studies, we usually set $\varepsilon = \beta_0$ \citep{le2000three}.

\subsection{Eulerian stability equations}
We closely follow the mathematical derivation of \citet{kirillov2017short}, extending it to the hydromagnetic case. 
Substituting expansions (\ref{appendixWKB:WKBforms}) in incompressible condition (\ref{appendixWKBdiv}) and collecting terms of order $\mathrm{i}/\varepsilon$ and $\varepsilon^0$ gives
\begin{subequations}
	\begin{align}
    	\mathrm{i}/\varepsilon:& \ \, \ \left [\boldsymbol{u}^{(0)}, \boldsymbol{b}^{(0)} \right ] \boldsymbol{\cdot} \boldsymbol{k} = 0, \label{appendixWKB:epsm1div} \\
        \varepsilon^{0}:& \ \, \ \boldsymbol{\nabla} \boldsymbol{\cdot} \left [ \boldsymbol{u}^{(0)}, \boldsymbol{b}^{(0)} \right ] = -\mathrm{i} \boldsymbol{k} \boldsymbol{\cdot} \left [ \boldsymbol{u}^{(1)}, \boldsymbol{b}^{(1)} \right ]  .
    \end{align}
\end{subequations}
The same procedure applied to governing equations (\ref{appendixWKBU})-(\ref{appendixWKBT}). Firstly, we have at the order $\mathrm{i}/\varepsilon$
\begin{equation}
	\frac{\mathrm{d} \varPhi}{\mathrm{d} t} \left [ \boldsymbol{u}^{(0)}, \boldsymbol{b}^{(0)}, \Theta^{(0)} \right ] = \left [- p^{(0)} \, \boldsymbol{k}, \boldsymbol{0}, 0 \right ]. 
	\label{appendixWKB:epsm1utc}
\end{equation}
The dot product of the first equation (\ref{appendixWKB:epsm1utc}) with $\nabla \varPhi$, under constraint (\ref{appendixWKB:epsm1div}), gives $p^{(0)} = 0$. Then, we obtain the Hamilton-Jacobi equation ${\mathrm{d} \varPhi}/{\mathrm{d}t} = 0$. 
Finally, taking the spatial gradient of the previous equation gives the eikonal equation and its initial condition \citep{lifschitz1991local}
\begin{equation}
	\frac{\mathrm{d} \boldsymbol{k}}{\mathrm{d}t} = - \left ( \boldsymbol{\nabla} \boldsymbol{U}_0 \right )^\top \boldsymbol{k}, \ \, \ \boldsymbol{k}(\boldsymbol{r},0) = \boldsymbol{k}_0, \ \, \ |\boldsymbol{k}(\boldsymbol{r},t)| = |\boldsymbol{k}_0|.
    \label{appendixWKB:eikonal}
\end{equation}

Now, by using the Hamilton-Jacobi equation and (\ref{appendixWKB:eikonal}), equations (\ref{appendixWKBU})-(\ref{appendixWKBT}) give at the next asymptotic order $\varepsilon^{0}$
\begin{subequations}
\begin{align}
	-\mathrm{i} \boldsymbol{k} &\left [ p^{(1)} + \alpha_B \, \widetilde{\boldsymbol{B}}_0 \boldsymbol{\cdot} \boldsymbol{b}^{(0)} \right ] = \left ( \frac{\mathrm{d}}{\mathrm{d}t} + \boldsymbol{\nabla} \boldsymbol{U}_0 + 2 \, \boldsymbol{\Omega} \, \times \right ) \boldsymbol{u}^{(0)} \label{appendixWKB:p1temp} \\ 
	{} & - \alpha_T \, \Theta^{(0)} \, \boldsymbol{g} - \mathrm{i} \alpha_B \, (\widetilde{\boldsymbol{B}}_0 \boldsymbol{\cdot} \boldsymbol{k}) \, \boldsymbol{b}^{(0)}, \nonumber \\
	\frac{\mathrm{d} \boldsymbol{b}^{(0)}}{\mathrm{d}t} &= \mathrm{i} \,  (\widetilde{\boldsymbol{B}}_0 \boldsymbol{\cdot} \boldsymbol{k}) \, \boldsymbol{u}^{(0)} + (\boldsymbol{\nabla} \boldsymbol{U}_0) \, \boldsymbol{b}^{(0)}, \label{appendixWKB:transportb}\\
	\frac{\mathrm{d} \Theta^{(0)}}{\mathrm{d}t} &= - \boldsymbol{u}^{(0)} \boldsymbol{\cdot} \nabla T_0. \label{appendixWKB:transportt}
    \end{align}
\end{subequations}
Equations (\ref{appendixWKB:transportb})-(\ref{appendixWKB:transportt}) are transport equations for the magnetic field and the temperature amplitudes. 
Applying the dot product of $\boldsymbol{k}$ with equation (\ref{appendixWKB:p1temp}) gives the first order pressure variable
\begin{multline}
	-\mathrm{i} \left [ p^{(1)} + \alpha_B \widetilde{\boldsymbol{B}}_0 \boldsymbol{\cdot} \boldsymbol{b}^{(0)} \right ] = \frac{\boldsymbol{k}}{|\boldsymbol{k}|^2} \boldsymbol{\cdot}  \left ( \frac{\mathrm{d}}{\mathrm{d}t} + \boldsymbol{\nabla} \boldsymbol{U}_0 + 2 \, \boldsymbol{\Omega} \, \times \right ) \boldsymbol{u}^{(0)} \\ 
     - \frac{\boldsymbol{k}}{|\boldsymbol{k}|^2} \boldsymbol{\cdot}  \left ( \alpha_T \,  \Theta^{(0)} \, \boldsymbol{g} \right ).
     \label{appendixWKB:p1temp2}
\end{multline}
Then, we differentiate equation (\ref{appendixWKB:epsm1div}) to get the identity \citep{lifschitz1991local}
\begin{equation}
	\frac{\mathrm{d}}{\mathrm{d}t} \left ( \boldsymbol{u}^{(0)} \boldsymbol{\cdot} \boldsymbol{k} \right ) = \frac{\mathrm{d} \boldsymbol{k}}{\mathrm{d}t} \boldsymbol{\cdot} \boldsymbol{u}^{(0)} + \boldsymbol{k} \boldsymbol{\cdot} \frac{\mathrm{d} \boldsymbol{u}^{(0)}}{\mathrm{d}t} = 0.
    \label{appendixWKB:identity}
\end{equation}
Finally, we use identity (\ref{appendixWKB:identity}) to simplify equation (\ref{appendixWKB:p1temp2}), then we substitute the resulting expression into equation (\ref{appendixWKB:p1temp}). After some algebra, we get the transport equation for the velocity amplitude
\begin{multline}
	\frac{\mathrm{d} \boldsymbol{u}^{(0)}}{\mathrm{d} t} = \left[ \left( \frac{2\, \boldsymbol{k} \boldsymbol{k}^{\top}}{|\boldsymbol{k}|^2}-\boldsymbol{I} \right) \boldsymbol{\nabla} \boldsymbol{U}_0 + 2 \left( \frac{ \boldsymbol{k} \boldsymbol{k}^{\top}}{|\boldsymbol{k}|^2}-\boldsymbol{I} \right) \boldsymbol{\Omega} \, \times  \right] \,  \boldsymbol{u}^{(0)}  \\
	- \alpha_T \, \Theta^{(0)} \, \left(  \boldsymbol{I} - \frac{ \boldsymbol{k} \boldsymbol{k}^{\top}}{|\boldsymbol{k}|^2} \right) \boldsymbol{g} + \mathrm{i} \alpha_B \, (\widetilde{\boldsymbol{B}}_0 \boldsymbol{\cdot} \boldsymbol{k} ) \, \boldsymbol{b}^{(0)}.
\label{appendixWKB:transportu}
\end{multline}
The stability equations, given by equations (\ref{appendixWKB:transportu}) and (\ref{appendixWKB:transportb})-(\ref{appendixWKB:transportt}), are dominant for the stability behaviour of WKB expansions (\ref{appendixWKB:WKBforms}) for long enough times in the limit $\varepsilon \ll 1$ \citep{lifschitz1991local,friedlander1991instability,lebovitz1992short,lifschitz1993local}.
The next order terms are only responsible for transient behaviours \citep{rodrigues2017second}. Thus, sufficient conditions for local instability are obtained by solving transport equations (\ref{appendixWKB:transportu}) and (\ref{appendixWKB:transportb})-(\ref{appendixWKB:transportt}).

\subsection{Lagrangian equations along fluid trajectories}
WKB stability equations are partial differential equations in Eulerian description. However, they are generally solved in Lagrangian description. The WKB perturbations are advected along the fluid trajectories $\boldsymbol{X}(t)$ of the background flow $\boldsymbol{U}_0$, passing through the initial point $\boldsymbol{X}_0$ at initial time $t=0$. In Lagrangian formalism, the WKB stability equations are
\begin{subequations}
\label{appendixWKB:transportxkutc2}
\allowdisplaybreaks
\begin{align}
	\frac{\mathrm{D} \boldsymbol{X}}{\mathrm{D}t} &= \boldsymbol{U}_0 (\boldsymbol{X}(t)), \ \, \ \boldsymbol{X}(0) = \boldsymbol{X}_0, \label{appendixWKB:transportx} \\
	\frac{\mathrm{D} \boldsymbol{k}}{\mathrm{D}t} &= - ( \boldsymbol{\nabla} \boldsymbol{U}_0)^\top \, \boldsymbol{k}, \ \, \ \boldsymbol{k}(0) = \boldsymbol{k}_0, \label{appendixWKB:transportk} \\
	\frac{\mathrm{D} \boldsymbol{u}^{(0)}}{\mathrm{D} t} &= \left [ \left( \frac{2\, \boldsymbol{k} \boldsymbol{k}^{\top}}{|\boldsymbol{k}|^2}-\boldsymbol{I} \right) \boldsymbol{\nabla} \boldsymbol{U}_0 + 2 \left( \frac{ \boldsymbol{k} \boldsymbol{k}^{\top}}{|\boldsymbol{k}|^2} - \boldsymbol{I} \right) \boldsymbol{\Omega} \, \times  \right ] \,  \boldsymbol{u}^{(0)} \label{appendixWKB:transportu2}  \\
	&-  \alpha_T \, \Theta^{(0)} \, \left (  \boldsymbol{I} -\frac{ \boldsymbol{k} \boldsymbol{k}^{\top}}{|\boldsymbol{k}|^2} \right) \boldsymbol{g} +  \mathrm{i} \alpha_B \, (\widetilde{\boldsymbol{B}}_0 \boldsymbol{\cdot} \boldsymbol{k} ) \, \boldsymbol{b}^{(0)},  \nonumber  \\
	\frac{\mathrm{D} \boldsymbol{b}^{(0)}}{\mathrm{D}t} &= \mathrm{i} \,  (\widetilde{\boldsymbol{B}}_0 \boldsymbol{\cdot} \boldsymbol{k}) \, \boldsymbol{u}^{(0)} + (\boldsymbol{\nabla} \boldsymbol{U}_0) \, \boldsymbol{b}^{(0)}, \label{appendixWKB:transportb2} \\
	\frac{\mathrm{D} \Theta^{(0)}}{\mathrm{D}t} &= - \boldsymbol{u}^{(0)} \boldsymbol{\cdot} \nabla T_0, \label{appendixWKB:transportt2}
\end{align}
\end{subequations}
with $\mathrm{D}/\mathrm{D}t$ the Lagrangian derivative. Therefore, equations (\ref{appendixWKB:transportxkutc2}) are interpreted as ordinary differential equations along the fluid trajectories of the background flow $\boldsymbol{U}_0$ for the amplitudes $(\boldsymbol{u}^{(0)}, \Theta^{(0)}, \xi^{(0)})$. In addition, the initial conditions satisfy
\begin{equation}
\boldsymbol{u}^{(0)} (0) \boldsymbol{\cdot} \boldsymbol{k}_0 = 0, \ \, \ \boldsymbol{b}^{(0)} (0) \boldsymbol{\cdot} \boldsymbol{k}_0 = 0,
\end{equation}
such the solenoidal conditions for the velocity and the magnetic field hold at any time. Sufficient conditions for instability are obtained when \citep[e.g.][]{lifschitz1991local,lebovitz1992short,lifschitz1993local}
\begin{equation}
\lim_{t\to \infty} \left ( |\boldsymbol{u}^{(0)}| + |\boldsymbol{b}^{(0)}| +  |\Theta^{(0)}| \right ) = \infty 
\end{equation}
for given $[\boldsymbol{X}_0,\boldsymbol{k}_0]$ and with suitable initial conditions for $[\boldsymbol{u}^{(0)}, \boldsymbol{b}^{(0)}, \Theta^{(0)}]$. 

\section{MAC modes in triaxial ellipsoids}
\label{appendix:MAC}
We present a method to compute the three-dimensional hydromagnetic eigenmodes of stratified Boussinesq fluids contained within rigid triaxial ellipsoids. This approach relies on a fully global, explicit spectral method in ellipsoids, in which the velocity field is described by polynomial finite-dimensional Galerkin bases \citep{vidal2017inviscid}. 
The algorithm has been benchmarked successfully against the Coriolis modes in ellipsoids \citep{vantieghem2014inertial}, while the fast and slow hydromagnetic solutions have been validated for the Malkus field in spheres \citep{malkus1967hydromagnetic,zhang2003nonaxisymmetric} and spheroids \citep{kerswell1994tidal}.

\subsection{Assumptions}
We work in dimensional variables for the sake of generality, and use the notations introduced in the main text. 
We consider a diffusionless, incompressible electrically conducting fluid, contained within a triaxial ellipsoid of semi-axes $(a,b,c)$. The fluid is stratified under the gravity field $\boldsymbol{g}^*$ in the Boussinesq approximation. 
The fluid is contained within an ellipsoidal container, which is rotating at the angular velocity $\boldsymbol{\Omega}$ in the inertial frame. 
We expand the velocity, the temperature and the magnetic field as small perturbations $[\boldsymbol{u}^*, \Theta^*, \boldsymbol{b}^*](\boldsymbol{r},t)$ around an equilibrium state of rest $[\boldsymbol{0}, T_0^*, \boldsymbol{B}_0^*](\boldsymbol{r})$. 

In the linear approximation, the dimensional governing equations are
\begin{subequations}
\label{eq:UTBtotWaves}
\begin{align}
	\frac{\partial \boldsymbol{u}^*}{\partial t} &= - 2 \boldsymbol{\Omega} \times \boldsymbol{u}^* -\nabla p^* - \alpha_T \, \Theta^* \boldsymbol{g}^* \\
	&+ \alpha_B \left [ (\boldsymbol{\nabla} \times \boldsymbol{b}^*) \times \boldsymbol{B}_0^* + (\boldsymbol{\nabla} \times \boldsymbol{B}_0^*) \times \boldsymbol{b}^* \right ],  \nonumber \\
	\frac{\partial \Theta^*}{\partial t} &= - (\boldsymbol{u}^* \boldsymbol{\cdot} \nabla) \, T_0^*, \\
	\frac{\partial \boldsymbol{b}^*}{\partial t} &= \boldsymbol{\nabla} \times (\boldsymbol{u}^* \times \boldsymbol{B}_0^*), \\
	\boldsymbol{\nabla} \boldsymbol{\cdot} \boldsymbol{u}^* &= \boldsymbol{\nabla} \boldsymbol{\cdot} \boldsymbol{b}^* = 0,
\end{align}
\end{subequations}
with $\alpha_B = (\rho_M \mu_0)^{-1}$ and $p^*$ the hydrodynamic pressure. 
By taking the time derivative of equations (\ref{eq:UTBtotWaves}), we can obtain a single wave-like equation of second order in time for the velocity perturbation $\boldsymbol{u}^*$. This reads
\begin{equation}
	\frac{\partial^2 \boldsymbol{u}^*}{\partial t^2} + 2 \boldsymbol{\Omega} \times \frac{\partial \boldsymbol{u}^*}{\partial t} = - \frac{\partial \nabla p^*}{\partial t}  + \alpha_T (\boldsymbol{u}^* \boldsymbol{\cdot} \nabla) \, T_0^* \, \boldsymbol{g}^* + \boldsymbol{f}_m^*,
	\label{eq:MACmodes}
\end{equation}
with the Lorentz force
\begin{multline}
	\boldsymbol{f}_m^* = \alpha_B \, (\boldsymbol{\nabla} \times \boldsymbol{B}_0^*) \times \left [ \boldsymbol{\nabla} \times (\boldsymbol{u}^* \times \boldsymbol{B}_0^*) \right ] \\
	+ \alpha_B \left [ \boldsymbol{\nabla} \times (\boldsymbol{\nabla} \times (\boldsymbol{u}^* \times \boldsymbol{B}_0^*)) \right ] \times \boldsymbol{B}_0^*.
\end{multline}
Note that equations (\ref{eq:UTBtotWaves}) cannot be recast into a single equation for the velocity perturbation $\boldsymbol{u}^*$ in the presence of a basic flow  $\boldsymbol{U}_0^*$. In this case, the problem must be formulated for the displacement vector \citep[e.g.][]{chandrasekhar1969ellipsoidal,lebovitz1989stability}. 

Finally, equation (\ref{eq:MACmodes}) is supplemented by the non-penetration boundary conditions
\begin{equation}
	\boldsymbol{u}^* \boldsymbol{\cdot} \boldsymbol{1}_n = 0, \ \, \ \boldsymbol{B}_0^* \boldsymbol{\cdot} \boldsymbol{1}_n = 0,
\end{equation}
with $\boldsymbol{1}_n$ the unit outward vector normal to the ellipsoidal boundary. 
We emphasise that alternative boundary conditions for the background magnetic field cannot be considered with the polynomial Galerkin description, at least to investigate consistently all the hydromagnetic modes. 
Allowing a non-zero normal magnetic field at the boundary would create a surface electrical density current, generating a Lorentz force $\boldsymbol{f}_m^*$ in the form of a discontinuous Dirac function distributed on the boundary \citep{friedlander1990nonlinear}. This would lead to spurious diffusionless solutions for the slow hydromagnetic modes. However, we would expect the fast hydromagnetic modes (that is Coriolis modes) to be only barely affected by the magnetic boundary condition, because the Lorentz force in momentum equation (\ref{eq:MACmodes}) has only second-order effects on the fast modes.

\subsection{Galerkin method}
We employ a Galerkin method to describe the velocity field. We seek a Galerkin expansion of the modes in the form
\begin{equation}
	\left [ \boldsymbol{u}^*, p^* \right ] (\boldsymbol{r},t) = \left [ \widehat{\boldsymbol{u}}^*, \widehat{p}^*  \right ] (\boldsymbol{r}) \exp(\mathrm{i} \omega_i t), \ \, \ \widehat{\boldsymbol{u}}^* = \sum \limits_{l=1}^\infty \gamma_l \, \widehat{\boldsymbol{u}}_l^*,
	\label{eqappend:Galerkin}
\end{equation}
where $\omega_i$ is the angular frequency, $\{\gamma_l\}$ modal complex coefficients and $\{\widehat{\boldsymbol{u}}_l^* (\boldsymbol{r})\}$ are real-valued basis Galerkin elements. 
Firstly, we rewrite equation (\ref{eq:MACmodes}) in the symbolic form
\begin{equation}
	\left ( -\omega_i^2 + \mathrm{i} \omega_i \, \boldsymbol{\mathcal{A}}_1 + \boldsymbol{\mathcal{A}}_0 \right ) \widehat{\boldsymbol{u}}^* = - \mathrm{i} \omega_i \, \nabla \widehat{p}^*,
	\label{eqappend:canonical}
\end{equation}
where $[\boldsymbol{\mathcal{A}}_1, \boldsymbol{\mathcal{A}}_0]$ are two linear operators. 
The basis elements $\{\widehat{\boldsymbol{u}}_l^* (\boldsymbol{r})\}$ are made of linear combinations of Cartesian monomials $\{x^i y^j z^k \}_{i+j+k < \infty}$, satisfying 
\begin{equation}
	\boldsymbol{\nabla} \boldsymbol{\cdot} \widehat{\boldsymbol{u}}_l^* = 0, \ \, \ \widehat{\boldsymbol{u}}_l^* \boldsymbol{\cdot}  \boldsymbol{1}_n = 0 \ \, \ \text{at the boundary}.
	\label{eqappend:GalerkinBC}
\end{equation} 
Several Cartesian expansions have been proposed \citep[see a comparison in][]{vidal2017inviscid}. 
Expansion (\ref{eqappend:Galerkin}) is similar to expansions used in the finite-element method (FEM). 
However, compared to the traditional FEM, our basis elements $\{\widehat{\boldsymbol{u}}_l^* (\boldsymbol{r})\}$ are global polynomials, infinitely differentiable in ellipsoids. The mathematical completeness of the polynomial expansion for incompressible fluids is then ensured by using the Weierstrass approximation theorem \citep{backus2017completeness,ivers2017enumeration}. Hence, this method
is a rigorous spectral method in ellipsoids. 

Then, we truncate series (\ref{eqappend:Galerkin}) at a given polynomial degree $n$ (such that  $i+j+k\leq n$).
In the absence of any stratified or magnetic effect, the Coriolis operator is exactly closed within the considered polynomial bases \citep[e.g.][]{kerswell1993instability,backus2017completeness}. Thus, the Coriolis modes are exactly described by the polynomial description \citep{vantieghem2014inertial,backus2017completeness}. Note that fast and slow MC modes also admit exact polynomial descriptions for some background magnetic fields that are linear in the Cartesian space coordinates \citep{malkus1967hydromagnetic,zhang2003nonaxisymmetric,kerswell1994tidal}. 
For any other practical configuration, we have to choose a maximum polynomial degree $n$ to ensure a good convergence of the desired modes (higher-order bases are excited by the buoyancy and Lorentz forces). 
We substitute the truncated expansion into equation (\ref{eqappend:canonical}), yielding the quadratic eigenvalue problem
\begin{equation}
	\left ( -\omega_i^2 \, \boldsymbol{A}_2 + \mathrm{i}\omega_i \, \boldsymbol{A}_1 + \boldsymbol{A}_0 \right ) \cdot \boldsymbol{\gamma} = \boldsymbol{0}, \\
	\label{eqappend:canonical2}
\end{equation}
where $\boldsymbol{\gamma} = (\gamma_1, \gamma_2, \dots)^\top$ is the eigenvector and $[\boldsymbol{A}_2, \boldsymbol{A}_1, \boldsymbol{A}_0]$ are three real-valued matrices. Their elements are given by the Galerkin projections over the ellipsoidal domain
\begin{subequations}
\label{eq:volGalerkin}
\begin{align}
	A_{2,ij} &= \int_\mathcal{V} \widehat{\boldsymbol{u}}_i^* \boldsymbol{\cdot} \widehat{\boldsymbol{u}}_j^* \, \mathrm{d} \mathcal{V}, \\
	A_{1,ij} &= \int_\mathcal{V} \widehat{\boldsymbol{u}}_i^* \boldsymbol{\cdot} (\boldsymbol{\mathcal{A}}_1 \widehat{\boldsymbol{u}}_j^*) \, \mathrm{d} \mathcal{V},\\
	A_{0,ij} &= \int_\mathcal{V} \widehat{\boldsymbol{u}}_i^* \boldsymbol{\cdot} (\boldsymbol{\mathcal{A}}_0 \widehat{\boldsymbol{u}}_j^*) \, \mathrm{d} \mathcal{V}.
\end{align}
\end{subequations}
The projection of the pressure term in equation (\ref{eqappend:canonical2}) vanishes by virtue of the divergence theorem, such that an explicit decomposition for the pressure is not required. 
If the background state can be written by using Cartesian monomials $x^i y^j z^k$, then volume integrals (\ref{eq:volGalerkin}) can be computed analytically \cite[see formula 50 in][]{lebovitz1989stability}.

\subsection{Hydromagnetic modes}
\begin{figure}
	\centering
	\begin{tabular}{c}
	\includegraphics[width=0.45\textwidth]{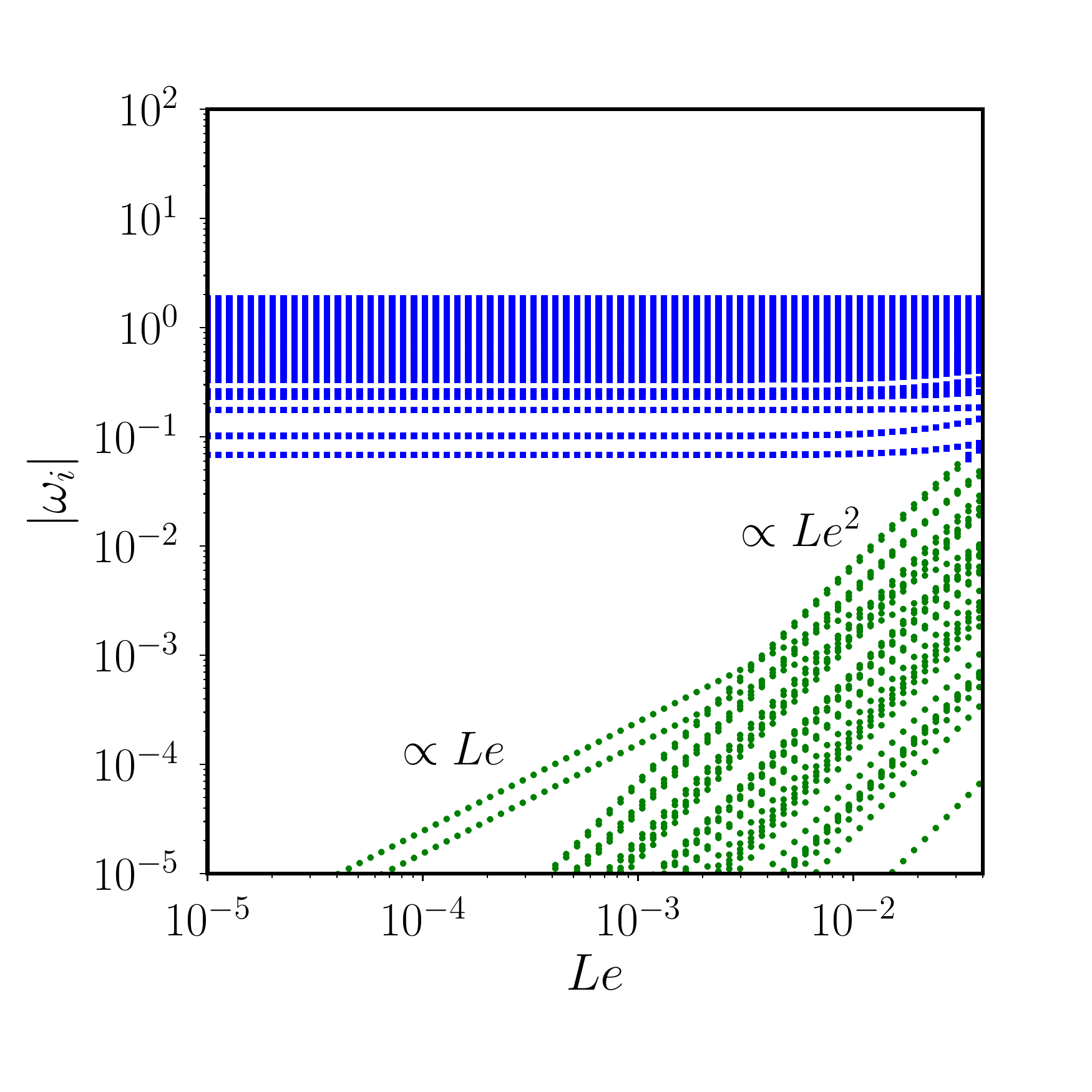} \\
	\includegraphics[width=0.45\textwidth]{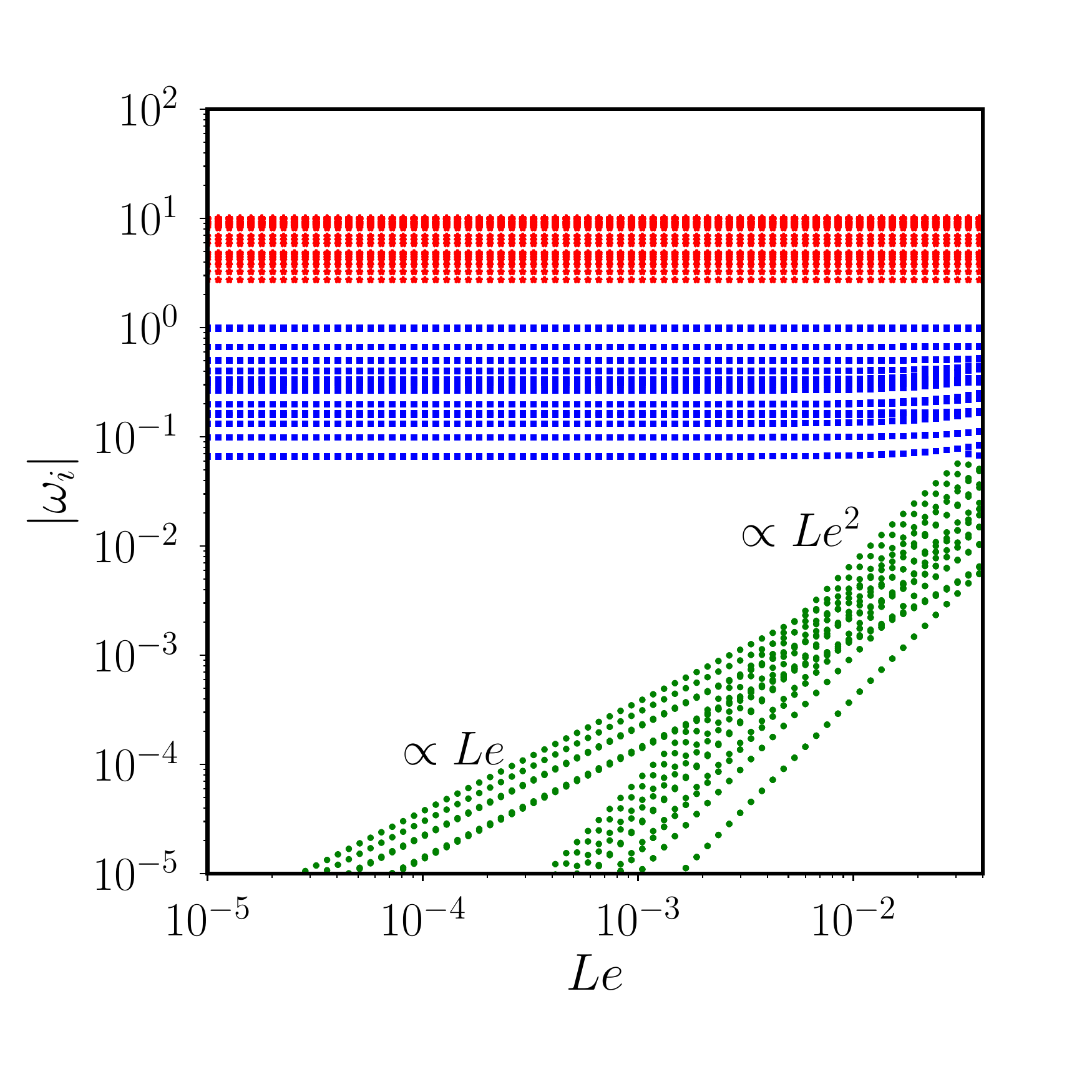} \\
	\end{tabular}
	\caption{Angular frequency $|\omega_i|$ of MAC modes, as a function of $Le$ in spheres ($\beta_0=0$), stratified under gravitational potential (\ref{eq:PhigravRef}). The background (toroidal) magnetic field is $\boldsymbol{B}_0 = 0.1 \left [ -z \, \boldsymbol{1}_y + y \, \boldsymbol{1}_z \right ] + \left [-y \, \boldsymbol{1}_x + x \, \boldsymbol{1}_y \right ]$ in dimensionless form. \emph{From bottom to top}: green circles are slow MC and torsional modes (respectively $\omega_i \propto Le^2$ and $\omega_i \propto Le$), blue squares represent fast MC modes and red stars are gravito-inertial modes. The truncation polynomial degree is $n=5$. \emph{Top panel}:  Neutral fluid ($N_0 /\Omega_\text{s} = 0$). \emph{Bottom panel}: Stratified fluid ($N_0 /\Omega_\text{s} = 10$).}
	\label{fig:MACmodes}
\end{figure}

We show in Fig. \ref{fig:MACmodes} the dimensionless eigenfrequency $\omega_i$ of MAC modes, for the relevant weak field regime $Le \leq 10^{-1}$. We have considered an arbitrary reference configuration to illustrate several representative properties of the modes. We identify three families of waves in neutrally stratified fluids (top panel of Fig. \ref{fig:MACmodes}), in agreement with investigations in spherical geometries \citep[e.g.][]{schmitt2010magneto,labbe2015magnetostrophic}. Firstly, the high frequency branch represents fast Magneto-Coriolis (MC) modes \citep{malkus1967hydromagnetic,labbe2015magnetostrophic}. They are similar to pure Coriolis (or inertial) modes \citep{greenspan1968theory,vantieghem2014inertial,backus2017completeness}, with a dimensionless spectrum bounded by $|\omega_i| \leq 2$ in the weak field regime $Le \ll 1$.  
These modes are regular in space and only weakly affected by large-scale magnetic fields in weakly deformed spheres \citep[e.g.][]{schmitt2010magneto,labbe2015magnetostrophic}. This is consistent with the weak frequency dependence on $Le$ observed in Fig. \ref{fig:MACmodes}. 
Note that they have a different behaviour compared to the singular modes localised on attractors \citep[e.g.][]{rieutord1997inertial,rieutord2018axisymmetric}, which only exist in shells because the mathematical problem is ill-posed \citep{rieutord2000wave}. 
Secondly, the low frequency branch represents slow Magneto-Coriolis (MC) modes. Their typical (dimensionless) frequency scales according to $|\omega_i| \propto Le^2$. 
In addition, the third intermediate branch represents torsional Alfv\'en modes \citep{labbe2015magnetostrophic}, scaling as $|\omega_i| \propto Le$. They are usually filtered out in reduced models, such as in local models considering uniform fields. They exist when the current direction $\boldsymbol{\nabla} \times \boldsymbol{B}_0$ of the basic state is misaligned with the spin rotation axis.

Then, we show the spectrum of MAC modes in stratified fluids in the bottom panel of Fig. \ref{fig:MACmodes}. 
The aforementioned hydromagnetic modes still exist in stably stratified interiors, yielding fast and slow MAC waves. However, their properties in the presence of buoyancy and magnetic fields are rather complex in spherical-like domains \citep{friedlander1987hydromagnetic}. 
On the one hand, fast MAC modes and gravito-inertial modes are barely modified by magnetic fields, as illustrated in Fig. \ref{fig:MACmodes} (bottom panel) when $Le \ll 1$. However, they strongly depend on stratification \citep{friedlander1982internala}.
On the other hand, slow MC modes can be strongly affected by the magnetic field and stratification \citep{friedlander1987hydromagnetic}.
Finally, the buoyancy force also sustains high frequency internal gravity modes. They can be affected by rotation, yielding gravito-inertial modes \citep{friedlander1982internala}.

\section{Mixed resonances of MAC waves}
\label{appendix:MixedMHDResonances}
We investigate the possible nonlinear couplings of hydromagnetic waves for tidal instability. We use the same dimensionless variables as in the main text.
Resonance condition (\ref{eq:resonancecond1}) can only be satisfied if tidal instability involves fast MAC waves (that is inertial or gravito-inertial waves), coupled with either fast or Magneto-Coriolis (slow MC) waves \citep{kerswell1993elliptical,kerswell1994tidal}. 
Indeed, in the astrophysical regime $Le \ll 1$, the illustrative spectrum in Fig. \ref{fig:MACmodes} clearly shows that no triadic couplings are effective in ellipsoids between two slow MC waves when $1 \leq \Omega_0 \leq 3$. Thus, the couplings of slow MC waves with the equilibrium tidal flow cannot be advocated in stellar interiors.

Secondly, the mixed couplings between slow and fast hydromagnetic waves is not forbidden in diffusionless fluids. In the weak field regime $Le \ll 1$, \citet{kerswell1993elliptical,kerswell1994tidal} showed that the typical diffusionless growth rate of tidal instability involving mixed couplings scales as (in dimensionless form)
\begin{equation}
	\sigma \propto Le^4 \beta_0.
\end{equation}
However, this diffusionless growth rate must be larger than the (laminar) Joule damping rate of the slow MC waves, that is $\tau_\Omega \propto -Em \, |\boldsymbol{k}_0|^2$ in the local theory \citep{rincon2003oscillations,sreenivasan2017damping}. 
This gives the typical upper bound on the wave vector
\begin{equation}
	|\boldsymbol{k}_0|^2 \ll \frac{Le^4}{Em} \, \beta_0.
	\label{eq:kSlowMCdamp}
\end{equation}
In short-period binaries, the typical value for the equatorial ellipticity is $\beta_0 \sim 10^{-3} - 10^{-2}$ (see Table \ref{table:databinamics}). As given in Table \ref{table:DimNumbers}, we have also the typical numbers $Em \leq 10^{-10}$ and $Le \leq 10^{-4}$. 
Then, condition (\ref{eq:kSlowMCdamp}) gives the upper bound $|\boldsymbol{k}_0| \ll 1$. This is incompatible with the short-wavelength stability theory, which requires $|\boldsymbol{k}_0| \gg 1$. 
Physically, this shows that the (laminar) Joule damping rate is always larger than the diffusionless growth rate in non-ideal fluids, for any resonance involving slow MC waves in the regime $Le \ll 1$. 
Therefore, mixed couplings of fast/slow waves can be discarded for tidal instability in realistic stellar interiors.

\section{Weakly eccentric synchronised orbits}
\label{appendix:ldei}
\subsection{Libration forcing}
We consider synchronous stratified binary systems moving on weakly eccentric coplanar orbits. 
Note that the following results are also relevant for (stratified) moons or gaseous planets orbiting around a massive central body \citep[e.g.][]{kerswell1998tidal,cebron2012elliptical,lemasquerier2017libration}.
We consider a diffusionless tidal model of the tidally deformed fluid body, characterised by an equatorial ellipticity $\beta_0$. 
The fluid body is rotating at the uniform angular velocity $\Omega_\text{s}$, aligned in the inertial frame with the orbital angular velocity of the companion along $\boldsymbol{1}_z$. 
We use the dimensionless variables introduced in Sect. \ref{sec:model}, that is taking $(\Omega_\text{s})^{-1}$ as the relevant timescale. Due to the weak orbital eccentricity $e \ll 1$, the orbital angular velocity has periodic time variations. For the sake of generality, we assume that the tidal forcing has the following (dimensionless) expression, at the leading order in the eccentricity
\begin{equation}
	\Omega_0 (t) = 1 + \epsilon_l \cos \left ( f t \right ),
	\label{eq:librations}
\end{equation}
where $f$ is the dimensionless frequency of the forcing and $\epsilon_l \leq 2e$ the dimensionless amplitude. Forcing (\ref{eq:librations}) is known as longitudinal librations.
For this tidal forcing, the equilibrium tidal velocity field has the following form in the central frame 
\begin{equation}
	\boldsymbol{U}_0 (\boldsymbol{r},t) = - \epsilon_l \cos(ft) \, \left [-(1+\beta_0) y \, \boldsymbol{1}_x + (1-\beta_0)x \, \boldsymbol{1}_y \right ].
	\label{eq:U0LDEI_Orb}
\end{equation}
Tidal flow (\ref{eq:U0LDEI_Orb}) is prone to libration-driven elliptical instability (LDEI), which is quite similar to tidal instability in non-synchronised systems \citep[e.g.][]{kerswell1998tidal,cebron2012elliptical,vidal2017inviscid,reun2019experimental}.

\subsection{Resonance condition of the LDEI}
\begin{figure}
	\centering
	\includegraphics[width=0.45\textwidth]{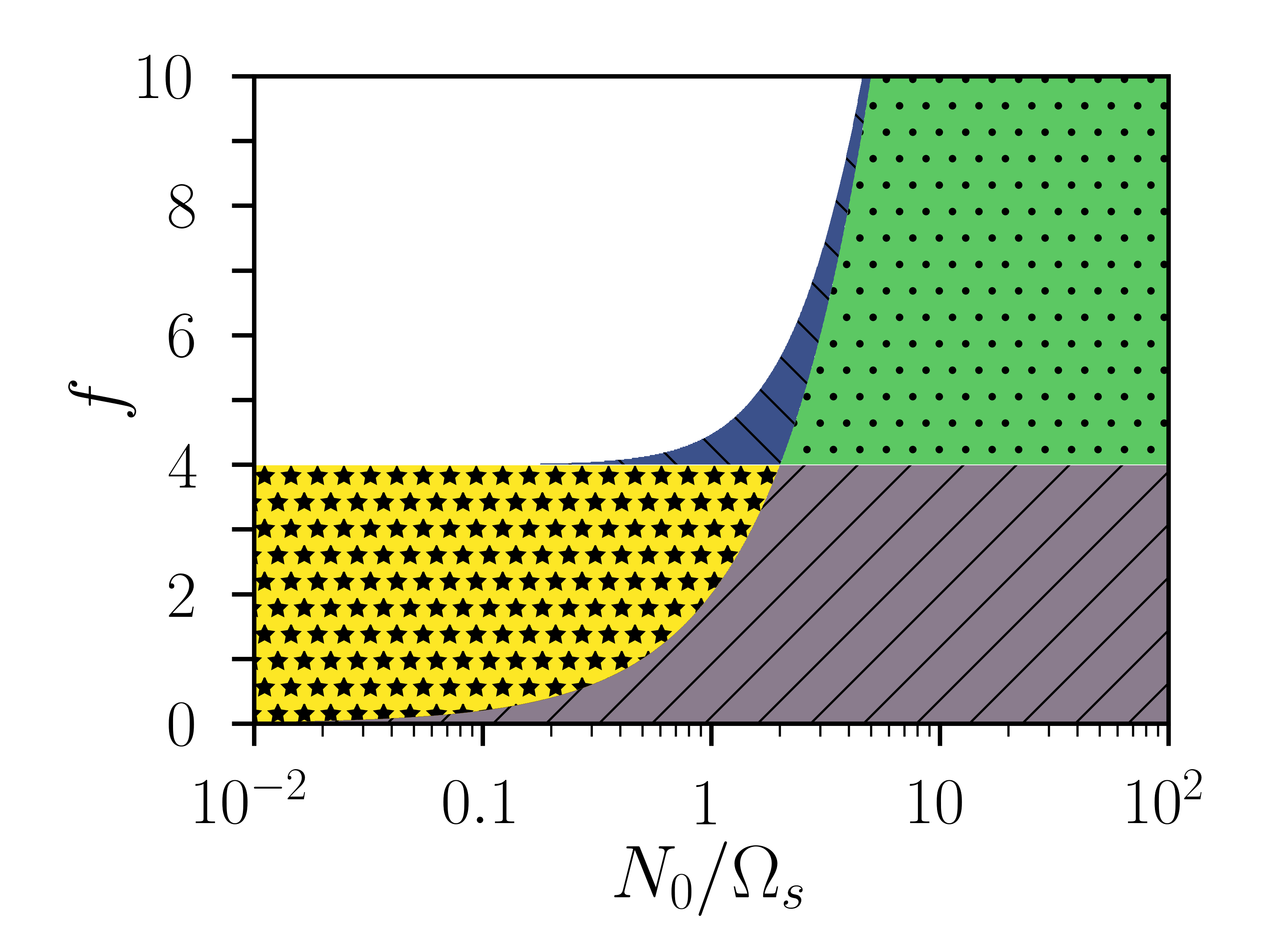}
	\caption{Waves at sub-harmonic resonance condition (\ref{eq:ResonanceLDEI}) for synchronised systems, as a function of (dimensionless) forcing frequency $f$ and $N_0/\Omega_s$. The other notations are identical to the ones introduced in the main text. White regions: no compatible waves satisfying (\ref{eq:ResonanceLDEI}). Stars (yellow area): hyperbolic waves $\mathcal{H}_1$. Right slash (purple area): hyperbolic waves $\mathcal{H}_2$. Dots (green area): elliptic waves $\mathcal{E}_1$. Back slash (blue area): elliptic waves $\mathcal{E}_2$. The classical allowable region of the instability is $0 \leq f \leq 4$ in neutral fluids. }
	\label{fig:LDEIStrat}
\end{figure}

LDEI is a fluid instability due to sub-harmonic resonances between two waves of angular frequency $|\omega_i|$ interacting with basic flow (\ref{eq:U0LDEI_Orb}). 
By analogy with formula (\ref{eq:resonancecond2}) in non-synchronised systems, the sub-harmonic resonance condition becomes 
\begin{equation}
	|\omega_i| = f/2.
	\label{eq:ResonanceLDEI}
\end{equation}
The four kinds of waves $[\mathcal{H}_1, \mathcal{H}_2, \mathcal{E}_1, \mathcal{E}_2]$, introduced Sect. \ref{subsec:theory}, can be nonlinearly coupled in the instability mechanism. We show the nature of the waves satisfying condition (\ref{eq:ResonanceLDEI}) in Fig. \ref{fig:LDEIStrat}. 

The classical allowable range of LDEI is $0 \leq f\leq 4$ \citep[e.g.][]{cebron2012libration}, in which only triadic couplings of inertia-gravity waves $[\mathcal{H}_1, \mathcal{H}_2,]$ are involved. 
In this frequency range, the instability is trapped along critical latitudes for strong enough stratification when $N_0/\Omega_\text{s} \gg 1$. Similar to the non-synchronised configurations, it turns out that the largest growth rate is unaffected by the ratio $N_0/\Omega_\text{s}$ on these critical latitudes. Thus, they are predicted by the diffusionless formula obtained in neutral fluids \citep[see formula 4 in][]{cebron2012libration}. 

In the other frequency range $f>4$, LDEI is only due to triadic couplings of internal-gravity waves $[\mathcal{E}_1, \mathcal{E}_2]$ modified by rotation. Moreover, the instability only exists for strong enough stratification ($N_0/\Omega_\text{s} \gg 1$). 

\subsection{Asymptotic growth rates of the LDEI}
As in Sect. \ref{subsec:analyticalsigma} and Sect. \ref{subsec:analyticalsigma_polar}, the local stability analysis provides analytical expressions of the diffusionless growth rates in the equatorial plane and on the rotation (polar) axis. In the equatorial plane, the resonance condition (\ref{eq:ResonanceLDEI}) becomes
  \begin{equation}
\sqrt{4+ \widetilde{N}_0^2 x_0^2} \, \, \cos \theta_0 = \pm \frac{f}{2}, \label{eq:jc21}
 \end{equation}
whereas on the rotation axis we have
  \begin{equation}
\sqrt{4\,  \cos^2 \theta_0+ \widetilde{N}_0^2 x_0^2 \sin^2 \theta_0} \, \, = \pm \frac{f}{2}.
 \end{equation}
Then, the diffusionless growth rate in the equatorial plane is
\begin{equation}
\sigma=  \left(1+\frac{f^2}{16} \right) \frac{|\beta_0-\widetilde{N}_0^2 x_0^2 (\beta_0-\beta_1)|  }{4+ \widetilde{N}_0^2 x_0^2 }\, \epsilon_l
\label{eq:gLDI}
\end{equation}
for a general baroclinic background state $\beta_0 \neq \beta_1$. On the rotation axis, the diffusionless growth rate is given by
     \begin{equation}
\sigma = \frac{(16+f^2)(1-4 \widetilde{N}_0^2 x_0^2 f^{-2})}{16 \, (4- \widetilde{N}_0^2 x_0^2  )} \,  \beta_0 \epsilon_l . 
\label{eq:ze56t2}
 \end{equation}
Naturally, we recover equation (4) of \citet{cebron2012libration}, obtained for neutral fluids ($\widetilde{N}_0=0$). Note that equation (25) of \citet{cebron2013elliptical}, obtained in the equatorial plane for a buoyancy force of the order $\beta_0$, is not recovered by equation (\ref{eq:gLDI}). Indeed, their equation (25) is approximate because they artificially set $\theta_0$ to its hydrodynamic value $2 \cos \theta_0= \pm f/2$, instead of using its exact value given by equation (\ref{eq:jc21}). Finally, by analogy with the arguments given in the main text for non-synchronised systems, the largest diffusionless growth rate in the stellar interior will be insensitive to the strength of stratification, yielding the value for neutral fluids \citep{cebron2012libration,cebron2013elliptical,vidal2017inviscid} recovered in formula (\ref{eq:gLDI}) when $\widetilde{N}_0=0$.

Note finally that formula (\ref{eq:DecayFMC_TDEI}) also provides exactly the Joule damping rate of the LDEI in neutral fluids ($\widetilde{N}_0=0$). Besides, formulas of \citet{cebron2012magnetohydrodynamic,cebron2012elliptical} are recovered in the limit $|\boldsymbol{k}_0| \gg 1 $ by using the LDEI resonance condition to set $\theta_0$, that is $\cos \theta_0=\pm f/4$ when $\widetilde{N}_0=0$.
 
\subsection{Mixing-length theory}
We can build a mixing-length theory to get a phenomenological prescription of the turbulent mixing in weakly eccentric synchronised orbits, by analogy with non-synchronised orbits. The main difference with non-synchronised systems is that the typical turbulent velocity $u_\text{t}$ should scale as \citep{favier2015generation,grannan2016tidally}
\begin{equation}
	u_\text{t} \propto \alpha_1 \epsilon_l \beta_0 r_\text{l} \, \Omega_\text{s}. 
	\label{eq:uturbmlt_LDEI}
\end{equation}
Then, the turbulent prescription becomes
\begin{equation}
	\tau_\text{t} \propto \frac{K_\alpha}{\epsilon_l^2 \beta_0^2 \, \Omega_\text{s}}
	\label{eq:timeturmlt_LDEI}
\end{equation}
with the numerical pre-factor $K_\alpha \sim 30-50$ as in expression (\ref{eq:timeturmlt}), which is based on the numerical pre-factors of formulas (\ref{eq:etaturmlthighN}). 
Hence, the timescale for the turbulent Ohmic diffusion of the fossil field ought to be reduced in synchronised systems (compared to non-synchronised ones) by using formula  (\ref{eq:timeturmlt_LDEI}).

\end{appendix}

\end{document}